\documentclass[12pt,a4paper]{article}
\usepackage{jheppub}
\usepackage[utf8]{inputenc}
\usepackage{amsfonts}
\usepackage{slashed}
\usepackage{xcolor}

\usepackage[english]{babel}

\usepackage{changes}

\newcommand{\inc}{\textrm{inc}}
\newcommand{\coh}{\textrm{coh}}

\preprint{CPHT-RR056.102022}

\title{Zero sound and higher-form symmetries \\in compressible holographic phases}

\author[a]{Richard A. Davison}
\emailAdd{r.davison@hw.ac.uk}
\affiliation[a]{Department of Mathematics and Maxwell Institute for Mathematical Sciences, Heriot-Watt University, Edinburgh EH14 4AS, U.K.}
\author[b]{Blaise Gout\'eraux}
\emailAdd{blaise.gouteraux@polytechnique.edu}
\affiliation[b]{CPHT, CNRS, Ecole polytechnique, IP Paris, F-91128 Palaiseau, France}
\author[c]{Eric Mefford}
\emailAdd{ericmefford@uvic.ca}
\affiliation[c]{Department of Physics and Astronomy, University of Victoria, Victoria, BC V8W 3P6, Canada}

\arxivnumber{2210.14802}
\date{\today}

\abstract{
Certain holographic states of matter with a global U(1) symmetry support a sound mode at zero temperature, caused neither by spontaneous symmetry breaking of the global U(1) nor by the emergence of a Fermi surface in the infrared. In this work, we show that such a mode is also found in zero density holographic quantum critical states. We demonstrate that in these states, the appearance of a zero temperature sound mode is the consequence of a mixed `t Hooft anomaly between the global U(1) symmetry and an emergent higher-form symmetry. At non-zero temperatures, the presence of a black hole horizon weakly breaks the emergent symmetry and gaps the collective mode, giving rise to a sharp Drude-like peak in the electric conductivity. A similar gapped mode arises at low temperatures for non-zero densities when the state has an emergent Lorentz symmetry, also originating from an approximate anomalous higher-form symmetry. However, in this case the collective excitation does not survive at zero temperature where, instead, it dissolves into a branch cut due to strong backreaction from the infrared, critical degrees of freedom. We comment on the relation between our results and the application of the Luttinger theorem to compressible holographic states of matter.}
\begin{document}

\maketitle

\newpage

\section{Introduction and summary of results}

\subsection{Introduction}

The Landau paradigm classifies phases of matter depending on their pattern of spontaneous symmetry breaking: an ordered phase is separated from a disordered phase by a critical point, where the dynamics are governed by fluctuations of the order parameter. Over the years, many phases have been characterized as falling outside the remit of the Landau paradigm, including quasi-long-ranged ordered two-dimensional superfluids \cite{Lubensky}, quantum topological phases of matter \cite{Wen:2017}, and the deconfined quantum critical points mediating a continuous transition between two phases with different order parameters, for instance in quantum antiferromagnets \cite{Senthil:2003eed,Senthil:2003bis}. In the latter case, the quantum critical theory features an emergent topological conservation law and deconfined gapless degrees of freedom with fractional quantum numbers. 

{}In recent developments, the Landau paradigm has been extended to capture such situations using the concept of higher-form (or `generalized') global symmetries \cite{Gaiotto:2014kfa} (see \cite{McGreevy:2022oyu} for a review). Familiar global continuous symmetries generate a charge that counts point-like objects and give rise to a conserved $1$-form current. These are $0$-form symmetries. Higher-form symmetries are instead related to spatially extended objects: a $p$-form symmetry gives rise to a $(p+1)$-form conserved current. Such symmetries arise, for example, in $(d+1)$-dimensional superfluids where a continuous global $U(1)$ $0$-form symmetry is spontaneously broken. In the absence of mobile topological defects, the conservation of the winding of the superfluid phase can be reformulated as an emergent $(d-1)$-form symmetry.\footnote{Higher-form symmetries also arise in the description of magnetohydrodynamics \cite{Grozdanov:2016tdf,Grozdanov:2017kyl,Hofman:2018lfz,Grozdanov:2018fic,Armas:2018ibg,Armas:2018atq,Armas:2018zbe} and crystalline solids \cite{Grozdanov:2018ewh,Armas:2019sbe}.}

{}In fact, the emergent $(d-1)$-form symmetry of superfluids is not exact \cite{Hofman:2017vwr}: it exhibits a mixed `t Hooft anomaly with the $0$-form $U(1)$ symmetry. This has profound consequences: the anomaly is ultimately responsible for the emergence of a gapless, propagating excitation -- the superfluid sound mode \cite{Delacretaz:2019brr} -- which in turn is responsible for a dissipationless $\delta(\omega)$ contribution to the electric conductivity $\sigma(\omega)$. An alternative to Goldstone's theorem can be established that does not require spontaneous symmetry breaking \cite{Delacretaz:2019brr} -- all that is required is the emergence of the anomalous higher-form symmetry described above. 

{}Emergent anomalous symmetries also arise at zero temperature in Luttinger and Fermi liquids, as recently emphasized in \cite{Else:2020jln}. In a Luttinger liquid, the left- and right-moving charge densities are separately conserved. In the presence of an external electric field, the axial combination is anomalous which in turn directly implies the existence of a propagating mode. In a Fermi liquid, the anomaly is only present at the linearized level \cite{Delacretaz:2022ocm}, but plays a similar role. The anomaly is also crucial in order for the state to be compressible, i.e.~for its density per unit lattice cell to be continuously tunable rather than an integer.

{}Similar collective propagating modes exist in $(d+1)$-dimensional holographic phases of quantum matter described by Dirac-Born-Infeld (DBI) actions and generalized Maxwell actions at zero density \cite{Karch:2008,Nickel:2010pr,Hoyos-Badajoz:2010ckd,Davison:2011ek,WitczakKrempa:2013ht,Witczak-Krempa:2013aea,Edalati:2013tma,Tarrio:2013tta,Davison:2014lua,Chen:2017dsy,Grozdanov:2018fic,Gushterov:2018spg}. These holographic modes were dubbed `holographic zero sound' \cite{Karch:2008}, in analogy to the collective excitation present in a zero temperature Fermi liquid \cite{landau1980course}. In this work, we investigate Maxwell actions with a running coupling, to linear order in perturbations. Building in particular on \cite{Nickel:2010pr}, we will demonstrate that at zero density, an anomalous $(d-1)$-form conservation law emerges in the infrared description of these states and is ultimately responsible for the presence of a propagating mode in their spectrum at zero temperature.

{}The emergent higher-form conservation law of a superfluid is violated in the presence of mobile vortices. These are gapped in the infrared and so only affect the dynamics at high enough energies. For example, in $(2+1)$-dimensions, they give a finite lifetime to the superfluid Goldstone above the Berezenski-Kosterlitz-Thouless temperature and destroy the associated superfluid sound mode of the infrared theory \cite{Bardeen:1970}. Traditional gapless superfluid hydrodynamics \cite{Lubensky} can be augmented to capture the relaxation due to vortices \cite{Davison:2016hno,Delacretaz:2019brr}. This transition falls within the Landau paradigm after its extension to incorporate higher-form symmetries and their anomalies \cite{Delacretaz:2019brr}.

{}Similarly, at any non-zero temperature, the presence of a black hole horizon in the holographic phases explicitly breaks the emergent $(d-1)$-form symmetry and removes the holographic zero sound mode from the low energy spectrum. At low temperatures the symmetry is broken weakly, which leaves a strong imprint in the electric conductivity of the state in the form of a sharp Drude-like peak. The fact that such Drude-like peaks are caused by an approximate higher-form symmetry was previously demonstrated in \cite{Chen:2017dsy,Grozdanov:2018fic}.

{}It is important to ask whether the higher-form symmetry persists for holographic states with a non-zero charge density. Indeed, it is known that, as long as the state exhibits an emergent Lorentz symmetry, the electrical conductivity will exhibit an extra Drude-like contribution that is characteristic of an approximate symmetry (in addition to the delta function arising from momentum conservation) \cite{Davison:2018ofp,Davison:2018nxm}. Here, we demonstrate that this Drude-like conductivity is in fact the consequence of an approximate anomalous higher-form conservation law that gives rise to a long-lived collective excitation at low temperatures, though, for reasons we illustrate, the excitation does not survive at zero temperature. 

{} Finally, compressibility and the presence of anomalous higher form symmetries are deeply related to charge fractionalization and the Luttinger theorem{, which in its simplest instance states that the microscopic charge  of a metal with a Fermi liquid fixed point is equal the volume of the Fermi surface. The anomaly of the emergent loop group of the Fermi liquid provides the link between the microscopic charge per unit cell and properties of the infrared effective theory, \cite{Else:2020jln}. A similar result is obtained in a superfluid, where there is also an emergent anomalous (higher-form) symmetry. Motivated by extensions of the Luttinger theorem to phases with fractionalized degrees of freedom, where the microscopic charge is equal the sum of the volume of the Fermi surface and a contribution from the fractionalized excitations, \cite{Oshikawa_2000,Senthil:2003sqj,PhysRevB.70.245118} and the connection of horizons to deconfinement \cite{Witten:1998zw}, such charged horizons have been argued to be composed of fractionalized excitations (see \cite{Hartnoll:2011fn} and references therein). We point out that in compressible holographic states, whether or not they support a long-lived excitation at low temperatures due to an approximate emergent higher-form symmetry of the kind discussed here, the boundary charge density remains carried entirely by the black hole horizon. Nevertheless, identifying an emergent anomalous symmetry in holographic compressible states would allow to write down their low-energy effective theory as well as illuminate the connection between the Luttinger theorem and the nature of the degrees of freedom of charged black holes.}

 A full understanding of emergent higher-form symmetries in holographic compressible states would therefore illuminate the connection between the Luttinger theorem and the nature of the degrees of freedom of charged black holes.

{}In the remainder of this Section, we proceed to give a more detailed summary of our results, followed by a discussion and outlook, where amongst other things, we offer more comments on the relation between our results and the Luttinger theorem. Technical derivations are given in the subsequent sections and appendices.

\subsection{Summary of results}

\paragraph{Holographic setup:}We will study compressible phases of quantum matter that are described by gauge/gravity duality \cite{Ammon:2015wua,Zaanen:2015oix,Hartnoll:2016apf}. This duality provides a way of modeling metallic phases without long-lived quasiparticles that is complementary to other treatments such as theories of non-Fermi liquid quantum critical metals \cite{Lee:2017njh} or the Sachdev-Ye-Kitaev model and its generalizations \cite{Chowdhury:2021qpy}. In the large-$N$ limit, the gravitational description of these phases makes manifest the renormalization group flow and allows the Lorentzian signature correlation functions to be directly computed. While in some examples the microscopic quantum field theory degrees of freedom can be explicitly identified \cite{Maldacena:1997re,Aharony:2008ug}, we will extrapolate beyond just these particular examples and study a more general class of phases with the same type of infrared symmetries.

{}We focus in particular on $(d+1)$-dimensional compressible states of quantum matter that arise when a holographic conformal field theory with a global $0$-form $U(1)$ symmetry is deformed by a relevant scalar operator. The holographic duals to these states are asymptotically Anti de Sitter (AdS) solutions to Einstein-Maxwell-scalar theories of gravity, characterized by the emergence of a scale-covariant metric
\cite{Charmousis:2010zz,Huijse:2011ef,Gouteraux:2011ce}
\begin{equation}
\label{scalingEMDgeometries}
ds^2=R^{\frac{2\theta}{d}}\left(-\frac{dt^2}{R^{2z}}+\frac{d\vec{x}^2+dR^2}{R^2}\right),
\end{equation}
in the infrared of the spacetime, where $R$ is the radial coordinate and $(t,\vec{x})$ the coordinates of the $(d+1)$-dimensional Minkowski spacetime where the state lives. This infrared spacetime is the gravitational representation of infrared quantum critical degrees of freedom, characterized by the dynamical critical exponent $z$ and hyperscaling violation exponent $\theta$. These scaling exponents control the temperature dependence of thermodynamic observables, for instance the entropy density $s\sim T^{(d-\theta)/z}$. It can be helpful to think of $\theta$ as setting the effective dimensionality $d_{\text{eff}}=(d-\theta)$ of these degrees of freedom. These infrared spacetimes are not the result of fine-tuning: each one typically arises for a continuous range of values of the ultraviolet couplings and so the corresponding states constitute a critical line \cite{Hartnoll:2011pp,Adam:2012mw,Gouteraux:2012yr,Gouteraux:2019kuy,Gouteraux:2020asq}.

{}Our main result is to demonstrate that, at low temperatures, an approximate global $(d-1)$-form symmetry emerges for linearized perturbations around large classes of these states. As in superfluids, this emergent symmetry exhibits a mixed `t Hooft anomaly with the $0$-form $U(1)$ symmetry. As a consequence, these states share many of the properties of superfluids despite the fact that the $0$-form $U(1)$ symmetry is not spontaneously broken. This occurs for states with both zero and non-zero density of the $0$-form charge, as we will now describe. We also discuss the impact of this emergent symmetry on the zero temperature spectrum, which is very different for zero and non-zero density states.

{}Throughout this work, Greek indices $\mu,\nu,\ldots=1,\ldots,d+1$ run over all field theory spacetime coordinates, while Latin indices $i,j,\ldots=1,\ldots,d$ run over spatial field theory coordinates. At non-zero wavevector $\vec{k}$, Latin indices $a,b,\ldots$ run over the field theory spatial coordinates transverse to $\vec k$. Finally, capital Latin indices $M,N,\ldots=1,\ldots,d+2$ run over the bulk spacetime coordinates.

\paragraph{Zero density:}For the zero density states, at zero temperature and at the lowest energies the linearized dynamics of the charges is governed by the anomalous conservation equations \begin{equation}
\label{effectiveeomszerodensity}
\partial_\mu j^\mu=0\,,\qquad\quad\quad\quad\quad \partial_\mu K^{\nu_1\ldots\nu_{d-1}\mu}= \frac12\epsilon^{\nu_1\ldots\nu_{d-1}\kappa\lambda}\bar{F}_{\kappa\lambda}\,.
\end{equation}
Here $j^\mu$ is the conserved $1$-form current that derives from the $0$-form $U(1)$ symmetry, while $K^{\mu\nu_1\ldots\nu_{d-1}}$ is the $d$-form current that derives from the $(d-1)$-form symmetry and $\bar{F}_{\kappa\lambda}=\partial_{\kappa}\bar{A}_{\lambda}-\partial_\lambda\bar{A}_\kappa$ is the field strength of the external gauge field that couples to $j^\mu$. These anomalous conservation equations are identical to those of a superfluid with frozen temperature and velocity fluctuations \cite{Delacretaz:2019brr}. These states support a propagating, Goldstone-like mode with dispersion relation
\begin{equation}
\label{eq:zeroTdispersionintro}
\omega(k)=\pm v k-i \gamma k^{1+\alpha}+\ldots.
\end{equation}
This can be understood as following from the aforementioned alternative to Goldstone's theorem \cite{Delacretaz:2019brr}. In contrast to zero temperature superfluids, the velocity $v$ is non-universal \cite{landaufluid}. Similarly, the exponent $\alpha$ governing the attenuation can be continuously tuned, unlike in a superfluid where the attenuation scales like $k^{d+2}$ due to phonon scattering.

{}The attenuation in \eqref{eq:zeroTdispersionintro} arises due to a deformation of the universal infrared theory \eqref{effectiveeomszerodensity} that explicitly breaks the $(d-1)$-form symmetry, and which originates from coupling to the quantum critical degrees of freedom associated to the infrared spacetime. This correction is irrelevant when $\alpha>0$, which is therefore the condition for the emergence of this symmetry in the infrared. In this sense, the state is less robust than a superfluid, where the explicit breaking is exponentially suppressed at small temperatures. This deformation governs the leading dissipative part of the $T=0$ optical conductivity
\begin{equation}
\label{conductivityzerodensityzeroT}
\text{Im}\,\sigma(\omega)=\frac{\chi_{JJ}}{\omega}+\ldots,\quad\quad\quad\quad\quad \text{Re}\,\sigma(\omega)\propto\omega^{\alpha-1}+\ldots,
\end{equation}
where $\chi_{JJ}$ is the $0$-form current susceptibility and the constant of proportionality in the second expression is related to a universal scaling function of the infrared quantum critical degrees of freedom. We emphasize that the irrelevant deformation is crucial for certain low energy properties like the dissipative part of the conductivity \eqref{conductivityzerodensityzeroT}, as it breaks a symmetry, and so is `dangerously' irrelevant.

{}At small non-zero temperatures, the state is governed by a theory similar to superfluid hydrodynamics. The weak explicit breaking of the $(d-1)$-form symmetry due to the irrelevant deformation modifies the effective conservation equations \eqref{effectiveeomszerodensity} to
\begin{equation}
\label{eomszerodensity}
\partial_\nu  K^{\mu_1\ldots\mu_{d-1}\nu}=\frac12\epsilon^{\mu_1\ldots\mu_{d-1}\kappa\lambda}\bar{F}_{\kappa\lambda}+\frac1\tau u_\nu K^{\mu_1\ldots\mu_{d-1}\nu},
\end{equation}
in the absence of an external magnetic field, where $u^\mu=(1,\vec{0})$ is a fixed timelike unit vector. The relaxation timescale $\tau$ is controlled by the irrelevant deformation and is parametrically long compared to the inverse temperature $T\tau\sim T^{-\alpha}\gg1$. This is the same approximate conservation law in phase-relaxed superfluids due to the presence of free vortices above the BKT temperature \cite{Davison:2016hno,Delacretaz:2019brr}. Correspondingly, as in a phase-relaxed superfluid, there is a crossover between slowly attenuating sound modes at frequencies $\omega\tau\gtrsim1$ with dispersion relations
\begin{equation}
\omega=\pm vk-\frac{i}{2}\tau^{-1}+\ldots,
\end{equation}
and diffusive and relaxational modes at frequencies $\omega\tau\lesssim1$ with dispersion relations
\begin{equation}
\omega=-\frac i{\tau}+\ldots\,,\quad\quad\quad\quad\quad\quad\quad\quad \omega=-iv^2\tau k^2+\ldots,
\end{equation}
where $\ldots$ denote subleading terms in a small $\omega\sim k\sim \tau^{-1}\ll T$ expansion. The electrical conductivity
\begin{equation}
\label{acconductivityzerodensityTnonzero}
\sigma(\omega)=\frac{\sigma_{dc}}{1-i\omega\tau}\,,\qquad\qquad \sigma_{dc}=\chi_{JJ}\tau,\qquad\qquad T\tau\sim T^{-\alpha}\,,
\end{equation}
has a sharp Drude-like peak of width $\tau^{-1}\ll T$. We emphasize that this is completely unrelated to the translational symmetry of the state -- in these zero density states there is no overlap between the electric current and momentum operators. Instead, it is a consequence of the (approximate) higher-form conservation law \eqref{eomszerodensity}. The susceptibility is $\chi_{JJ}\sim T^0$ and so the $T$ dependence of the dc conductivity is entirely controlled by the timecale $\tau$. Specifically $\sigma_{dc}\sim T^{-\alpha-1}$ and so there is an emergent symmetry for states with a large dc conductivity $\sigma_{dc}\gg 1/T$ at low $T$. In the limit $\omega\tau\gg1$, \eqref{acconductivityzerodensityTnonzero} reproduces the dissipationless part of the $T=0$ conductivity \eqref{conductivityzerodensityzeroT}.  

{}These properties should be contrasted with the $\alpha<0$ cases, where the infrared symmetry is just the usual $0$-form $U(1)$ symmetry and no higher-form symmetry emerges. In these cases the electrical conductivity at zero temperature has no $i/\omega$ contribution but instead the low frequency form $\sigma(\omega)\sim\omega^{-\alpha-1}$. At small non-zero temperatures, the conductivity is a universal scaling function of $\omega/T$, with $\sigma_{dc}\sim T^{-\alpha-1}\ll 1/T$ and no Drude-like peak.

{}These infrared dynamics are captured by the effective action
\begin{equation}
\label{IRaction}
S=\int d^{d+1}x\left[-\frac{\chi_{\rho\rho}}{2}\left(\partial_t\varphi+a_t-\bar{A}_t\right)^2+\frac{\chi_{JJ}}{2}\left(\partial_i\varphi+a_i-\bar{A}_i\right)^2\right]+\int d\omega d^dk \frac{f_{ti}^2}{2g(\omega,T)},
\end{equation}
where $\chi_{\rho\rho}$ is the 0-form charge susceptibility and $\varphi$ is the Goldstone-like field associated to the emergent $(d-1)$-form symmetry. It is coupled to an external gauge field $\bar{A}_\mu$ (the source for the $0$-form $U(1)$ current) and an emergent dynamical gauge field $a_\mu$ with $f=da$. The coupling to the emergent gauge field explicitly breaks the higher-form symmetry: in superfluid language it acts as an electric field in the Josephson equation and relaxes $\varphi$. The Maxwell-like term for the emergent gauge field arises from integrating out the near-horizon spacetime \eqref{scalingEMDgeometries} representing the quantum critical degrees of freedom and in general is neither local nor Lorentz invariant. The effective electromagnetic coupling $g(\omega,T)=-\omega^2 G^{-1}_{IR}(\omega,T)$ is controlled by the retarded Green's function $G_{IR}(\omega,T)$ of an operator in the critical theory. This is a universal scaling function $G_{IR}(\omega,k,T)=\omega^{-\alpha}h(T/\omega)$, where $\alpha$ is related to the dimension of the operator. The precise value of this dimension (and hence whether the higher-form symmetry emerges in the infrared) depends on the details of the holographic theory: essentially, the higher-form symmetry emerges when the bulk electromagnetic coupling is small enough near the horizon. The relation of effective couplings to lifetimes was emphasized in \cite{Ghosh:2020lel}.

{}In cases with an emergent symmetry, the interpretation of the quantum critical degrees of freedom represented by the infrared spacetime \eqref{scalingEMDgeometries} requires some care. The correct infrared theory is obtained by imposing mixed boundary conditions on the bulk Maxwell field in this spacetime. From the perspective of holographic renormalization, this means that the identification of operator dimensions in this spacetime requires alternate quantization, and that the naive action must be supplemented by relevant double-trace deformations of these operators. The importance of mixed boundary conditions for higher-form symmetries (in the ultraviolet) was discussed in \cite{Hofman:2017vwr,Grozdanov:2017kyl,Grozdanov:2018ewh}. Alternate boundary conditions also play an important role in the holographic description of phases with spontaneous symmetry breaking, where they are satisfied by the bulk field dual to the phase of the order parameter -- see e.g. \cite{Amoretti:2017frz,Alberte:2017oqx} for the case of broken translations. 

\paragraph{Non-zero density:}We now turn to holographic states with a non-zero density of the $0$-form $U(1)$ charge, which have important differences to those described above. In the infrared, the corresponding spacetimes still have the form \eqref{scalingEMDgeometries} and are solutions to equations of motion that either neglect terms involving the bulk Maxwell field or include such terms. It is the former case that will be of interest to us. The corresponding infrared spacetimes necessarily have dynamical exponent $z=1$, and $\theta<0$. Furthermore, changing the density of $0$-form charge corresponds to deforming the infrared theory by an irrelevant operator whose coupling has dimension $\Delta_A<0$. 

{}At low temperatures, these states support long-lived excitations that carry $0$-form $U(1)$ charge. The conductivity of this charge at low temperatures and frequencies is \cite{Davison:2018ofp,Davison:2018nxm}
\begin{equation}
\label{acconductivitynonzerodensity}
\sigma(\omega)=\frac{\sigma_{\inc}^{dc}}{1-i\omega\tau}+\sigma_{\coh}(\omega)\,,\quad\quad\quad\quad \sigma_{\inc}^{dc}=\chi_{J_\inc J_\inc}\tau\,,\quad\quad\quad\quad T\tau\sim T^{2\Delta_A}\,.
\end{equation}
$\sigma_\coh(\omega)=\left(\chi_{JP}^2/\chi_{PP}\right)\left(i/\omega\right)$ is the contribution of coherent processes which drag momentum, where $\chi$ denote the static susceptibilities of the current $J$ and momentum $P$ operators. In the zero density states discussed before, $\chi_{JP}=0$ and so this contribution vanishes. The remaining Drude-like term arises from processes that do not drag momentum. It originates from the `incoherent' part of the current $J_\inc\equiv\chi_{PP} J-\chi_{JP}P$, so named because it does not overlap with the momentum: $\chi_{J_\inc P}=0$. In \cite{Davison:2018ofp,Davison:2018nxm} we argued that \eqref{acconductivitynonzerodensity} followed from slow relaxation of $J_{\text{inc}}$ over a timescale $\tau\gg T^{-1}$ governed by the irrelevant coupling $\Delta_A$. We showed that at low frequencies, there is thermal diffusion with diffusivity $D_T=c_{IR}^2\tau/(d-\theta)$, and conjectured that when $\omega\tau\gtrsim 1$ long-lived propagating modes should emerge with velocity $v_\inc^2=c_{IR}^2/(d-\theta)$ where $c_{IR}$ is the infrared speed of light.

{}Here we show that these expectations are borne out by deriving the linearized effective theory governing the low temperature dynamics of these states. Similarly to the zero density states above, at low but non-zero temperatures, the theory features an emergent, approximate $(d-1)$-form symmetry, mixing with the $0$-form $U(1)$ symmetry through a `t Hooft anomaly. The associated $d$-form current is the Hodge dual of the incoherent current density. The emergent higher-form symmetry is only approximate as it is weakly broken by the irrelevant coupling. At intermediate times and distances $\tau^{-1}\ll\omega, k\ll T$ these states indeed support (in addition to the normal sound modes) a pair of propagating modes which attenuate slowly at a rate governed by the dangerously irrelevant coupling. At the longest times and distances $\omega, k\ll\tau^{-1}$ the effects of symmetry breaking become important and these modes mutate into diffusive and relaxational modes, while at shorter times and distances $\omega, k\gg T$ the effects of other excitations become important.

{}There are key differences between zero and non-zero density states. Firstly, the velocity of the propagating mode is now universal: $v_\inc^2=c_{IR}^2/(d-\theta)$. Accounting for the effective dimensionality $d_{\text{eff}}=d-\theta$ of our states, this is analogous to the velocity of second sound in a superfluid, which takes the universal value $v_\inc^2=1/d$ at low temperatures. This reflects the fact that this subsector of the effective theory is approximately described by a superfluid Goldstone-like action in this regime of energy scales. 

{}Secondly, the static susceptibility of the $0$-form current $\chi_{J_\inc J_\inc}\sim T^{d+1-\theta}$ now vanishes at zero temperature. Equivalently, the static susceptibility of the $(d-1)$-form charge $\chi_{KK}$ diverges at low temperature as $T^{\theta-(d+1)}$. This is also in stark contrast with superfluids, where the superfluid density $\sim \chi_{KK}$ is non-vanishing at zero temperature. This important difference is rooted in the linearized constitutive relation for the higher-form current, which reads $(\star K)^t=\delta\mu-(\mu/T) \delta T$ in our case, while $(\star K)^t=\delta \mu$ for a superfluid \cite{Delacretaz:2019brr}. The divergent susceptibility of the $(d-1)$-form charge constitutes an example of critical drag{, which has been recently argued in \cite{Else_2021,Else:2021dhh} to be relevant for strange metallic transport in high $T_c$ superconductors.} The emerging symmetry tends to produce a large $\sigma_{\inc}^{dc}$ due to the diverging relaxation time, while the diverging $(d-1)$-form susceptibility tends to produce a small $\sigma_{\inc}^{dc}$. The result of this is that $\sigma_{\inc}^{dc}\sim T^{d-\theta+2\Delta_A}$, which may vanish or diverge at low temperatures depending on whether $\theta-d<2\Delta_A<0$ or $2\Delta_A<\theta-d<0$. This is in contrast to the zero density case where the emergent symmetry is always correlated with a diverging dc conductivity. 

{}However, the most important difference from the zero density states, and from superfluids, is that the emergent propagating mode does not survive at zero temperature.\footnote{We thank Dominic Else for discussions on the fate of the emergent higher-form symmetry at zero temperature.} Instead, at zero temperature the infrared dynamics is dominated by the quantum critical degrees of freedom associated to the near-horizon spacetime. This is consistent with the absence of a delta function contribution in the $T=0$ optical conductivity \cite{Davison:2018ofp,Davison:2018nxm}
\begin{equation}
\sigma_\inc(\omega,T=0)\sim \omega^{d-\theta-2\Delta_A}+\ldots,
\end{equation}
and we present evidence that the low energy response function of the $0$-form charge density at $T=0$ is characterized by branch points at $\omega=\pm c_{IR}k$.

\subsection{Discussion and outlook}

\paragraph{Other holographic examples of `zero sound':}There are further examples of holographic theories that exhibit a slowly relaxing current, to which our results could be extended: those with higher-derivative \cite{Myers:2010pk,Witczak-Krempa:2013aea} or probe DBI \cite{Karch:2008,Chen:2017dsy,Gushterov:2018spg} actions for the gauge field. For the probe DBI cases, it was argued in \cite{Chen:2017dsy} using the square root form of the action that the non-linear effective theory is qualitatively different than the linearized one. Our zero density states exhibit a slowly relaxing current without a square root action and so it would be very interesting to determine whether the linearized effective theory we have derived can be smoothly extended to include non-linear effects, and whether the emergent higher-form conservation law persists. For probe DBI examples, incorporating an external magnetic field in the effective theory we have described should allow one to interpret the results obtained for the ac conductivity in \cite{Jokela:2017ltu,Jokela:2021uws}. Higher-derivative and massive gravity theories \cite{Grozdanov:2016vgg,Alberte:2017cch} also support emergent long-lived modes when a higher-derivative coupling is made comparable to the leading Einstein term. We expect that such modes can be understood by a suitable extension of our results.

\paragraph{Effective theories of holographic matter:} Elucidating the effective theories governing the low energy, low or zero temperature dynamics of holographic matter is an essential step in order to connect to non-holographic phases of matter. This program was initiated in \cite{Faulkner:2010tq,Nickel:2010pr,Faulkner:2010jy}, mostly in the probe limit where the backreaction of scalar, gauge or fermionic probes on the spacetime geometry is neglected. There is good reason for this, as with this approximation much analytical control is gained. This is clear from our results as well, as the probe limit allowed us to write down very explicitly the effective field theory governing the zero density holographic states at low and zero temperature. On the other hand, we saw that when backreaction is included (at non-zero density), the coupling to the infrared, critical degrees of freedom cannot be neglected at zero temperature, even when it is weak at small, non-zero temperatures. This had an important impact on the spectrum: the would-be propagating mode observed at non-zero temperatures dissolves into a branch cut at zero temperature.

{}Here we have focussed on the case of non-zero density states with an emergent Lorentz-invariant, hyperscaling-violating infrared. Holographic states with an emergent AdS$_2\times R^d$ infrared metric play a prominent role in applications of holography to strongly coupled condensed matter systems \cite{Zaanen:2015oix,Hartnoll:2016apf} and are closely related to Sachdev-Ye-Kitaev models of non-Fermi liquids \cite{Chowdhury:2021qpy}. These states exhibit gapless collective modes at zero temperature \cite{Edalati:2009bi,Edalati:2010hk,Edalati:2010pn,Davison:2011uk,Davison:2013bxa,Gushterov:2018spg,Moitra:2020dal,Arean:2020eus}, with sound velocities and diffusivities given by the naive $T\to0$ limit of the corresponding coefficients entering in their $T\neq0$ hydrodynamics. Given the results we obtained, we expect that this feature can be explained by constructing the zero temperature effective holographic theory of these states, including the backreaction of the critical degrees of freedom (see \cite{Nickel:2010pr,Maldacena:2016upp,Engelsoy:2016xyb,Jensen:2016pah} for related work on coupling AdS$_2$ degrees of freedom to holographic matter).

{}Fermionic probes of holographic states also reveal the existence of Fermi surfaces and associated (non-Fermi liquid) gapless collective modes, depending on the details of the underlying spacetime and of the fermionic action \cite{Liu:2009dm,Cubrovic:2009ye,Iizuka:2011hg}. Studying whether these features survive when including backreaction would also be very interesting. However, fully backreacting the fermion fields in the bulk on the spacetime geometry remains an open challenge \cite{Hartnoll:2010gu,Cubrovic:2010bf,Hartnoll:2011dm,Cubrovic:2011xm,Allais:2012ye,Allais:2013lha,Medvedyeva:2013rpa,Chagnet:2022ykl}.

\paragraph{Deconfined quantum critical points:} The infrared physics we have described in the previous section bears some similarity to that near deconfined quantum critical points, which are also characterized by an emergent global symmetry. In (2+1)-dimensions and at zero density, our Goldstone-like mode can be dualized into a $U(1)$ gauge field for which the $0$-form symmetry is exact and the emergent $1$-form symmetry is broken by a dangerously irrelevant deformation. These symmetries resemble those at the deconfined phase transition between two valence bond solid phases, which is described by a theory of emergent spinons coupled to a gauge field. This critical point is characterized by an emergent $0$-form $U(1)$ symmetry that is broken by dangerously irrelevant monopole operators in the valence bond solid phase. The spinons are gapped at low energies and so the spectrum exhibits a (quadratically-dispersing) Goldstone-like mode described by the critical quantum Lifshitz model \cite{Vishwanath:2003}.

{}In contrast to this, at the deconfined quantum critical point separating the N\'eel anti-ferromagnetic phase from the resonating valence bond phase in (2+1)-dimensions \cite{Senthil:2003eed,Senthil:2003bis} the deconfined spinons are gapless. The coupling between the spinons and the emergent gauge field breaks the emergent electric $1$-form symmetry in the infrared. This destroys the would-be gapless mode, and correlation functions instead display a branch cut. These properties bear a resemblance to those of the non-zero density holographic quantum critical phases described above.

\paragraph{Holography, Luttinger theorem and fractionalized degrees of freedom:}At non-zero density, it has been emphasized that emergent anomalous symmetries have a deep relation to the Luttinger theorem \cite{Else:2020jln}. The Luttinger theorem states that the filling (the density per unit cell) of a quantum phase of matter on a lattice is given by the volume of the Fermi surface if the ground state is a Fermi liquid. It is thus a non-perturbative statement directly connecting a microscopic property of the state (the filling) and its infrared properties (the volume of the Fermi surface). A topological proof relying on threading a unit flux of magnetic field through one of the cycles of a periodic lattice was given in \cite{Oshikawa_2000}. When the infrared theory includes a non-trivial topological sector, such as in phases of matter featuring fractionalized degrees of freedom, there is no longer a one-to-one correspondence between the filling and the Fermi volume \cite{Senthil:2003sqj,PhysRevB.70.245118}. 

{}The applicability of the Luttinger theorem to non-zero density, compressible holographic phases of matter was investigated in a series of works \cite{Sachdev:2010um,Hartnoll:2010gu,Hartnoll:2010xj,Huijse:2011hp,Hartnoll:2011dm,Hartnoll:2011fn,Hartnoll:2011pp,Iqbal:2011bf,Huijse:2011ef}. In the absence of spontaneous breaking of the $0$-form $U(1)$ symmetry, the total density is the sum of the charge density in the bulk and of the black hole horizon. The Luttinger relation is most obviously recovered in cases when no charge is left on the horizon at zero temperature and all the density is carried by Fermi surface-forming bulk fermions \cite{Hartnoll:2010gu,Hartnoll:2010xj}. The charge of the horizon was correspondingly interpreted as `fractionalized', further motivated by the fact that the presence of a horizon is a holographic signature of the deconfinement of gauge fields \cite{Witten:1998zw,Aharony:2003sx}.

{}The total boundary charge density in the family of holographic states considered in this work is always equal to the charge of the horizon, independently of whether the state exhibits a long-lived excitation at low temperatures or not. On the other hand, it remains unclear how the charge behind the horizon should be understood from the Luttinger perspective. In particular, one would like to ascertain whether an emergent symmetry can explicitly be identified (different than the one discussed in the context of this work) that would lend further support to the interpretation of such degrees of freedom as fractionalized \cite{Hartnoll:2011fn}.

{} This would also presumably shed some light on the scaling theories that describe the infrared, low temperature dynamics of these states and the presence of large, anomalous dimensions for the charge density and current in the infrared effective theory \cite{Gouteraux:2014hca,Karch:2014mba,Davison:2018nxm} (see also \cite{LaNave:2019mwv} on the interpretation of these states as fractional electromagnetism). There is a long-standing debate on the origin of various scaling behaviours in transport observable in so-called strange metals, which are incompatible with conventional quantum critical scenarios and simple scale invariance (see \cite{Phillips:2022nxs} for a recent review). Identifying fractionalized excitations in holographic quantum critical metals would bring them a step closer to the unconventional quantum criticality of strange metals and theoretical models thereof based on fractionalized degrees of freedom and topological order \cite{Sachdev:2016qwg,Sachdev:2018ddg}.

\section{Equilibrium properties of the holographic states}
\label{sec:NonZeroDensityEMD}

{}We still study classical solutions of the Einstein-Maxwell-scalar theories with action
\begin{equation}
\label{eq:bulkEMDaction}
S=\int d^{d+2}x\sqrt{-g}\left(\mathcal{R}-\frac{1}{2}\partial_M\Phi\partial^M\Phi+V(\Phi)-\frac{Z(\Phi)}{4}F_{MN}F^{MN}\right),
\end{equation}
where $\mathcal{R}$ is the Ricci scalar, $F_{MN}=\nabla_M A_N-\nabla_N A_M$ is the field strength of the bulk $U(1)$ gauge field $A_M$ and $\Phi$ is a neutral scalar field. Specifically, we are interested in asymptotically AdS (Anti de Sitter) planar solutions supported by a running scalar field $\Phi=\Phi(r)$ and a radial electric field $A=A_t(r)dt$.  This allows a very large family of solutions dual to field theories with an ultraviolet (UV) fixed point that exhibit non-trivial renormalization group flow towards the infrared (IR), driven by a relevant scalar operator and a chemical potential for a conserved $0$-form $U(1)$ charge. We will consider cases where there are flows to a class of IR spacetimes characterized by a scaling symmetry that is not necessarily relativistic nor satisfies hyperscaling. The infrared theories are interpreted as a class of strongly-interacting quantum critical states of matter \cite{Goldstein:2009cv,Charmousis:2010zz,Gouteraux:2011ce,Huijse:2011ef,Gouteraux:2012yr,Gouteraux:2013oca,Davison:2018nxm} as we will now review. 

{}More specifically we consider spacetimes of the form
\begin{equation}
ds^2=-D(r)dt^2+C(r)d\vec{x}^2+B(r)dr^2,
\end{equation}
where $r$ is the radial coordinate. At the asymptotically AdS boundary $r\rightarrow0$ we require that
\begin{equation}
\begin{aligned}
\label{eq:generalFGexpansioneqm}
&\,B(r)\rightarrow r^{-2}+\ldots,\quad C(r)\rightarrow r^{-2}+\ldots,\\
&\,D(r)\rightarrow r^{-2}+\ldots,\quad A_t(r)\rightarrow \mu-\rho r+\ldots,
\end{aligned}
\end{equation}
where ellipses denote terms subleading in the $r\to0$ limit. $\mu$ and $\rho$ are the chemical potential and density of the conserved $U(1)$ charge. There are solutions of this type for potentials of the form $V(\Phi\rightarrow0)\rightarrow6+O(\Phi^2)$ and $Z(\Phi\rightarrow0)\rightarrow 1+O(\Phi^2)$, where we have set the UV AdS radius and the UV gauge coupling to unity without loss of generality. The asymptotic form of $\Phi(r)$ depends on the quadratic term in the potential $V$ (i.e.~on the scaling dimension of the relevant scalar operator). 

{}We consider solutions that furthermore have a planar horizon at $r=r_0>0$, near which they are of the form
\begin{equation}
\begin{aligned}
\label{eq:BlackHoleNHExpansion}
&\,B(r\rightarrow r_0)\rightarrow(4\pi T(r_0-r))^{-1}+\ldots,\quad C(r\rightarrow r_0)\rightarrow \left(\frac{s}{4\pi}\right)^{2/d}+\ldots,\\
&\,D(r\rightarrow r_0)\rightarrow 4\pi T(r_0-r)+\ldots,\quad\quad\,\,\, A_t(r\rightarrow r_0)\rightarrow A_h(r_0-r)+\ldots,\\
&\,\Phi(r\rightarrow r_0)\rightarrow \Phi_0+\ldots,
\end{aligned}
\end{equation}
where ellipses denote terms subleading in $(r_0-r)$ as $r\to r_0$. $T$ and $s$ correspond to the temperature and entropy density of the field theory state. Integrating the $t$-component of the bulk Maxwell equation between the horizon and the boundary gives a relation for $A_h$ in terms of the charge density
\begin{equation}
\label{bulkchargedensity}
0=\left(\frac{ZC^{d/2}A_t'}{\sqrt{BD}}\right)'\quad\Rightarrow\quad \rho=-\frac{ZC^{d/2}A_t'}{\sqrt{BD}}\underset{r=r_0}{=}\frac{s}{4\pi}Z(\Phi_0) A_h\,.
\end{equation}

{}The scaling symmetry of the infrared theory manifests itself in a scaling symmetry of the near-horizon metric at low temperatures. In order to see this, it is convenient to introduce an alternative radial coordinate $R(r)$, for which the horizon is located at $R(r_0)=R_0$. We assume that we can appropriately define such a coordinate in order that our solution is
\begin{equation}
\begin{aligned}
\label{eq:IRmetricFiniteT}
&\,ds^2=-f(R)\frac{L_t^2L^{2(z-\theta/d)}}{R^{2(z-\theta/d)}}dt^2+\frac{L_x^2L^{2(1-\theta/d)}}{R^{2(1-\theta/d)}}d\vec{x}^2+\frac{\tilde{L}^2L^{-2\theta/d}}{R^{2(1-\theta/d)}}\frac{dR^2}{f(R)},\\
&\,f(R)=1-\left(\frac{R}{R_0}\right)^{d+z-\theta},\\
\end{aligned}
\end{equation}
for $R_0>R\gg R_{UV}$. In other words, we consider solutions which, sufficiently close to the horizon, take the form \eqref{eq:IRmetricFiniteT}. $R_{UV}$ is typically set by the chemical potential and can be thought of as the UV cut-off of the IR scaling region. Its precise value is state-dependent and is not important for our purposes. The temperature $T$ may be written as
\begin{equation}
\label{eq:temp}
T=\frac{L_t}{\tilde{L}}\frac{(d+z-\theta)}{4\pi}\left(\frac{R_0}{L}\right)^{-z},
\end{equation}
and hence in the zero temperature limit ($f\rightarrow1$), the near-horizon metric manifestly exhibits {covariance under the scaling transformation $(t,R,\vec x)\mapsto(\lambda^z, \lambda R, \lambda \vec x)$,} characterized by the two exponents $z$ and $\theta$. $L$ is a length scale related to $R_{UV}$ and defined more precisely below in \eqref{runningPhi}. We will restrict to $d\geq2$ as well as $z\geq1$. In order to satisfy the null energy condition and to ensure positivity of heat capacity, we also require $d(z-1)-\theta\geq0$, $(z-1)(d+z-\theta)\geq0$ and $d-\theta\geq0$.  

{}Spacetimes of the form \eqref{eq:IRmetricFiniteT} have been studied intensely. Their scaling symmetries are manifestations of the scaling properties of the dual field theory: $z$ is the dynamical critical exponent of the fixed point and $\theta$ parameterises the violation of hyperscaling (such that the effective dimensionality of the fixed point is $(d-\theta)$). Note that we will always be considering cases in which the metric \eqref{eq:IRmetricFiniteT} is only realised in the IR i.e.~for $R\gg R_{UV}$, the UV cutoff of the near-horizon spacetime. $R_{UV}$ denotes the scale at which irrelevant deformations to the infrared theory (that ultimately lead to the flow to a CFT in the ultraviolet) become important. The precise values of $R_{UV}$, $z$ and $\theta$ depend on the particular gravitational solution. For a given choice of $V$ and $Z$, varying $\mu$ and the UV source for the scalar operator typically changes the length scales $L_t$ and $L_x$ in the IR spacetime \eqref{eq:IRmetricFiniteT} but not the values of $z$ or $\theta$. The zero temperature states should therefore be considered as comprising quantum critical lines, rather than points. When $z=1$ the IR metric is conformal to that of planar AdS$_d$, with a speed of light $c_{IR}$ given by
\begin{equation}
\label{eq:cIRdefn}
c_{IR}=\frac{L_t}{L_x}.
\end{equation}

{}The near-horizon spacetimes \eqref{eq:IRmetricFiniteT} are supported by matter fields. An exponential potential $V(\Phi\rightarrow\infty)\rightarrow V_0e^{-\delta\Phi}$ and a logarithmically running scalar
\begin{equation}
\label{runningPhi}
\Phi(R)=\kappa\log\left(\frac{R}{L}\right),\quad\quad\quad\kappa^2=\frac{2}{d}(d-\theta)(d(z-1)-\theta),\quad\quad\quad \kappa\delta=\frac{2\theta}{d},
\end{equation}
are responsible for the violation of hyperscaling in the infrared ($\theta\ne0$).

{}The gauge field is instead responsible for the fate of relativistic symmetry in the IR. We will consider theories which have an exponential gauge coupling $Z(\Phi\rightarrow\infty)\rightarrow Z_0e^{\gamma\Phi}$, and there are two qualitatively different cases to consider. When $\kappa\gamma$ is sufficiently large, the backreaction of the gauge field is small in the IR, and so the IR metric preserves relativistic symmetry ($z=1$) but violates hyperscaling ($\theta<0$). In this case, the gauge field in the IR region $R_0\ll R \ll R_{UV}$ is
\begin{equation}
A_t(R)=L_tA_0\left(\frac{R}{L}\right)^{\theta-d-1+2\Delta_A},
\end{equation}
where $\Delta_A=d-\theta+\theta/d-\kappa\gamma/2<0$. The backreaction of the gauge field gives small corrections to the zero temperature metric that are suppressed in the IR by powers of $R^{2\Delta_A}$. On the other hand, when $\kappa\gamma$ is sufficiently small, the backreaction becomes important and destroys the relativistic symmetry of the IR metric ($z\ne1$).\footnote{In this case we have $A_t(R)\sim R^{\theta-d-z}$.} Note that scale invariance is restored only if $\kappa=0$, not just if $\theta=0$, due to the matter field profiles.

{}In the language of the quantum criticality, $A_0$ is a coupling with dimension $\Delta_A$ that deforms the IR fixed point. The corresponding operator, dual to the field $A_t$, has scaling dimension $d+1-\theta-\Delta_A$. For fixed points with $z=1$ this is an irrelevant deformation while for fixed points with $z\ne1$ it is marginal and has scaling dimension $d+z-\theta$. 

{}The following relation between $A_0$ and the charge density is found upon using equation \eqref{bulkchargedensity}
\begin{equation}
\label{bulkchargedensityzeroT}
\rho=(d+1-\theta-2\Delta_A)\tilde L^{-1}L_x^d Z_0 A_0\,.
\end{equation}

\section{Dynamics of zero density states}
\label{sec:ZeroDensityHolography}

{}In this Section, we will illustrate the emergence of an anomalous $(d-1)$-form global symmetry in a simple context. We will consider the equilibrium states described in Section \ref{sec:NonZeroDensityEMD}, but in theories with an additional $0$-form $U(1)$ global symmetry. The states have zero density of the additional $U(1)$ charge density, and we still study the dynamics of small amplitude perturbations of this additional charge. This is a helpful technical simplification as these decouple from the dynamics of the energy and momentum of the state, as well as from perturbations of the original $U(1)$ charge. In the gravitational language, these dynamics are captured by the Maxwell action
\begin{equation}
\label{eq:ZeroDensityMaxwellAction}
S=-\frac{1}{4}\int d^{d+2}x\sqrt{-g}Z(\Phi)F_{MN}F^{MN},
\end{equation}
in the `probe' limit. In other words, the spacetime metric $g_{MN}(r)$ and dilaton $\Phi(r)$ are fixed functions corresponding to the black hole solutions of the Einstein-Maxwell-dilaton theories described in Section \ref{sec:NonZeroDensityEMD}. We are abusing notation here by labelling the additional $U(1)$ gauge field $F$ and its coupling $Z$ with the same name as those of the original $U(1)$. Everywhere where they appear in this Section, it should be understood that these refer to the additional $U(1)$ field. We will take the coupling to be of the power law form
\begin{equation}
Z=Z_0\left(\frac{R}{L}\right)^{d-\theta-z+2\frac{\theta}{d}-\Delta_\chi},
\end{equation}
in the IR region of the spacetime $R_0\gg R\gg R_{UV}$ where the metric has the scaling form \eqref{eq:IRmetricFiniteT}. We will show later that the constant $\Delta_\chi$ sets the infrared scaling dimension of the additional $U(1)$ charge operator.

{}The main result of this Section is that a $(d-1)$-form global symmetry emerges in the deep infrared when $\Delta_\chi+2(z-1)<0$. There is a mixed 't Hooft anomaly between this $(d-1)$-form symmetry and the original $0$-form $U(1)$ symmetry, so that the symmetries are the same as those of a superfluid \cite{Delacretaz:2019brr}. These states therefore exhibit many of the properties of a superfluid -- such as a gapless Goldstone-like mode -- but without the spontaneous breaking of a $0$-form $U(1)$ symmetry. We start by considering the hydrodynamic equations of the additional $0$-form charge, showing that they exhibit an approximate higher-form conservation law at low temperatures, broken by a dangerously irrelevant deformation. We then give a complementary description in terms of a superfluid-like action, where the weak breaking of the $(d-1)$-form symmetry is realised by a coupling to an emergent $U(1)$ gauge field. Finally we show that the symmetry persists in the infrared of the zero temperature state, clarifying the nature of the dangerously irrelevant coupling that weakly breaks the emergent symmetry. 

\subsection{Relaxed hydrodynamics of $0$-form charge density}
\label{sec:zerodensityholoquasihydro}

{}We begin by deriving the hydrodynamic-like equations governing small amplitude perturbations of the additional $0$-form $U(1)$ charge and current density. We will show that for the holographic states just described, this current density relaxes parametrically slowly at low temperatures. This is the first indication of the emergence of the (anomalous) $(d-1)$-form symmetry. Throughout this Section we will use $j^\mu$ to denote linear perturbations of the additional $0$-form $U(1)$ charge and current densities, and $\bar{A}_\mu$ to denote the corresponding external sources.

{}The $U(1)$ charge density identically obeys the local conservation equation
\begin{equation}
\label{eq:zerodensityconservationeq}
\partial_\mu j^\mu=0.
\end{equation}
A hydrodynamic theory for this charge is obtained by supplementing this with a constitutive relation for $j^i$, obtained by finding the ingoing solution to the linearized bulk Maxwell equations $\partial_M(\sqrt{-g}Z(\Phi)F^{MN})=0$ in a derivative expansion. At leading order, this is simply \cite{Iqbal:2008by}
\begin{equation}
\label{eq:zerodensityconstitutive}
j^i = -\sigma_{dc}\left(\chi_{\rho\rho}^{-1}\partial_i j^t+\bar{F}_{ti}\right)+O(\partial^2),
\end{equation}
where $\sigma_{dc}$ and $\chi_{\rho\rho}$ are the dc conductivity and susceptibility of the $0$-form charge and $\bar{F}=d\bar{A}$ is the field strength of the external source. In terms of bulk fields, they are given by
\begin{equation}
\label{eq:zerodensitychirhorho}
\sigma_{dc}=Z(\Phi_0)C(r_0)^{d/2-1},\quad\quad\quad\quad\quad \chi_{\rho\rho}^{-1}=\int^{r_0}_0dr\frac{\sqrt{B(r)D(r)}}{Z(\Phi(r))C(r)^{d/2}}.
\end{equation}
Note that the specific combination that appears in the dissipative term in \eqref{eq:zerodensityconstitutive} is fixed by the requirement that it vanishes in static equilibrium \cite{Banerjee:2012iz,Jensen:2012jh}, and so there is only one independent dissipative coefficient at this order in the derivative expansion.

{}Higher derivative corrections to the constitutive relation \eqref{eq:zerodensityconstitutive} become important at shorter scales. We will focus on one such correction and move it to the left hand side to give
\begin{equation}
\label{eq:zerodensityalmostcons}
\left(\tau\partial_t+1\right) j^i =-\sigma_{dc}\left(\chi_{\rho\rho}^{-1}\partial_i j^t+\bar{F}_{ti}\right)+O(\partial^2).
\end{equation}
Physically this correction accounts for the fact that perturbations of the current do not relax instantaneously, but rather over the timescale $\tau$. Following the same steps as in \cite{Grozdanov:2018fic}, the relaxation timescale is related to the conductivity by
\begin{equation}
\label{eq:zerodensitytaudefn}
\tau=\sigma_{dc}\chi_{JJ}^{-1},
\end{equation}
where 
\begin{equation}
\label{eq:zerodensitychiJJ}
\chi_{JJ}^{-1}=\int^{r_0}_{0}dr\left(\sqrt{\frac{B(r)}{D(r)}}\frac{1}{Z(\Phi(r))C(r)^{d/2-1}}-\frac{1}{4\pi T\sigma_{dc}(r_0-r)}\right).
\end{equation}
$\chi_{JJ}$ corresponds to the static susceptibility of the current in the theory obtained by taking $\tau\partial_t\gg1$ in equation \eqref{eq:zerodensityalmostcons} (i.e.~in the theory where the current is exactly conserved).

{}At this stage the expression \eqref{eq:zerodensityalmostcons} is formal: we have chosen to retain just one of the many corrections. This is only sensible if this correction is parametrically larger than those we are ignoring, i.e.~when the current relaxes parametrically slowly compared to generic quantities. This should be the case when $\tau T\gg 1$, and the equation \eqref{eq:zerodensityalmostcons} should be understood as being valid at leading order in a generalised derivative expansion where $\tau^{-1}\sim\partial\ll T$. This derivative expansion is a generalization of hydrodynamics that accounts for the dynamics of slowly relaxing quantities, in addition to exactly conserved ones \cite{Grozdanov:2018fic}. We will refer to these equations as those of `relaxed hydrodynamics'.

{}By calculating $\tau$ directly from the expressions above, we now determine whether the relaxed hydrodynamic equation \eqref{eq:zerodensityalmostcons} is meaningful for a particular gravitational state. By focusing on the low temperature limit, we can obtain the $T$-scaling of $\tau$ from properties of the scaling region near the horizon. We can evaluate the integral \eqref{eq:zerodensitychiJJ} in this limit using a similar strategy to the evaluation of $\chi_{\rho\rho}$ \cite{Blake:2016wvh}. We split the integral into two parts: one over the IR region ($R\geq R_{UV}$) of the spacetime and one over the UV region ($R\leq R_{UV}$). At low temperatures the latter integral is insensitive to the presence of the small horizon in the IR region and so is $T$-independent. The value of $\Delta_\chi$ determines whether this contribution from the UV region, or the contribution from the IR region, dominates the integral. For $\Delta_\chi+2(z-1)>0$ the contribution from the IR region to the integral is cutoff independent and diverges at low temperatures while for $\Delta_\chi+2(z-1)<0$ it is temperature independent but cutoff dependent. Therefore
\begin{equation}
\label{eq:zerodensitysusceptibility}
\chi_{JJ}(T\rightarrow0)\sim\begin{cases}T^{\frac{\Delta_\chi+2(z-1)}{z}} & \Delta_\chi+2(z-1)>0\\ T^0,&\Delta_{\chi}+2(z-1)<0\end{cases}.
\end{equation}
At low temperatures, the dc conductivity $\sigma_{dc}\sim T^{\frac{\Delta_\chi+z-2}{z}}$ \cite{Blake:2016wvh} and so
\begin{equation}
\label{eq:zerodensitylongtimescale}
T\tau(T\rightarrow0)\sim\begin{cases}T^0 & \Delta_\chi+2(z-1)>0\\ T^{\frac{\Delta_\chi+2(z-1)}{z}},&\Delta_{\chi}+2(z-1)<0\end{cases}.
\end{equation}

{}Equation \eqref{eq:zerodensitylongtimescale} is the key result: for states with $\Delta_\chi+2(z-1)<0$ the current relaxes parametrically slowly and so these states are governed by the relaxed hydrodynamic equations \eqref{eq:zerodensityconservationeq} and \eqref{eq:zerodensityalmostcons} for times much longer than the inverse temperature. Naively one might expect that the strong interactions in a holographic theory cause all non-conserved quantities to relax quickly, but when $\Delta_\chi+2(z-1)<0$ the current does not. $\Delta_\chi$ controls the $U(1)$ gauge coupling near the horizon, and smaller $\Delta_\chi$ corresponds to a smaller gauge coupling. From this perspective, the slow relaxation of the current is a consequence of it coupling weakly to the thermal bath represented by the horizon. The relation of lifetimes to near-horizon couplings was emphasize in \cite{Ghosh:2020lel}.

{}The slow relaxation of the current in the relaxed hydrodynamic equation \eqref{eq:zerodensityalmostcons}  has an important impact on the low energy properties of the state. Firstly, the conductivity of the corresponding charge has a sharp Drude-like peak with a width much narrower than the inverse temperature
\begin{equation}
\label{eq:zerodensityhydroconductivity}
\sigma_{JJ}(\omega)=\frac{i}{\omega}G^R_{JJ}=\frac{\sigma_{dc}}{1-i\omega\tau}.
\end{equation}
This is unrelated to the translational symmetry of the state -- in these zero density states there is no overlap between the $0$-form $U(1)$ current and the momentum operators. It stands in contrast to the incoherent low frequency conductivity found when the current relaxes quickly, $\sigma_{JJ}(\omega)=\sigma_{dc}$. Note that from \eqref{eq:zerodensitychiJJ}, \eqref{eq:zerodensitysusceptibility} and \eqref{eq:zerodensitylongtimescale}, the dc conductivity $\sigma_{dc}$ is a direct indicator of the relaxation time of the current: when the current decays slowly, the state is highly conductive $\sigma_{dc} T\gg1$ (and vice versa).

{}Secondly, the spectrum of collective excitations that carry the $0$-form $U(1)$ charge displays a characteristic crossover when the current relaxes slowly. Their dispersion relations are given by solutions to the quadratic equation
\begin{equation}
\label{eq:zerodensityhydrodispersions}
\omega^2+i\omega\tau^{-1}-v^2k^2=0,\quad\quad\quad\quad\quad v^2=\frac{\chi_{JJ}}{\chi_{\rho\rho}}.
\end{equation}
At very low frequencies $\omega,k\ll \tau^{-1}\ll T$ there is diffusion of the charge density
\begin{equation}
\omega(k)=-iDk^2+\ldots,\quad\quad\quad\quad\quad D=v^2\tau,
\end{equation}
and relaxation of the current $\omega(k)=-i\tau^{-1}+\ldots$. At higher frequencies $\tau^{-1}\ll\omega,k\ll T$ the current is approximately conserved and the modes propagate coherently at speed $v$
\begin{equation}
\label{eq:ordinaryhydropropmode}
\omega(k)=\pm vk-\frac{i}{2}\tau^{-1}+\ldots.
\end{equation} 
As the charge susceptibility of the states with a slowly relaxing current is temperature-independent at low temperatures \cite{Blake:2016wvh}, slow relaxation of the current is associated with a parametrically large diffusivity that depends sensitively on $\Delta_\chi$
\begin{equation}
D\sim T^{\frac{\Delta_\chi+z-2}{z}}.
\end{equation}
The propagating speed $v\sim T^0$ at low temperatures, and in Appendix \ref{app:zerodensityspeedbound} we show that $v$ is bounded from above by the speed of light.

\begin{figure}[h]
\begin{center}
\includegraphics[scale=.5]{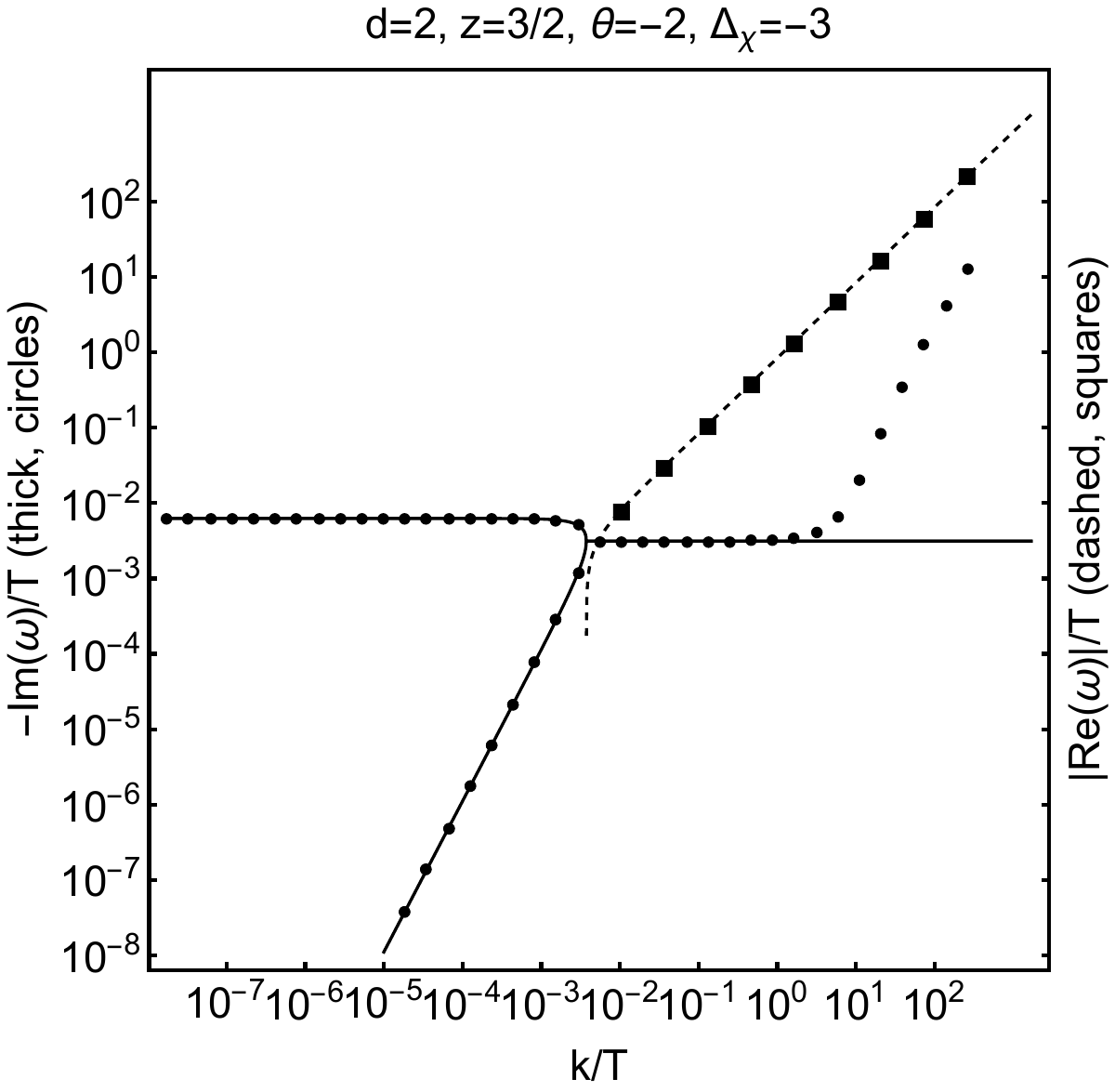}
\caption{\label{zerodensitydispersion_fig} Real (squares) and imaginary (circles) parts of the dispersion relations of the long-lived collective modes of a state with $d=2$, $\theta=-2$, $z=3/2$ and $\Delta_\chi=-3$ at $T/\mu= 7.5\times10^{-4}$. There is excellent agreement with the real (dashed line) and imaginary (solid line) parts of the solutions of the equation \eqref{eq:zerodensityhydrodispersions} of relaxed hydrodynamics. For clarity, the dispersion relation of only one of the propagating modes is shown.}
\end{center}
\end{figure}

{}Figure \ref{zerodensitydispersion_fig} shows a comparison between the numerically-determined dispersion relations of a collective excitation of a low temperature state with $\Delta_\chi+2(z-1)<0$ and the expressions \eqref{eq:zerodensitychirhorho}, \eqref{eq:zerodensitychiJJ} and \eqref{eq:zerodensityhydrodispersions}. There is extremely good agreement for $\omega,k\ll T$ where the relaxed hydrodynamic theory should apply, including a crossover from diffusive and relaxational modes to propagating modes as the wavevector is increased. For large wavevectors $k\gtrsim T$, the imaginary part of the frequency $\omega$ deviates from the predictions of the relaxed hydrodynamic theory and instead is governed by the zero temperature dynamics of the state, as we explain in Section \ref{sec:ZeroDensityZeroTemperature}. Details of the numerical calculations can be found in Appendix \ref{app:numericaldetails_zerodensity}.

\subsection{Higher-form formulation of relaxed hydrodynamics}
\label{sec:ZeroDensityEffectiveTheory}

{}We will now show that the relaxed hydrodynamic theory governing the state at low temperatures can be recast as the relaxed hydrodynamics of an approximately conserved higher-form charge. The same hydrodynamic theory governs a phase-relaxed superfluid.

{}{From \eqref{eq:zerodensityconservationeq}, \eqref{eq:zerodensityalmostcons} and \eqref{eq:zerodensitytaudefn},} the two equations of the relaxed hydrodynamic theory are
\begin{equation}
\label{eq:zerodensityhydroeq1}
\partial_t j^t+\partial_i j^i=0,\quad\quad\quad\quad\quad \partial_t j^i+\frac{\chi_{JJ}}{\chi_{\rho\rho}}\partial_i j^t=-\frac{1}{\tau} j^i+\chi_{JJ}\bar{F}_{it},
\end{equation}
and these are valid in the generalised derivative expansion $\partial\ll T$ with $\tau\partial\sim1$. We first consider the limit $\tau^{-1}\ll\omega,k\ll T$, in which the current relaxation term can be neglected. In terms of the $d$-form $K$ defined by $(\star\,K)_\mu=\left(-j^t/\chi_{\rho\rho}, j^i/\chi_{JJ}\right)$, the relaxed hydrodynamic equations in this limit can be written as the closure of the 1-form $(\star\,K+\bar{A})$
\begin{equation}
\label{eq:higherformEoMintermediate}
\partial_{\mu}(\star\,K)_{\nu}-\partial_\nu(\star\,K)_\mu=-\bar{F}_{\mu\nu}\quad\quad\Longleftrightarrow\quad\quad\partial_{[\mu}(\star\,K+\bar{A})_{\nu]}=0.
\end{equation}
Assuming that this $1$-form is also smooth then it must be exact and so can be written as the gradient of a smooth single-valued scalar field $\varphi$
\begin{equation}
\label{eq:Goldstonelikemode}
(\star\,K)_\mu+\bar{A}_\mu=\partial_\mu\varphi\quad\quad \Longleftrightarrow\quad\quad (\star\,K)_\mu=D_\mu\varphi,
\end{equation}
where the $U(1)$ covariant derivative is $D_\mu\varphi=\partial_\mu\varphi-\bar{A}_\mu$. $\varphi$ is reminiscent of the Goldstone mode of a superfluid.

{}The similarity to a superfluid in this limit can be made precise by writing the equation of motion \eqref{eq:higherformEoMintermediate} as
\begin{equation}
\label{eq:anomaloushigherformeq}
d\star K=-\bar{F}.
\end{equation}
When $\bar{F}=0$, this equation signifies the local conservation of the $d$-form current $K$ and therefore the existence of a $(d-1)$-form symmetry. The state also retains the $0$-form $U(1)$ symmetry associated to the conservation law \eqref{eq:zerodensityhydroeq1}. The source term $\bar{F}$ on the right hand side of \eqref{eq:anomaloushigherformeq} is a mixed anomaly of these two symmetries: the $d$-form current is no longer conserved in the presence of an external source for the $1$-form current. This anomalous symmetry is precisely that of a superfluid with frozen temperature and velocity fluctuations \cite{Hofman:2017vwr,Delacretaz:2019brr}. In the superfluid case, the higher-form symmetry describes the conservation of winding in the absence of free vortices. 

{}Unlike in a superfluid, however, in the holographic states we have described, the anomalous symmetry emerges without the spontaneous breaking of a $0$-form $U(1)$ symmetry. Nevertheless, as many of the properties of a superfluid follow from the anomalous symmetry alone \cite{Delacretaz:2019brr}, they will also be valid for the holographic states. The most basic of these is the existence of a gapless degree of freedom which we will refer to as Goldstone-like. In the hydrodynamic regime, this gapless mode is guaranteed by the following argument \cite{Delacretaz:2019brr}. The hydrodynamic variables are the charge densities $j^t$ and $(\star\,K)^i$ and the constitutive relations for the corresponding currents are
\begin{equation}
j^i=\tilde{\mu}^i+\ldots,\quad\quad\quad\quad\quad (\star\,K)^t=\mu+\ldots,
\end{equation}
where $(\mu,\tilde{\mu}^i)$ are the chemical potentials for the charges $(j^t, (\star\,K)^i)$. The anomaly term in equation \eqref{eq:anomaloushigherformeq} (alongside Onsager reciprocity and consistency with the static limit) fixes these constitutive relations at leading order in the derivative expansion and so is responsible for ensuring the existence of gapless hydrodynamic modes with dispersion relations
\begin{equation}
\label{eq:anomalyhydromodes}
\omega(k)=\pm\frac{1}{\sqrt{\chi_{\rho\rho}\chi_{KK}}}k+O(k^2),
\end{equation}
where $\chi_{\rho\rho}\equiv\partial j^t/\partial\mu$ and $\chi_{KK}\equiv \partial(\star\,K)^i/\partial\tilde{\mu}^i$ (no sum) are the static susceptibilities of the charges. The propagating modes \eqref{eq:ordinaryhydropropmode} that we identified earlier in the regime $\tau^{-1}\ll\omega,k\ll T$ are precisely those \eqref{eq:anomalyhydromodes} necessitated by the anomaly. The susceptibility of the $d$-form charge is related to the parameters in the original relaxed hydrodynamic equations \eqref{eq:zerodensityhydroeq1} by $\chi_{KK}=\chi_{JJ}^{-1}$. These modes produce the dissipationless conductivity $\sigma(\omega)\rightarrow\chi_{KK}^{-1}(i/\omega)$ found in a superfluid, as can be seen by taking the appropriate limit of the Drude-like expression \eqref{eq:zerodensityhydroconductivity}.

{}However, unlike in a superfluid, in the holographic states, the anomalous conservation law is only valid in the strict $\omega\tau,k\tau\rightarrow\infty$ limit. Beyond this, the non-zero relaxation time of the current explicitly violates it. The small symmetry-breaking parameter $\tau^{-1}$ is important at low energies where it produces a large, but finite, dc conductivity \eqref{eq:zerodensitytaudefn}. In a superfluid, such explicit breaking occurs when the phase is relaxed by mobile vortices on which winding planes can end. When this breaking is weak, and in the absence of an external magnetic field, the corresponding anomalous conservation equation is modified to \cite{Grozdanov:2018fic,Delacretaz:2019brr}
\begin{equation}
\partial_\nu K^{\mu_1\ldots\mu_{d-1}\nu}=\frac12\epsilon^{\mu_1\ldots\mu_{d-1}\kappa\lambda}\bar{F}_{\kappa\lambda}+\frac1\tau u_\nu K^{\mu_1\ldots\mu_{d-1}\nu},
\end{equation}
where $u_\mu$ is a fixed timelike unit vector. The hydrodynamics of our holographic states therefore coincides with that of a phase-relaxed superfluid (with frozen temperature and velocity fluctuations) \cite{Davison:2016hno}: this exhibits the Drude-like conductivity in \eqref{eq:zerodensityhydroconductivity} and hydrodynamic modes with dispersion relations given by solutions to the equation \eqref{eq:zerodensityhydrodispersions}.

{}Unlike in a conventional superfluid -- where vortices are gapped at low temperatures -- in the holographic states the higher-form symmetry is explicitly broken at any non-zero $T$ and so the emergence of the higher-form symmetry in the infrared is less robust. We will later move beyond the hydrodynamic limit to the $T=0$ state, where the weak symmetry breaking term transforms from power law in $T$ to power law in $\omega$, and so the effects of the explicit symmetry breaking vanish as $\omega\rightarrow0$. In this sense, the holographic states are similar to the quantum Lifshitz model \cite{Else_2021}.

\subsection{Effective action for the Goldstone-like mode $\varphi$}
\label{sec:GoldstoneActionFiniteT}

{}We have argued using symmetries that the holographic states should support a superfluid Goldstone-like mode at low temperatures, despite the lack of spontaneous symmetry breaking. We will now show how to make this mode manifest at the level of the action.

{}A description of the low energy charge transport in a holographic theory can be obtained by integrating out the spacetime beyond a radial hypersurface. In \cite{Nickel:2010pr} it was shown that this description comprises a massless mode $\varphi$ -- the radial Wilson line -- coupled to the remaining spacetime, and this idea has been further developed in \cite{Faulkner:2010jy,deBoer:2015ija,Crossley:2015tka,deBoer:2018qqm,Glorioso:2018mmw,Bu:2020jfo}, connecting to Schwinger-Keldysh constructions of effective hydrodynamic actions for a conserved $U(1)$ charge \cite{Glorioso:2018wxw}. This description can be formally obtained whether the current relaxes slowly or not. The key distinction is in the coupling of $\varphi$ to the remaining spacetime. We will show that it is only when the current relaxes slowly that this coupling is unimportant and thus that the description of the system in terms of a superfluid-like Goldstone mode is useful.

{}Following \cite{Nickel:2010pr}, we first split the bulk action integral \eqref{eq:ZeroDensityMaxwellAction} into two pieces $S=S_{UV}+S_{IR}$ by dividing the spacetime along a radial hypersurface $r=r_*$ and imposing boundary conditions on this hypersurface such that $A_\mu(r_*)=a_\mu$. Ultimately the generating functional is (minus) the on-shell action as a functional of the external gauge field $A_\mu(r=0)=\bar{A}_\mu$, which we obtain by evaluating $S[\bar{A}_\mu,a_\mu]=S_{UV}[\bar{A}_\mu,a_\mu]+S_{IR}[a_\mu]$ for linearized solutions to Maxwell's equations and then putting $a_\mu$ on-shell. To see the Goldstone-like mode, we will do this in stages. We first evaluate $S_{UV}$ for solutions that obey only the $(t,\vec{x})$ components of Maxwell's equations and not the $r$ component. These solutions are
\begin{equation}
\label{eq:UVgaugefieldexps}
\begin{aligned}
A_t&\,=c_t\left(\left(\bar{A}_t+\partial_t\int^r_0 A_rdr\right)\int^{r_*}_r\frac{\sqrt{BD}}{C^{d/2}Z}dr+\left(a_t+\partial_t\int^r_{r_{*}}A_rdr\right)\int^r_0\frac{\sqrt{BD}}{C^{d/2}Z}dr\right)+\ldots,\\
A_i&\,=c_x\left(\left(\bar{A}_i+\partial_i\int^r_0 A_rdr\right)\int^{r_*}_r\sqrt{\frac{B}{D}}\frac{dr}{C^{d/2-1}Z}+\left(a_i+\partial_i\int^r_{r_{*}}A_rdr\right)\int^r_0\sqrt{\frac{B}{D}}\frac{dr}{C^{d/2-1}Z}\right)+\ldots,
\end{aligned}
\end{equation}
where
\begin{equation}
\label{eq:fintegraldefinitions}
c_t^{-1}=\int^{r_*}_0\frac{\sqrt{BD}}{C^{d/2}Z}dr,\quad\quad\quad\quad\quad c_x^{-1}=\int^{r_*}_0\sqrt{\frac{B}{D}}\frac{dr}{C^{d/2-1}Z},
\end{equation}
and $\ldots$ denote terms that are higher order in derivatives. The on-shell action for such solutions is
\begin{equation}
\label{eq:gold}
S_{UV}[\varphi,\bar{A}_\mu,a_\mu]=\frac{1}{2}\int d^{d+1}x\left(-c_t\left(\partial_t\varphi-\bar{A}_t+a_t\right)^2+c_x\left(\partial_i\varphi-\bar{A}_i+a_i\right)^2+\ldots\right),
\end{equation}
where the Goldstone-like field $\varphi$ is the radial Wilson line
\begin{equation}
\varphi=\int^{0}_{r_*}A_rdr.
\end{equation}
Physically, we have integrated out the high energy degrees of freedom associated to the UV region of the spacetime. In doing so we have explicitly retained the massless field $\varphi$: this is the hydrodynamic degree of freedom associated to the conserved $0$-form $U(1)$ charge, and is important at low energies. Ultimately we will want to integrate over this field, as this corresponds to imposing the remaining $r$ component of Maxwell's equations. Doing so will yield the local conservation of $0$-form $U(1)$ charge
\begin{equation}
\label{eq:zerodensitychargeconservationholo}
\partial_\mu j^\mu=0,
\end{equation}
where
\begin{equation}
\label{eq:zerodensityconstitutiveholo}
j^t=c_t\left(\partial^t\varphi-\bar{A}^t+a^t\right)+\ldots,\quad\quad\quad\quad\quad j^i=c_x\left(\partial^i\varphi-\bar{A}^i+a^i\right)+\ldots.
\end{equation}

{}At this stage, the low energy theory $S_{UV}[\varphi,\bar{A}_\mu,a_\mu]+S_{IR}[a_\mu]$ is that of a Goldstone-like mode coupled via $a_\mu$ to the low energy degrees of freedom associated to the IR region of the spacetime (and to an external gauge field $\bar{A}_\mu$). If we were to turn off this coupling -- for example by introducing a hard wall at $r=r_*$ such that $a_\mu=0$ -- then the action would be exactly that of a superfluid Goldstone mode with characteristic speed $v^2=c_x/c_t$, paralleling the Schwinger-Keldysh construction of effective actions for superfluid hydrodynamics \cite{Delacretaz:2021qqu}. 

{}To determine to what extent this superfluid Goldstone-like mode survives in genuine black hole solutions (where $a_\mu$ is a dynamical field), we now turn to $S_{IR}$. This is a $(d+2)$-dimensional holographic action for $A_M(r,x^\mu)$,  that represents a set of strongly coupled low energy degrees of freedom of the state. It is helpful to integrate out the interior region $r>r_*$ to obtain a $(d+1)$-dimensional action for $a_\mu(x^\mu)$. Considering perturbations whose wavevector is aligned with the $x$-axis (without loss of generality, due to rotational symmetry), this procedure yields the Fourier space action\footnote{We will implicitly impose ingoing boundary conditions on solutions at the horizon so that this action produces the retarded Green's function, in the sense described in \cite{Nickel:2010pr}.}
\begin{equation}
\label{eq:GoldstoneIRaction}
S_{IR}[a_\mu]=-\frac{1}{2}\int d\omega dk\left(\frac{\tilde{f}_{tx}(-\omega,-k)\tilde{f}_{tx}(\omega,k)}{g_\parallel(\omega,k)}+\frac{\tilde{f}_{tb}(-\omega,-k)\tilde{f}_{tb}(\omega,k)}{g_\perp(\omega,k)}\right),
\end{equation}
where tildes denote Fourier transforms, $\tilde{f}_{tx}(\omega,k)=\omega\tilde{a}_x(\omega,k)+k\tilde{a}_t(\omega,k)$, $\tilde{f}_{tb}(\omega,k)=\omega\tilde{a}_b(\omega,k)$ and the index $b$ here runs over all spatial coordinates $\vec{x}$ except the longitudinal direction $x$. Integrating out the interior region of spacetime corresponds to integrating out low energy degrees of freedom, and the price to be paid for this is that $S_{IR}[a_\mu]$ is generally non-local. This integration can be done order by order in a derivative expansion $\omega,k\ll T$, along the lines described in \cite{Kovtun:2005ev,Davison:2018nxm}. The result of this calculation is that the effective gauge couplings are
\begin{equation}
\begin{aligned}
\label{geq}
g_\parallel(\omega,k)&\,=\sigma_{dc}^{-1}\left(i\omega-k^2\sigma_{dc}(c_t^{IR})^{-1}-\omega^2\sigma_{dc}(c_x^{IR})^{-1}+\ldots\right),\\
g_\perp(\omega,k)&\,=\sigma_{dc}^{-1}\left(i\omega-\omega^2\sigma_{dc}(c_x^{IR})^{-1}+\ldots\right),
\end{aligned}
\end{equation}
where
\begin{equation}
\begin{aligned}
(c_t^{IR})^{-1}&\,=\int^{r_*}_{r_0}dr\frac{\sqrt{BD}}{ZC^{d/2}},\\
(c_x^{IR})^{-1}&\,=\int^{r_*}_{r_0}dr\left(\sqrt{\frac{B}{D}}\frac{1}{ZC^{d/2-1}}-\frac{1}{4\pi T\sigma_{dc}(r_0-r)}\right)-\frac{\log\left(1-\frac{r_*}{r_0}\right)}{4\pi T\sigma_{dc}},
\end{aligned}
\end{equation}
and $\sigma_{dc}$ is as defined in \eqref{eq:zerodensitychirhorho} above. 

{}After integrating out the UV and IR regions of the spacetime as described above, we have obtained a $(d+1)$-dimensional effective theory of a Goldstone-like field $\varphi$ coupled to an emergent $U(1)$ gauge field $a_\mu$. To make contact with a local hydrodynamic theory, we will first push the cutoff $r_*\rightarrow r_0$ so that the emergent gauge field $a_\mu$ corresponds to the thermal bath represented by the black hole horizon. With this choice
\begin{equation}
\begin{aligned}
\label{eq:gcouplingsnearhorizon}
g_\parallel(\omega,k)&\,=i\sigma_{dc}^{-1}\omega-(c_x^{IR})^{-1}\omega^2+\ldots,\\
g_\perp(\omega,k)&\,=i\sigma_{dc}^{-1}\omega-(c_x^{IR})^{-1}\omega^2+\ldots,\\
c_t&\,=\chi_{\rho\rho},
\end{aligned}
\end{equation}
and we will deal with the divergence in $(c_x^{IR})^{-1}$ in this limit shortly.\footnote{{Since $(c_t^{IR})^{-1}\to0$ when $r_\star\to r_0$, there is no $k$ dependence left to the order we are working.}} The Fourier space action \eqref{eq:GoldstoneIRaction} becomes rotationally invariant
\begin{equation}
\label{eq:finalIRactionzerodensity}
S_{IR}[a_\mu]=-\frac{1}{2}\int d\omega dk\frac{\tilde{f}_{ti}(-\omega,-k)\tilde{f}_{ti}(\omega,k)}{g(\omega)},
\end{equation}
where $g(\omega)=i\sigma_{dc}^{-1}\omega-(c_x^{IR})^{-1}\omega^2$ and the $i$ index now runs over all $\vec{x}$ coordinates. This is not Lorentz invariant because the near-horizon degrees of freedom that it represents are in a thermal state. After fixing the gauge $a_t=0$ we can put $a_i$ on-shell to obtain the following constitutive relations for the current defined in equation \eqref{eq:zerodensityconstitutiveholo}
\begin{equation}
\begin{aligned}
\label{eq:combinedconstitutivehydro}
j^t&\,=\chi_{\rho\rho}\left(\partial^t\varphi-\bar{A}^t\right)+\ldots,\\
j^i&\,=\frac{\sigma_{dc}\partial_t+\ldots}{\left(1+\chi_{JJ}^{-1}\sigma_{dc}\partial_t+\ldots\right)}\left(\partial^i\varphi-\bar{A}^i\right)+\ldots.
\end{aligned}
\end{equation}
Noting that the last two terms in $(c_x^{IR})^{-1}$ can be rewritten as a single integral of $1/(r_0-r)$ from $0$ to $r_0$, the terms in $c_x^{-1}$ and $(c_x^{IR})^{-1}$ that are divergent in the limit $r_*\rightarrow r_0$ cancel out in the expression for $j^i$, leaving behind $\chi_{JJ}^{-1}$, defined in \eqref{eq:zerodensitychiJJ}. As indicated above, putting the final dynamical field $\varphi$ on-shell ensures that $j^t$ and $j^i$ obey the local conservation equation \eqref{eq:zerodensitychargeconservationholo}. 

{}The constitutive relation \eqref{eq:combinedconstitutivehydro} for the current $j^i$ clearly has two different regimes. At energies $\omega\gg \chi_{JJ}\sigma_{dc}^{-1}$ it reduces to that of a superfluid Goldstone mode. However, at the lowest energies $\omega\ll \chi_{JJ}\sigma_{dc}^{-1}$ it is qualitatively different: the interactions of $\varphi$ with the emergent gauge field $a_\mu$ become important at these energies and destroy the superfluid-like physics. This latter limit is the more familiar one: the coupling of the current to the thermal bath is important and the constitutive relation reduces to that of diffusive hydrodynamics: $j^i=\sigma_{dc}\partial_t(\partial_i\varphi-\bar{A}_i)+\ldots=-\sigma_{dc}\chi_{\rho\rho}^{-1}\partial_i j^t-\sigma_{dc}\left(\partial_t\bar{A}_i-\partial_i\bar{A}_t\right)\ldots$. 

{}Our main focus is instead the former limit, where the effect of the emergent gauge field is small (i.e.~the current couples weakly to the thermal bath) and thus the superfluid Goldstone-like mode is long-lived. Since the corrections neglected in \eqref{eq:combinedconstitutivehydro} become important at the thermal energy scale, this superfluid-like regime exists in states with $\chi_{JJ}\sigma_{dc}^{-1}\ll T$. This is only the case for the holographic theories with $\Delta_\chi+2(z-1)<0$ at low temperatures, for which $\chi_{JJ}\sigma_{dc}^{-1}=\tau^{-1}\ll T$. From this perspective, the relaxation of the Goldstone-like mode when $\omega\tau\ll1$ in these states occurs due to its coupling to the emergent gauge field that represents the degrees of freedom at the black hole horizon.

\subsection{Explicit breaking of higher-form symmetry} 
\label{sec:higherformbreaking}

{}From now on we will focus on states that have $\Delta_\chi+2(z-1)<0$ and so exhibit a superfluid-like regime. In this subsection we will explore in more detail the role of the emergent gauge field for these states, explaining how it explicitly breaks the higher-form symmetry. 

{}The effective action for the Goldstone-like mode is given by the sum of \eqref{eq:gold} and \eqref{eq:GoldstoneIRaction}, and we can make the $(d-1)$-form symmetry manifest by defining the $d$-form $K$ in terms of the fields in our effective action by
\begin{equation}
\left(\star\,K\right)_t= \left(\partial_t\varphi-\bar{A}_t+a_t\right),\quad\quad\quad\quad \left(\star\,K\right)_i= \left(\partial_i\varphi-\bar{A}_i+a_i\right).
\end{equation}
The difference from the case of a superfluid is the $a_\mu$ terms required by gauge-invariance. By construction $K$ satisfies the equation
\begin{equation}
\label{eq:explicitbreakinghigherformeq}
d\star\,K=-\bar{F}+f.
\end{equation}
where $\bar{F}=d\bar{A}$ and $f=da$.

{}In a superfluid, where there is no emergent gauge field ($f=0$), the equation \eqref{eq:explicitbreakinghigherformeq} is that of an anomalous $(d-1)$-form symmetry. The anomaly is a mixed anomaly, as the right hand side of equation \eqref{eq:explicitbreakinghigherformeq} is proportional to the field strength of the external source $\bar{A}_\mu$ for the current of the $0$-form $U(1)$ symmetry. Physically, this anomaly means that an external electric field creates winding planes of the superfluid phase -- in more familiar words, it generates an electric current.

{}The coupling of the Goldstone-like mode to the emergent gauge field $a_\mu$ also results in non-conservation of the higher-form charge $K$. However, in this case it corresponds to an explicit breaking of the symmetry since $f$ is a dynamical field that is sourced by current flow, rather than an external one. In superfluid language, we would say that an external electric field creates winding planes of the phase causing a current to flow. However, this current sources an emergent electromagnetic field which then destroys the winding planes according to equation \eqref{eq:explicitbreakinghigherformeq} and relaxes the current. In this sense, the role of the emergent electromagnetic field is analogous to that of vortices in a superfluid.

{}The result of the explicit breaking of the symmetry is sensitive to the form of the effective coupling $g$ in equation \eqref{eq:finalIRactionzerodensity}. If the emergent gauge field had a Maxwell action, then the effective action would just be that of a superconductor, in which a mass is generated by the Higgs mechanism. In our case the action is not Lorentz invariant, non-local, and at the level of dimension counting the action has one less derivative than the Maxwell action. Furthermore, this coupling explicitly breaks time reversal invariance as the emergent gauge field represents the coupling to a thermal bath. As a consequence, the explicit breaking of the symmetry does not generate a mass but rather a small lifetime for the Goldstone-like mode. In this sense, our states are somewhat similar to phase-relaxed superfluids in which interactions with an emergent Chern-Simons gauge field slowly relax the phase \cite{Davison:2016hno}, but with an unbroken parity symmetry.

\subsection{Zero temperature dynamics}
\label{sec:ZeroDensityZeroTemperature}

{}Until now, we have focused on states at non-zero temperature $T$, illustrating the existence of a superfluid-like regime for $\tau^{-1}\ll\omega\ll T$. The lower cutoff $\tau^{-1}$ is set by the explicit symmetry breaking scale. As $\tau^{-1}$ vanishes as $T\rightarrow0$, it is conceivable that for zero temperature states, the superfluid-like regime extends all the way to zero frequency. The subtlety, of course, is that this involves exiting the hydrodynamic regime $\omega\ll T$. In this Section, we will study the holographic states at $T=0$. We will show that a truly gapless superfluid-like mode survives, with an unconventional attenuation due to an irrelevant coupling that explicitly breaks the higher-form symmetry.

{}To illustrate this, we will proceed as in Section \ref{sec:GoldstoneActionFiniteT}, integrating out the UV part of the spacetime ($r\leq r_*$) and leaving a Goldstone-like mode $\varphi$ with action $S_{UV}[\varphi,\bar{A}_\mu,a_\mu]$ coupled to the IR spacetime ($r>r_*$). At zero temperature, the metric deep in the interior of the spacetime has the scaling form \eqref{eq:IRmetricFiniteT} with $f=1$, and the natural choice for $r_*$ is close to the boundary of this scaling region: $R_*\equiv R(r_*)\gg R_{UV}$. With this choice, the physics is that of a Goldstone-like mode interacting with the infrared quantum critical degrees of freedom represented by the scaling spacetime \eqref{eq:IRmetricFiniteT} near the horizon, {as we show below in \eqref{eq:renormalisedgoldstoneaction} and \eqref{eq:renormalisedIRaction}}.

{}There is a subtlety associated with this choice. Suppose we temporarily ignore $S_{UV}$ and interpret just the IR action as a standalone holographic theory in its own right. In order to do this, we have to renormalize it, as the IR on-shell action is not finite in the limit $R_*\rightarrow0$ where we remove the cutoff. More precisely, in the IR spacetime the general solutions near the cutoff are
\begin{equation}
\label{eq:ZeroTIRgaugefieldexps}
A_i=a_i^{(1)}\left(\frac{R}{L}\right)^{\Delta_\chi+2(z-1)}+\ldots+a_i^{(0)}+\ldots,\quad\quad\quad A_t=a_t^{(1)}\left(\frac{R}{L}\right)^{\Delta_\chi}+\ldots+a_t^{(0)}+\ldots,
\end{equation}
where $a_\mu^{(0,1)}$ are independent functions of $x^\mu$, and the corresponding on-shell action is
\begin{equation}
\begin{aligned}
\label{eq:divergentIRaction}
S_{\text{IR}}=\frac{Z_0L_x^d}{2\tilde{L}L_t}\int d^{d+1}x\Biggl(&\,\Delta_\chi a_t^{(1)}a_t^{(1)}\left(\frac{R_*}{L}\right)^{\Delta_\chi}-c_{IR}^2(\Delta_\chi+2(z-1))a_i^{(1)}a_i^{(1)}\left(\frac{R_*}{L}\right)^{\Delta_\chi+2(z-1)}+\ldots\Biggr).
\end{aligned}
\end{equation}
The cutoff dependence of the IR action \eqref{eq:divergentIRaction} is not a fundamental problem: physical answers of course do not depend on our arbitrary choice of separating the system into two parts using a hard radial cutoff. However, by making a more refined separation into UV and IR contributions, we obtain a simple interpretation of the IR theory as we remove the cutoff. Specifically, we renormalize both the UV and IR parts of the action by counterterms of opposite sign $S=\left(S_{\text{UV}}-S_{\text{ct}}\right)+\left(S_{\text{IR}}+S_{\text{ct}}\right)$. The counterterm action is
\begin{equation}
\label{eq:countertermsabc}
S_{\text{ct}}=\left.\tilde{L}\int d^{d+1}x\sqrt{-\gamma}Z\left(\frac{R}{L}\right)^{\theta/d}\left(\frac{F^{At}n_AF^{B}_{\,\,\,t}n_B}{2\Delta_\chi}+\frac{F^{Ai}n_AF^B_{\,\,\,i}n_B}{2(\Delta_\chi+2(z-1))}\right)\right|_{r=r_*},
\end{equation}
where $n$ is the unit vector normal to the surface $r=r_*$ and $\gamma$ is the induced metric on this surface. 

{}We can now go partially on-shell as before, using the renormalized actions. {We start by considering the partially on-shell UV action \eqref{eq:gold} where the interior cutoff $R_*$ is chosen to be in the scaling region and we express the interior boundary fields $a_\mu$ using the expansions \eqref{eq:ZeroTIRgaugefieldexps}
\begin{equation}
\begin{aligned}
S_{UV}=\frac{1}{2}\int d^{d+1}x\Biggl(&\,-c_t\left(\partial_t\varphi-\bar{A}_t+a_t^{(1)}\left(\frac{R_*}{L}\right)^{\Delta_\chi}+a_t^{(0)}+\ldots\right)^2\\
&\,+c_x\left(\partial_i\varphi-\bar{A}_i+a_i^{(1)}\left(\frac{R_*}{L}\right)^{\Delta_\chi+2(z-1)}+a_i^{(0)}+\ldots\right)^2\Biggr).
\end{aligned}
\end{equation}}
{The contribution to the integrals $c_t^{-1}$ and $c_x^{-1}$ from this scaling region diverge at small $R_*$ such that $c_t\sim (R_*/L)^{-\Delta_\chi}$ and $c_x\sim (R_*/L)^{-(\Delta_\chi+2(z-1))}$ vanish as the cutoff is removed. The terms that survive this limit are
\begin{equation}
\begin{aligned}
S_{UV}=\frac{L_x^dZ_0}{L_t\tilde{L}}\int d^{d+1}x\Biggl(&\,-\Delta_\chi a_t^{(1)}\left(\partial_t\varphi-\bar{A}_t+a_t^{(0)}\right)+c_{IR}^2\left(\Delta_\chi+2(z-1)\right)a_x^{(1)}\left(\partial_x\varphi-\bar{A}_x+a_x^{(0)}\right)\\
&\,-\frac{\Delta_\chi}{2}a_t^{(1)}a_t^{(1)}\left(\frac{R_*}{L}\right)^{\Delta_\chi}+c_{IR}^2\frac{\Delta_\chi+2(z-1)}{2}a_x^{(1)}a_x^{(1)}\left(\frac{R_*}{L}\right)^{\Delta_\chi+2(z-1)}\Biggr).
\end{aligned}
\end{equation}
The terms on the second line diverge but are cancelled exactly by the counterterms \eqref{eq:countertermsabc}. So the renormalized UV action is given by the terms on the first line. }

{The constants $a_\mu^{(1)}$ can be related to $a_\mu^{(0)}$ by using \eqref{eq:ZeroTIRgaugefieldexps} and \eqref{eq:countertermsabc} by matching the solutions \eqref{eq:ZeroTIRgaugefieldexps} to \eqref{eq:UVgaugefieldexps} in the $R\to0$ region, leading to:
\begin{equation}
\begin{aligned}
\frac{L_x^dZ_0}{L_t\tilde{L}}\Delta_\chi a_t^{(1)}=\tilde{c}_t\left(\partial_t\varphi-\bar{A}_t+a_t^{(0)}\right),\\
\frac{L_x^dZ_0}{L_t\tilde{L}}c_{IR}^2\left(\Delta_\chi+2(z-1)\right) a_x^{(1)}=\tilde{c}_x\left(\partial_x\varphi-\bar{A}_x+a_x^{(0)}\right).\\
\end{aligned}
\end{equation}
The tildes on $\tilde{c}_\mu$ indicate that we have subtracted the terms that diverge as the cutoff is removed
\begin{equation}
\tilde{c}_t^{-1}=c_t^{-1}-\frac{L_t\tilde{L}}{L_x^dZ_0}\frac{1}{\Delta_\chi}\left(\frac{R_*}{L}\right)^{\Delta_\chi},\quad \tilde{c}_x^{-1}=c_x^{-1}-\frac{L_t\tilde{L}}{L_x^dZ_0}\frac{c_{IR}^{-2}}{\Delta_\chi+2(z-1)}\left(\frac{R_*}{L}\right)^{\Delta_\chi+2(z-1)}.
\end{equation}}
{}{As a result,} the partially on-shell renormalized UV action is again that of a superfluid Goldstone-like mode
\begin{equation}
\label{eq:renormalisedgoldstoneaction}
S_{\text{UV}}-S_{\text{ct}}=\frac{1}{2}\int d^{d+1}x\left(-\tilde{c}_t\left(\partial_t\varphi-\bar{A}_t+a_t^{(0)}\right)^2+\tilde{c}_x\left(\partial_i\varphi-\bar{A}_i+a_i^{(0)}\right)^2\right),
\end{equation}
where now $a_\mu^{(0)}$ are the constant terms in the expansions \eqref{eq:ZeroTIRgaugefieldexps} and the coefficients have been renormalized from those in equation \eqref{eq:fintegraldefinitions} to
\begin{equation}
\begin{aligned}
\tilde{c}_t&\,=\chi_{\rho\rho}(T=0)+\ldots,\quad\quad\quad\quad\quad \tilde{c}_x&\,=\chi_{JJ}(T=0)+\ldots,
\end{aligned}
\end{equation}
where $\chi_{\rho\rho}$ and $\chi_{JJ}$ are the $T=0$ ($r_0\to+\infty$) limits of the expressions defined in equations \eqref{eq:zerodensitychirhorho} and \eqref{eq:zerodensitychiJJ}, and $\ldots$ are terms that vanish as $R_*\rightarrow0$. Earlier, around \eqref{eq:zerodensitysusceptibility}, we noted that after a naive radial separation of the degrees of freedom, the IR contributions to the susceptibilities are cutoff-dependent. In this more refined separation, the role of the counterterms is to remove the cutoff dependence by shifting the full susceptibility into the UV (Goldstone-like) part of the action. 

{}We now obtain our low energy theory by formally taking the limit $R_*\rightarrow0$. The action consists of the Goldstone-like mode in \eqref{eq:renormalisedgoldstoneaction} coupled to the quantum critical degrees of freedom represented by the renormalized IR action in the zero temperature spacetime \eqref{eq:IRmetricFiniteT} (with $f=1$) with a boundary at $R=0$. As they are now described by a standalone holographic theory, we can now be more precise about the nature of these quantum critical degrees of freedom. Varying the renormalized IR action (and imposing the equations of motion in the bulk of the IR spacetime) yields
\begin{equation}
\delta S_{\text{IR}}+\delta S_{\text{ct}}=\frac{Z_0L_x^d}{\tilde{L}L_t}\int d^{d+1}x\left(\Delta_\chi a_t^{(1)}\delta a_t^{(0)}-c_{IR}^2(\Delta_\chi+2(z-1))a_i^{(1)}\delta a_i^{(0)}\right),
\end{equation}
which indicates that we should interpret $\mathcal{J}_\mu=a_\mu^{(0)}$ as the sources of operators in the quantum critical theory, and
\begin{equation}
\label{eq:IRoperatordefns}
\mathcal{O}^t=-\frac{Z_0L_x^d}{\tilde{L}L_t}\Delta_\chi a_t^{(1)},\quad\quad\quad\quad\mathcal{O}^i=\frac{Z_0L_x^{d-2}L_t}{\tilde{L}}\left(\Delta_\chi+2(z-1)\right)a_i^{(1)},
\end{equation}
as their expectation values. From this point of view, the role of the counterterms is to enforce alternate quantisation for these spacetimes, i.e. we identify the field theory sources as the subleading terms in the near-boundary expansions \eqref{eq:ZeroTIRgaugefieldexps}. By treating the $R$ direction as the energy scale in the usual way, and recalling that the quantum critical state has effective dimensionality $(d+z-\theta)$, we can then obtain the following scaling dimensions (with the conventions $[t]=-z$ and $[x]=-1$)
\begin{equation}
\begin{aligned}
\label{eq:IRscalingdimensionszerodensity}
\Delta_{\mathcal{O}_t}&\,=\frac{1}{2}(d+z-\theta+\Delta_\chi),\quad\quad\quad\quad\quad\quad \Delta_{\mathcal{O}_i}=\frac{1}{2}(d+z-\theta+\Delta_\chi+2(z-1)),\\
\Delta_{\mathcal{J}_t}&\,=\frac{1}{2}(d+z-\theta-\Delta_\chi),\quad\quad\quad\quad\quad\quad \Delta_{\mathcal{J}_i}=\frac{1}{2}(d+z-\theta-\Delta_\chi-2(z-1)).
\end{aligned}
\end{equation}
These scaling dimensions are the key properties of the infrared degrees of freedom represented by the scaling spacetime \eqref{eq:IRmetricFiniteT}. The quantity $\Delta_\chi$, which determines whether a higher-form symmetry emerges in the infrared or not, is a parameterisation of the scaling dimensions of the operators dual to the Maxwell field in the IR spacetime.

{}To understand the effect that the interactions with the infrared quantum critical degrees of freedom have on the Goldstone-like mode, it is again helpful to integrate out the IR spacetime. Considering perturbations whose wavevector is aligned with the $x$-axis (again without loss of generality, due to rotational symmetry), this yields a Fourier space effective action for an emergent $U(1)$ gauge field
\begin{equation}
\label{eq:renormalisedIRaction}
S_{\text{IR}}+S_{\text{ct}}=-\frac{1}{2}\int d\omega dk\left(\frac{\tilde{f}_{tx}(-\omega,-k)\tilde{f}_{tx}(\omega,k)}{g_\parallel(\omega,k)}+\frac{\tilde{f}_{tb}(-\omega,-k)\tilde{f}_{tb}(\omega,k)}{g_\perp(\omega,k)}\right),
\end{equation}
where the index $b$ here again runs over the spatial coordinates $\vec{x}$ except $x$, but now $\tilde{f}_{tx}=\omega\tilde{a}_x^{(0)}(\omega,k)+k\tilde{a}_t^{(0)}(\omega,k)$ and $\tilde{f}_{tb}(\omega,k)=\omega\tilde{a}^{(0)}_b(\omega,k)$. As before, this action will be non-local as we have integrated out gapless modes. Since this renormalized action is (minus) the generating functional for the infrared quantum critical theory, we can relate the effective gauge coupling $g_\parallel$ to $G_{\text{IR}}$, the retarded Green's function of $\mathcal{O}^x$ in the critical state (with appropriate normalisation), via
\begin{equation}
g_\parallel(\omega,k)=-\omega^2G_{\text{IR}}(\omega,k)^{-1}.
\end{equation}

{}The effect of the interactions with the gauge field is to modify the dispersion relations of the superfluid-like Goldstone mode to solutions of
\begin{equation}
\label{eq:zeroTdispersionlocus}
0=\frac{\chi_{\rho\rho}}{\chi_{JJ}}\omega^2-k^2+\chi_{\rho\rho}g_\parallel(\omega,k)=\chi_{\rho\rho}\left(\chi_{JJ}^{-1}-G_{IR}^{-1}\right)\omega^2-k^2.
\end{equation}
From the first equality, we see that a superfluid Goldstone-like mode will survive when the coupling $g$ is sufficiently small at low energies $\omega\sim k$. The second equality allows us to immediately deduce when this is the case. From equation \eqref{eq:IRscalingdimensionszerodensity}, the Fourier space IR Green's function $G_{\text{IR}}(\omega,k)$ has dimension $\Delta_\chi+2(z-1)$ and therefore can be written as $G_{\text{IR}}(\omega,k)=\omega^{\frac{\Delta_\chi+2(z-1)}{z}}h(k^z/\omega)$ for a universal scaling function $h$. Assuming that $h(0)$ is finite and $z>1$, then we see that a superfluid Goldstone-like mode survives at $T=0$ provided $\Delta_\chi+2(z-1)<0$.\footnote{An exact expression for $h(0)$ is derived in Appendix \ref{app:zerodensityIRGreens}, where it is also shown that the conclusions below continue to hold for the case $z=1$.} So in every case where there is a long-lived superfluid-like mode at small $T$, this mode survives -- and is gapless -- in the $T=0$ state.

{}By repeating the arguments of Section \ref{sec:higherformbreaking} we see that in the $T=0$ state an anomalous $(d-1)$-form symmetry emerges in the infrared, and the gapless mode is a consequence of this. As before, the coupling to the emergent electromagnetic field $f$ -- which in this case represents the coupling to the infrared quantum critical degrees of freedom -- explicitly breaks this symmetry
\begin{equation}
d\star K=-\bar{F}+f.
\end{equation}
The condition $\Delta_\chi+2(z-1)<0$ ensures that the coupling $g$ is small at low energies and so gives small, power law corrections to observables that vanish as $\omega\rightarrow0$.

{}Nevertheless, as this coupling breaks a symmetry, for some observables these corrections are important. For example, the zero temperature conductivity can be calculated by choosing the gauge $a_t=0$, putting $a_i$ on-shell, and then taking a variational derivative of the resulting action with respect to $\bar{A}_x$ to obtain an expression for the current $j^x$ as a function of $\bar{A}_x$. The result is
\begin{equation}
\label{eq:zeroTconductivitycorrection}
\sigma(\omega)=\frac{i}{\omega}\frac{j^x}{\bar{A}_x}=\chi_{JJ}\frac{i}{\omega}\left(1+\chi_{JJ}h{(0)}^{-1}\omega^{-\frac{\Delta_\chi+2(z-1)}{z}}+\ldots\right),
\end{equation}
which has a small dissipative part at low energies due to the weak explicit symmetry breaking. 

{}Similarly, this symmetry breaking causes non-uniform perturbations of the gapless mode to attenuate slowly. From equation \eqref{eq:zeroTdispersionlocus}, the attenuation is given quantitatively by $G_{\text{IR}}(\omega,k)$ in the limit $\omega\sim k\rightarrow 0$. When $z>1$, the scaling form of the Green's function in $k^z/\omega$ ensures this information is captured by $G_{\text{IR}}(\omega,0)$ and so the renormalized IR action in this limit is rotationally invariant
\begin{equation}
S_{\text{IR}}+S_{\text{ct}}=-\frac{1}{2}\int d\omega dk\frac{\tilde{f}_{ti}(-\omega,-k)\tilde{f}_{ti}(\omega,k)}{g(\omega)},
\end{equation}
where $g(\omega)=-\omega^2G_{\text{IR}}(\omega,0)^{-1}$ and the index $i$ now runs over all $\vec{x}$ coordinates. For the case $z=1$, the renormalized action becomes Lorentz invariant
\begin{equation}
S_{\text{IR}}+S_{\text{ct}}=-\frac{1}{2}\int d\omega dk\frac{\tilde{f}_{ti}(-\omega,-k)\tilde{f}_{ti}(\omega,k)-c_{IR}^2\tilde{f}_{ij}(-\omega,-k)\tilde{f}_{ij}(\omega,k)}{g(\omega^2-c_{IR}^2k^2)},
\end{equation}
where $g(\omega^2-c_{IR}^2k^2)=-\omega^2G_{\text{IR}}(\omega,k)^{-1}$. Using the results in Appendix \ref{app:zerodensityIRGreens} for $G_{\text{IR}}$, the leading real and imaginary terms in the dispersion relation are
\begin{equation}
\label{eq:zne1dispersionzeroT}
\omega(k)=\pm vk-i\frac{\pi L_x^{2-d}\chi_{JJ}}{2Z_0\Gamma\left(\frac{2-\Delta_\chi}{2z}\right)^2}\left(\frac{\tilde{L}v}{2zL_t}\right)^{-\frac{\Delta_\chi+z-2}{z}}k^{1-\frac{\Delta_\chi+2(z-1)}{z}}+\ldots,
\end{equation}
 for $z\ne1$, and
\begin{equation}
\label{eq:z1dispersionzeroT}
\omega(k)=\pm vk-i\frac{\pi L_x^{2-d}\chi_{JJ}}{2Z_0\Gamma\left(1-\frac{\Delta_\chi}{2}\right)^2}\left(1-\frac{c_{IR}^2}{v^2}\right)\left(\frac{\tilde{L}v}{2L_t}\right)^{1-\Delta_\chi}k^{1-\Delta_{\chi}}+\ldots,
\end{equation}
for $z=1$. Note that the imaginary part is always negative due to the lower bound on $v$ proven in Appendix \ref{app:zerodensityspeedbound}. These results should be contrasted with the collective modes in a zero-temperature superfluid, in which self-interactions of the Goldstone mode give rise to a $\sim k^{d+2}$ attenuation. Non-zero temperature corrections to these results could be calculated by replacing the universal scaling function by its thermal generalization $h(T/\omega,k^z/\omega)$.

{}The symmetry breaking terms vanish upon artificially removing the infrared quantum critical degrees of freedom, for example by replacing the IR spacetime with a hard wall. The IR spacetime acts like the `soft wall' of holographic models of QCD, and in this context, the possible emergence of a long-lived mode was pointed out in \cite{Karch:2010eg}. 

\begin{figure}[h]
\begin{center}
\includegraphics[scale=.31]{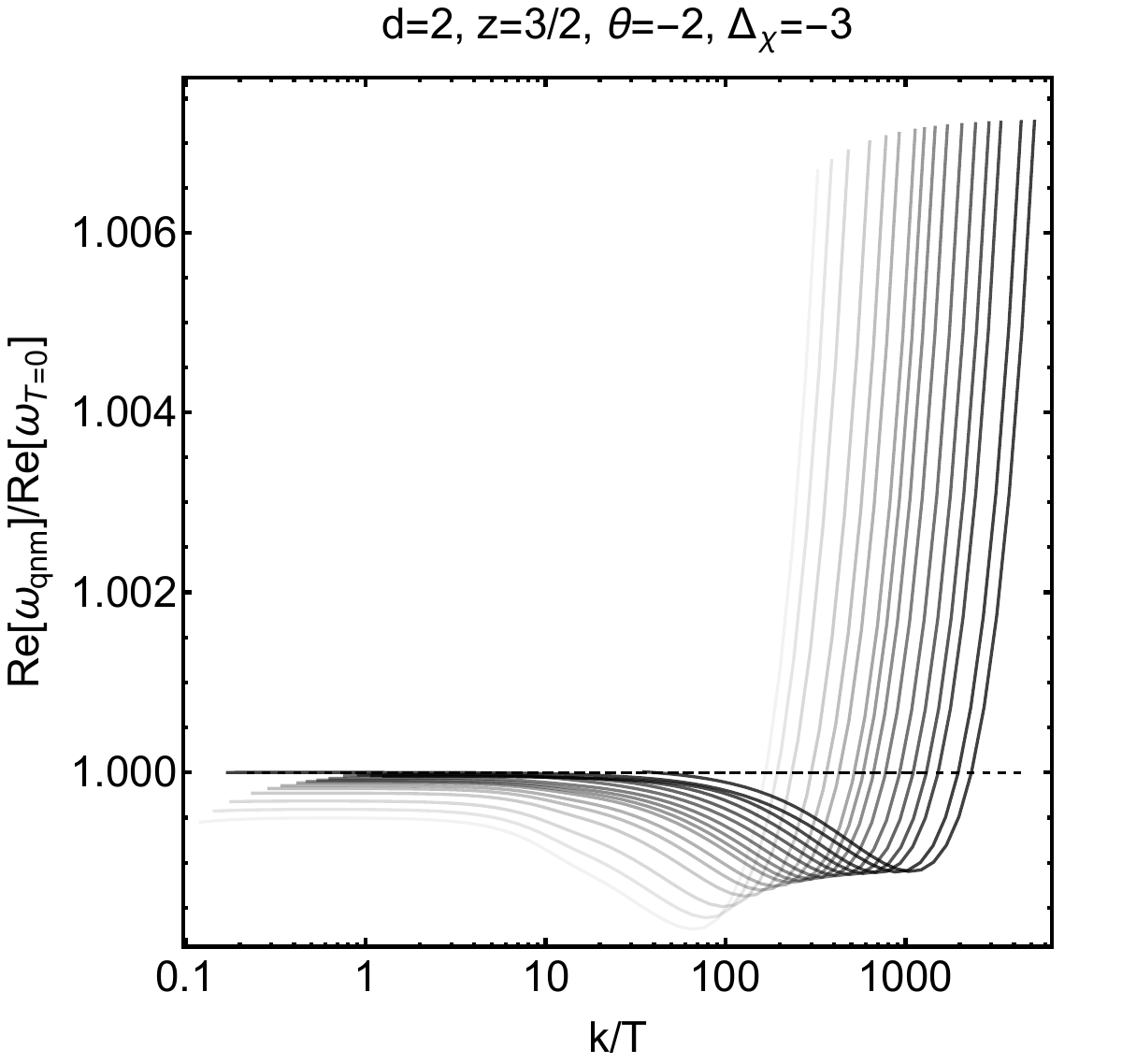}
\includegraphics[scale=.3]{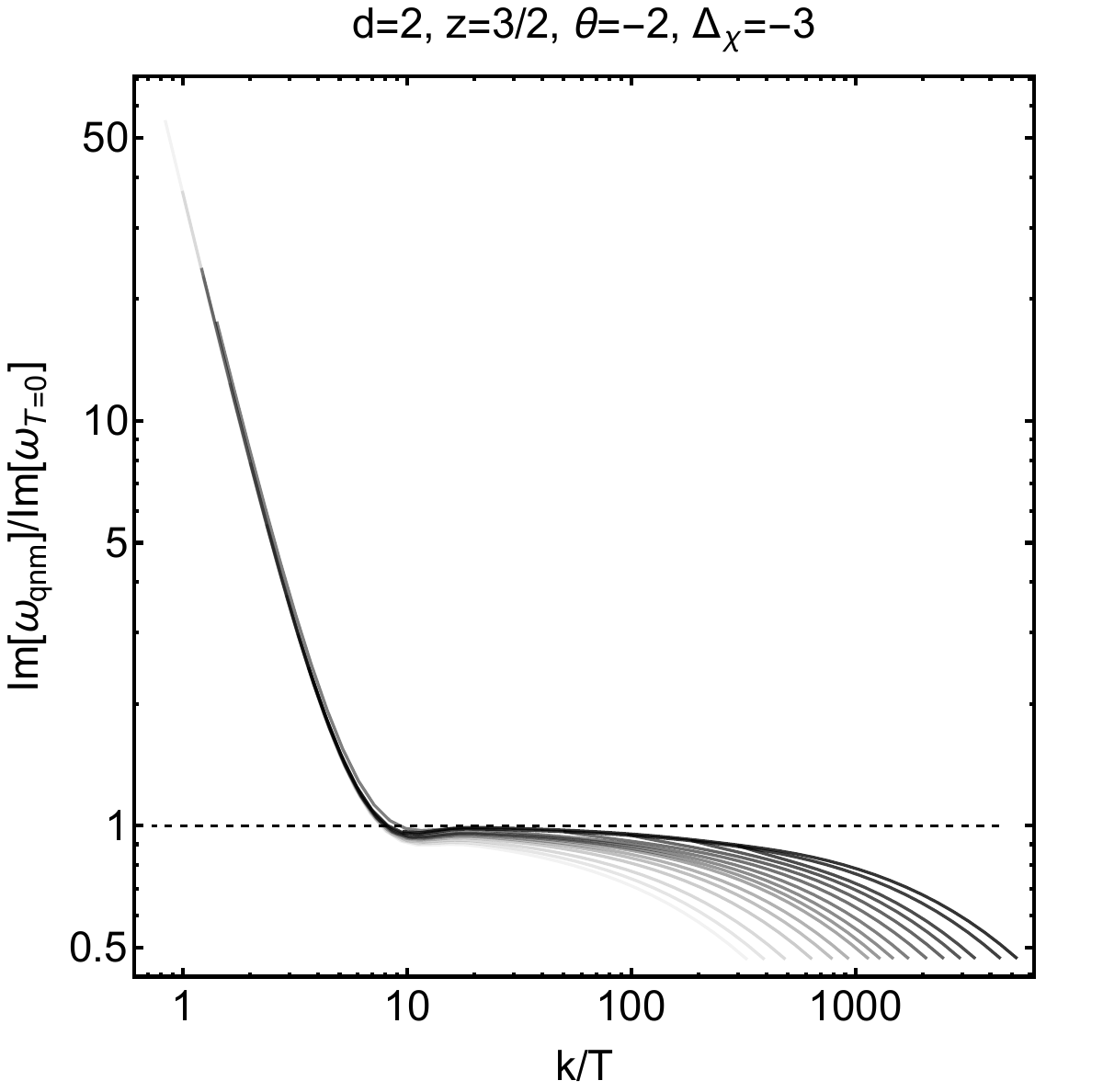}
\includegraphics[scale=.33]{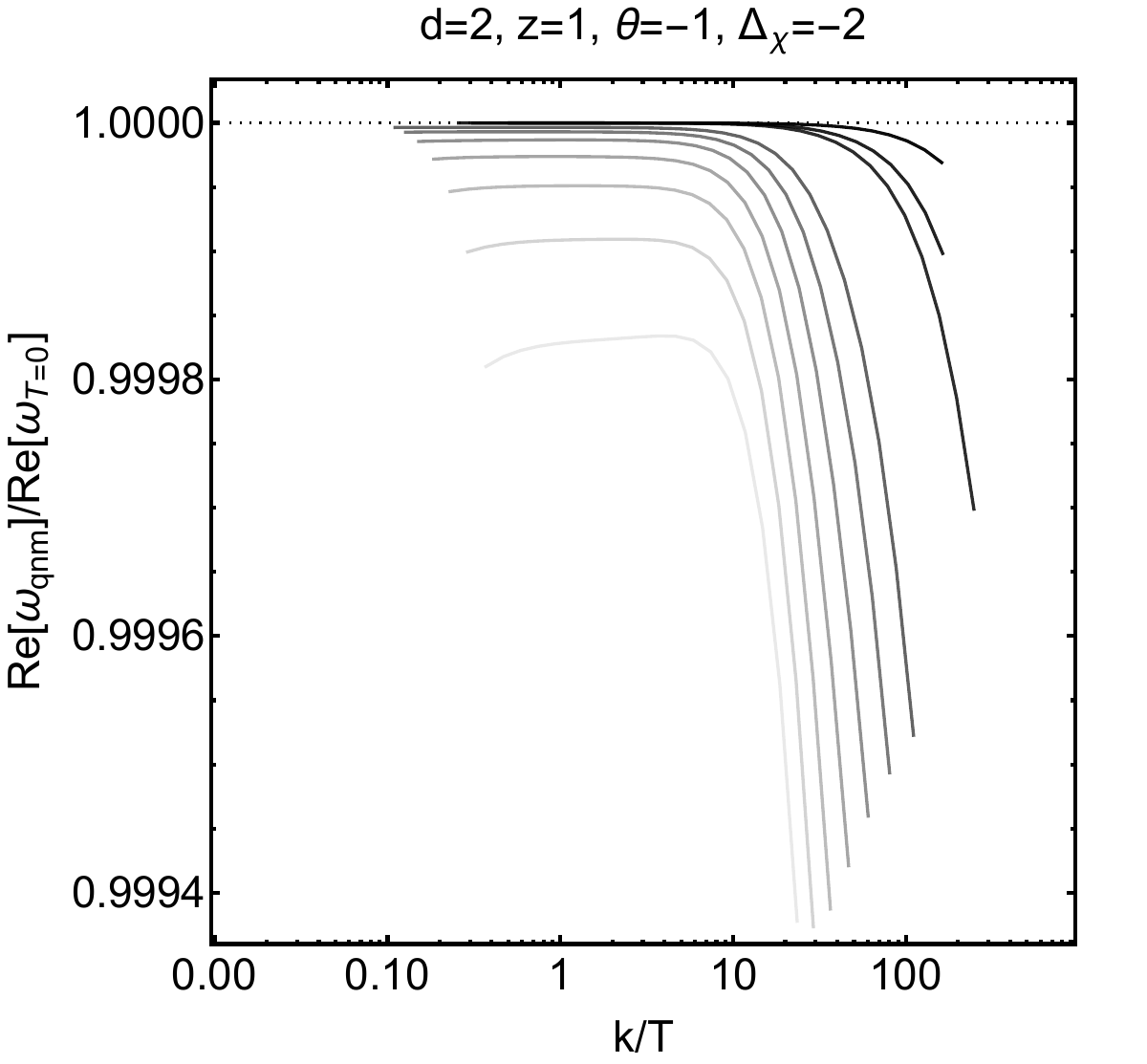}
\includegraphics[scale=.3]{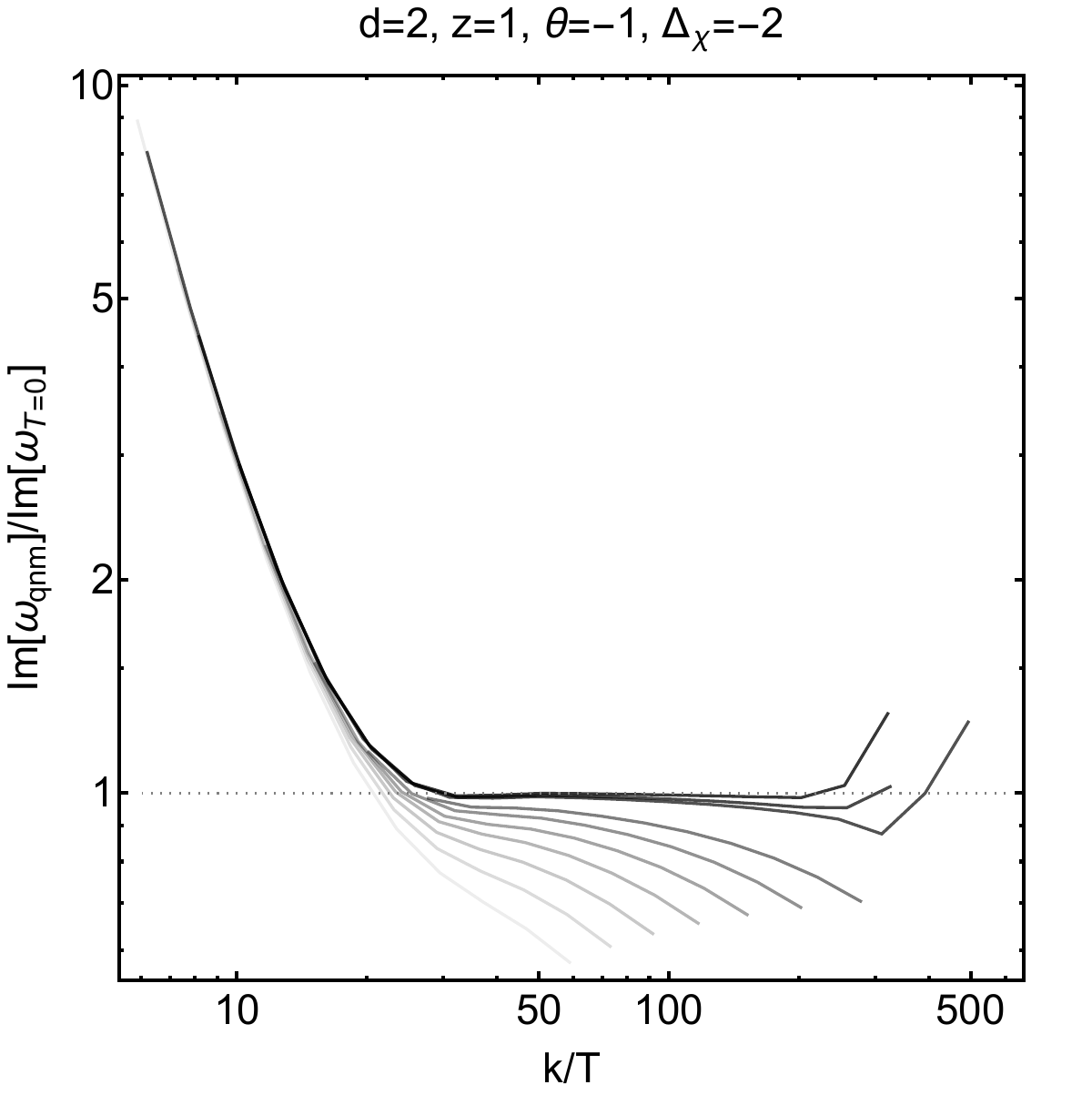}
\caption{\label{intermediatescaling_fig} Comparison of the dispersion relations obtained numerically (solid lines) to the analytic $T=0$ results \eqref{eq:zne1dispersionzeroT} and \eqref{eq:z1dispersionzeroT} (dashed lines) for two examples of states with $\Delta_\chi+2(z-1)<0$.{ Darker curves correspond to lower temperatures in the ranges $T/\mu=3.0\times 10^{-5}$ to $4.8\times 10^{-4}$ (top row, $v=.83$) and $T/\mu=1.2\times 10^{-4}$ to $6.9\times 10^{-3}$ (bottom row, $v=.96$). As $T$ is reduced, the dispersion relations in the range $T\ll(\omega,k)\ll\mu$ are increasingly well approximated by the analytic $T=0$ results, which only apply when $(\omega,k)\ll\mu$.} }
\end{center}
\end{figure}

{}In summary, we have shown that there is an emergent infrared symmetry in the $T=0$ state. Unlike at non-zero $T$ -- where the explicit symmetry breaking causes a crossover at low frequencies $\omega\tau\sim1$ -- the effects of the explicit symmetry breaking at $T=0$ vanish as $\omega,k\rightarrow0$. We can further verify this by numerically calculating the dispersion relations of collective modes of the holographic states in the low temperature regime $\tau^{-1}\ll T\ll\omega,k\ll\mu$, which should connect smoothly to those at $T=0$. This is confirmed in Figure \ref{intermediatescaling_fig}, which illustrates that these modes are becoming gapless and weakly attenuated as $T\rightarrow0$. Further details regarding the numerical calculations can be found in Appendix \ref{app:numericaldetails_zerodensity}.

\subsection{Mixed boundary conditions}

{}While it is often instructive to retain the Goldstone-like field explicitly in the effective action, it can always be gauged away. Doing this corresponds to using radial gauge. We will now show that in this gauge, integrating out the UV region of the spacetime generates boundary terms which are interpreted as double trace operators in the effective action of the infrared theory of the quantum critical degrees of freedom.

{}At zero temperature (and after renormalising the actions as described above), the effective action in radial gauge is
\begin{equation}
S=S_{\text{IR}}+S_{\text{ct}}+\frac{1}{2}\left.\int d^{d+1}x\left(-\chi_{\rho\rho}\left(a_t^{(0)}-\bar{A}_t\right)^2+\chi_{JJ}\left(a_i^{(0)}-\bar{A}_i\right)^2\right)\right|_{R=0}.\end{equation}
On-shell, the external sources $\bar{A}_\mu$ are related to the solutions in the IR region of the spacetime by
\begin{equation}
\bar{A}_t=a_t^{(0)}-\frac{\mathcal{O}^t}{\chi_{\rho\rho}},\quad\quad\quad\quad\quad\quad\bar{A}_i=a_i^{(0)}-\frac{\mathcal{O}^i}{\chi_{JJ}},
\end{equation}
and so the effect of integrating out the UV region is to generate two relevant double-trace terms in the effective action \cite{Faulkner:2010jy}
\begin{equation}
\label{eq:doubletraceeffectiveaction}
S=S_{\text{IR}}+S_{\text{ct}}+\frac{1}{2}\left.\int d^{d+1}x\left(-\frac{\mathcal{O}^t\mathcal{O}^t}{\chi_{\rho\rho}}+\frac{\mathcal{O}^i\mathcal{O}^i}{\chi_{JJ}}\right)\right|_{R=0}.\end{equation}
These terms are both relevant indicating that the low energy physics is not captured simply by $S_{\text{IR}}+S_{\text{ct}}$, but is sensitive to physics in the UV region of the spacetime. In this gauge, it is these double trace terms that incorporate the effects of the Goldstone-like mode. Indeed using \eqref{eq:doubletraceeffectiveaction} as (minus) the generating functional -- with the appropriate mixed boundary conditions on fields due to the double trace deformations \cite{Witten:2001ua,Berkooz:2002ug,Marolf:2006nd} -- correctly reproduces the low energy correlators of $j^\mu$.

{}In other words, if we wish to regard the quantum critical degrees of freedom represented by the IR spacetime as the only ones important at low energies then the sources and operators have to be identified using the mixed boundary conditions above. For holographic theories with higher-form symmetries in the UV, the importance of correctly identifying the appropriate mixed boundary conditions (near the AdS boundary) was emphasized in \cite{Hofman:2017vwr,Grozdanov:2017kyl,Grozdanov:2018ewh}. In the cases we are studying, the emergent higher-form symmetry in the infrared also requires such boundary conditions at the boundary of the IR spacetime. 

\section{Dynamics of non-zero density states}
\label{sec:NonZeroDensityHydroSection}

{}In this Section we will study the dynamics arising from the action \eqref{eq:bulkEMDaction} near equilibrium states of the type described in Section \ref{sec:NonZeroDensityEMD}, rather than the simpler zero density case studied in the previous Section. A preliminary investigation of this was performed in \cite{Davison:2018ofp,Davison:2018nxm}, where it was shown that these states support a parametrically long-lived excitation in cases where the $0$-form charge density operator is irrelevant in the IR. We will extend the results of \cite{Davison:2018ofp,Davison:2018nxm} and derive a complete theory of relaxed hydrodynamics for these states, valid at non-zero wave numbers and in the presence of an external electromagnetic field. We will then discuss how this relates to the hydrodynamics of a phase-relaxed superfluid, before studying the fate of the long-lived mode at zero temperature.

\subsection{Equilibrium states}
\label{sec:EMDblackholes}

{}For simplicity we will focus on a particular class of the equilibrium states described in Section \ref{sec:NonZeroDensityEMD}. Specifically, we will consider the case $d=2$ and where the asymptotic form of the potentials is
\begin{equation}
V(\Phi\rightarrow0)\rightarrow6+\Phi^2+O(\Phi^4),\quad\quad\quad\quad Z(\Phi\rightarrow0)\rightarrow1+O(\Phi^2).
\end{equation}
This corresponds to the bulk field $\Phi$ being dual to a conformal symmetry-breaking scalar operator $O_\Phi$ with UV scaling dimension $\Delta_\Phi=2$. The value of the UV scaling dimension is not important for the IR physics and so this is a very mild restriction. Furthermore, it is conceptually straightforward (but technically tedious) to extend the calculations below to the most general case, or to arbitrary space dimension $d>2$.

{}The equations of motion that follow from the action \eqref{eq:bulkEMDaction} and the background ansatz described in Section \ref{sec:NonZeroDensityEMD} can be compactly expressed as
\begin{equation}
\begin{aligned}
\label{eq:hydroBGeqns}
\frac{CZA_t'}{\sqrt{BD}}&\,=-\rho,\quad\quad\quad\quad\quad\quad\quad \frac{d}{dr}\log\frac{C'}{\sqrt{BCD}}=-\frac{1}{2}\frac{C{\Phi'}^2}{C'},\\
\frac{C^2(D/C)'}{\sqrt{BD}}&\,=-sT-\rho A_t,\quad\quad\quad\quad \frac{d}{dr}\left(\frac{(CD)'}{\sqrt{BD}}\right)=2\sqrt{BD}CV.
\end{aligned}
\end{equation}
Throughout, we will use primes to denote derivatives with respect to $r$ and dots to denote derivatives with respect to $\Phi$. For the case $\Delta_\Phi=2$, the asymptotically AdS solutions \eqref{eq:generalFGexpansioneqm} have the following near-boundary expansion in Fefferman-Graham coordinates
\begin{equation}
\begin{aligned}
\label{eq:BGbdyansatz}
A_t(r\rightarrow0)&\,=\mu-\rho r+O(r^2),\quad\quad\quad \Phi(r\rightarrow0)=Jr+\mathcal{O} r^2+O(r^3),\\
B(r\rightarrow0)&\,=\frac{1}{r^2}+O(r^2),\quad\quad\quad\quad\quad D(r\rightarrow0)=\frac{1}{r^2}-\frac{J^2}{8}-\frac{2p}{3}r+O(r^2),\\
C(r\rightarrow0)&\,=\frac{1}{r^2}-\frac{J^2}{8}+\frac{\left(\varepsilon-J\mathcal{O}\right)}{6}r+O(r^2),
\end{aligned}
\end{equation}
where $J$, $\mathcal{O}$, $\varepsilon$, $p$, $\mu$ and $\rho$ are constants. To give field theory interpretations to these constants we must supplement the action \eqref{eq:bulkEMDaction} with the following boundary terms on the $r=\epsilon$ surface \cite{Skenderis:2002wp,Caldarelli:2016nni}
\begin{equation}
S_{\text{bdy}}=\int d^{3}x\sqrt{-\gamma}\left.\left(2K-4-R[\gamma]-\frac{1}{2}\phi^2\right)\right|_{r=\epsilon},
\end{equation}
where $\gamma$ is the induced metric, $R$ is the corresponding Ricci scalar, and $K$ is the trace of the extrinsic curvature. Following this, we have a well-defined holographic dictionary: $J$ is the source of the relevant operator $O_\Phi$ with expectation value $\langle O_\Phi\rangle=\mathcal{O}$. $\mu$ is the chemical potential of a $U(1)$ charge operator and
\begin{equation}
\label{eq:equilibriumexpectationvalues}
\langle T^{tt}\rangle = \varepsilon,\quad\quad\quad\quad \langle T^{xx}\rangle=\langle T^{yy}\rangle = p,\quad\quad\quad\quad \langle j^t\rangle=\rho,
\end{equation}
are the expectation values of the $U(1)$ charge density $j^t$ and the diagonal components of the energy-momentum tensor. The constants are not all independent but are required by the equations of motion \eqref{eq:hydroBGeqns} to satisfy the equilibrium Ward identity for scale transformations
\begin{equation}
\label{eq:eqmTraceIdentity}
-\varepsilon+2p=J\mathcal{O}.
\end{equation}
Furthermore, evaluating the third equation of motion in \eqref{eq:hydroBGeqns} at the boundary yields the Smarr relation
\begin{equation}
\label{eq:Smarrrelation}
\varepsilon+p=sT+\rho\mu,
\end{equation}
where $s$ and $T$ are again the entropy density and temperature of the state. The equations of motion \eqref{eq:hydroBGeqns} and identities \eqref{eq:eqmTraceIdentity} and \eqref{eq:Smarrrelation} are used repeatedly to simplify expressions in the following calculations. The equation of state $p(T,\mu)$ of the family of black hole solutions cannot be determined by the asymptotic analysis we have done here, and instead requires more explicit knowledge of the solutions of the equations of motion \eqref{eq:BGbdyansatz}. 

\subsection{Near-equilibrium dynamics}

{}The black hole solutions we have just described are non-conformal, thermal, charged equilibrium states of a theory with microscopic Lorentz invariance but, in general, no conformal invariance. Therefore we expect that at low wavenumbers and frequencies their perturbations to be governed by the theory of relativistic, non-conformal charged hydrodynamics. The first element of our results is a general proof of the non-zero density, non-conformal, linearized fluid/gravity correspondence to first order in the hydrodynamic expansion (this is a generalization of the non-zero density conformal case proven in \cite{Erdmenger:2008rm,Banerjee:2008th} and the zero density, non-conformal case proven in \cite{Donos:2022uea}).\footnote{This part of our results could also be obtained from the recent work \cite{Donos:2022www} on non-conformal superfluid hydrodynamics by taking the limit where there is no condensate.} We also include the effects of one ostensibly higher-order relaxation term, which we show in fact becomes parametrically large in the low temperature limit. The result is a theory of relaxed hydrodynamics, analogous to that obtained in Section \ref{sec:zerodensityholoquasihydro} for the zero density states.

\subsubsection{Perturbation variables}

{}To derive the relaxed hydrodynamic theory, we study linearized perturbations around the equilibrium state. We label the field theory spatial directions $x$ and $y$ and use rotational invariance to consider perturbations that depend on $t$, $x$ and $r$ without loss of generality. With these conventions, we denote the metric perturbations $\delta g_{MN}$ as 
\begin{equation}
\begin{aligned}
&\,\delta g_{tt}\equiv-Dh_{tt},\quad \delta g_{xx}\equiv Ch_{xx},\quad \delta g_{yy}\equiv Ch_{yy},\quad \delta g_{rr}\equiv -2\sqrt{B}\partial_r\left(\frac{\zeta_r}{\sqrt{B}}\right),\\
&\,\delta g_{xt}\equiv Ch_{xt},\quad\,\,\,\,\,\delta g_{rt}\equiv -\partial_t\zeta_r-D\partial_r\left(\frac{\zeta_t}{D}\right),\quad\,\,\,\,\delta g_{xr}\equiv-\partial_x\zeta_r-C\partial_r\left(\frac{\zeta_x}{C}\right),\\
&\,\delta g_{yt}\equiv Ch_{yt},\quad\,\,\,\,\, \delta g_{xy}\equiv Ch_{xy},\quad\quad\quad\quad\quad\quad\quad\,\,\, \delta g_{yr}\equiv-C\partial_r\left(\frac{\zeta_y}{C}\right).
\end{aligned}
\end{equation}
The reparameterisation of the components $\delta g_{rM}$ in terms of the functions $\zeta_M$ is for later convenience. The linearized perturbations of the matter fields are labelled
\begin{equation}\
\delta A_t\equiv a_t,\quad\quad\delta A_i\equiv a_i,\quad\quad \delta A_r\equiv A_t\partial_r\left(\frac{\zeta_t}{D}\right)-\partial_r\Lambda,\quad\quad \delta\Phi\equiv \phi,
\end{equation}
where we have also reparameterised $\delta A_r$ in terms of a new function $\Lambda$ for later convenience.

{}It will be convenient to work not directly with the perturbations as listed above, but with the tilded fields defined as follows
\begin{equation}
\begin{aligned}
\label{eq:splitfields}
&\,h_{xx}+h_{yy}=\tilde{h}_+-\frac{2C'}{BC}\zeta_r-\frac{2}{C}\partial_x\zeta_x,\quad\quad\quad\quad\quad h_{tt}=\tilde{h}_{tt}+\frac{2}{D}\partial_t\zeta_t-\frac{D'}{BD}\zeta_r, \\
&\,h_{xx}-h_{yy}=\tilde{h}_--\frac{2}{C}\partial_x\zeta_x,\quad\quad\quad\quad\quad\quad\quad\quad\,\,\,\, h_{it}=\tilde{h}_{it}-\frac{1}{C}\left(\partial_t\zeta_i+\partial_i\zeta_t\right),\\
&\,h_{xy}=\tilde{h}_{xy}-\frac{1}{C}\partial_x\zeta_y,\quad\quad\quad\quad\quad\quad\quad\quad\quad\quad\quad\,\,\,\, \phi=\tilde{\phi}-\frac{\Phi'}{B}\zeta_r,\\
&\,a_t=\tilde{a}_t+\frac{A_t}{D}\partial_t\zeta_t-\frac{A_t'}{B}\zeta_r-\partial_t\Lambda,\quad\quad\quad\quad\quad\quad\,\,\, a_i=\tilde{a}_i+\frac{A_t}{D}\partial_i\zeta_t-\partial_i\Lambda.
\end{aligned}
\end{equation}
The field equations for these perturbations are given in Appendix \ref{app:perturbationeoms}. We will neglect terms of $O(\partial^2)$ in these equations, as these will produce only higher-order corrections to the theory of relaxed hydrodynamics. This means we only have to deal with 14 of the field equations. The tilded fields can be thought of as the metric perturbations in radial gauge, while the $\zeta_M$ and $\Lambda$ are gauge transformations (up to terms of $O(\partial^2)$). 

{}Due to the gauge invariance of the linearized equations of motion, the 14 equations involve only the 10 tilded variables. Once they are solved, the general solution for the linearized perturbations is obtained by combining these solutions with any functions $\zeta_M$ and $\Lambda$ as shown in equation \eqref{eq:splitfields}. In order to extract near-equilibrium properties of the field theory from the solutions, it is helpful to work in Fefferman-Graham gauge near the boundary. In order to achieve this, we restrict the pure gauge solutions to have the asymptotic form 
\begin{equation}
\begin{aligned}
&\,\zeta_t(r,t,x)=D(r)\xi_t(t,x)+O(r^2),\quad\zeta_i(r,t,x)=C(r)\xi_i(t,x)+O(r^2),\\
&\,\zeta_r(r,t,x)=O(r^3),\quad\quad\quad\quad\quad\quad\quad\,\, \Lambda(r,t,x)=\lambda(t,x)+O(r^2),
\end{aligned}
\end{equation}
as $r\rightarrow0$. The functions $\lambda$, $\xi_t$ and $\xi_i$ are large gauge transformations and are taken to be $O(\partial^{-1})$ in the derivative expansion. With this restriction, the holographic dictionary relating near-equilibrium properties of the field theory to solutions for the linearized perturbations is given in Appendix \ref{app:holodictionaryperturbations}. For simplicity, we will restrict ourselves to the case where the perturbations of the field theory metric and perturbations of the source for the scalar operator vanish.

\subsubsection{Bulk hydrodynamic variables}

{}The key step in deriving the relaxed hydrodynamic theory is solving the linearized equations to $O(\partial)$. To make this manageable, it is helpful to introduce one final set of perturbation variables as follows
\begin{equation}
\begin{aligned}
\label{eq:bulkhydrofieldsdefn}
\mathcal{E}(r)&\,= C\sqrt{\frac{D}{B}}\left(\sqrt{\frac{D}{C}}\frac{d}{dr}\left(\frac{\tilde{h}_+}{\sqrt{D/C}}\right)+\Phi'\tilde{\phi}\right),\quad\quad \mathcal{P}_i(r)= -\sqrt{\frac{D^3}{B}}\frac{d}{dr}\left(\frac{\tilde{h}_{it}}{D/C}\right),\\
\mathcal{T}_-(r)&\,= C\sqrt{\frac{D}{B}}\tilde{h}_{-}',\quad\quad\quad\quad\quad\quad\quad\quad\quad\quad\quad\quad\quad \mathcal{T}_{\times}(r)=C\sqrt{\frac{D}{B}}\tilde{h}_{xy}',\\
\mathcal{T}_+(r)&\,=-2C\sqrt{\frac{D}{B}}\left(\tilde{h}_{tt}'+\frac{1}{2}\tilde{h}_{+}'+\Phi'\tilde{\phi}\right)+(sT+\rho A_t)\tilde{h}_{tt}-2\rho\tilde{a}_t,\\
\mathcal{Q}(r)&\,= -\frac{CZ}{\sqrt{BD}}\left(\tilde{a}_t'-\frac{1}{2}A_t'(\tilde{h}_{tt}-\tilde{h}_+)+\frac{A_t'\dot{Z}}{Z}\tilde{\phi}\right),\\
\mathcal{J}_i(r)&\,= \sqrt{\frac{D}{B}}Z\left(\tilde{a}_i'+\frac{A_t'\tilde{h}_{it}}{(D/C)}\right).
\end{aligned}
\end{equation}
One reason that these variables are useful is that they allow 9 of the equations of motion to be written as radial conservation equations
\begin{equation}
\begin{aligned}
\label{eq:bulkradialconservationeqs}
&\frac{d}{dr}\left(\mathcal{E}(r)-A_t(r)\mathcal{Q}(r)\right)=0,\quad\quad\,\,\, \frac{d}{dr}\left(\mathcal{T}_\pm(r)\right)=0,\quad\quad\,\,\, \frac{d}{dr}\left(\mathcal{T}_{\times}(r)\right)=0,\\
&\frac{d}{dr}\left(\mathcal{P}_i(r)-A_t(r)\mathcal{J}_i(r)\right)=0,\quad\quad \frac{d}{dr}\left(\mathcal{Q}(r)\right)=0,\quad\quad\quad \frac{d}{dr}\left(\mathcal{J}_i(r)\right)=0,
\end{aligned}
\end{equation}
up to terms of $O(\partial^2)$. Furthermore, to this order each of these fields has a clear field theory interpretation. Using the holographic dictionary in Appendix \ref{app:holodictionaryperturbations}
\begin{equation}
\begin{aligned}
\label{eq:bulkhydroconstants}
&\,\mathcal{E}(r)-A_t(r)\mathcal{Q}(r)=\langle\delta T^{tt}\rangle-\mu\langle\delta j^t\rangle+sT\partial_x\xi_x,\quad\quad\,\,\, \mathcal{Q}(r)=\langle \delta j^t\rangle+\rho\partial_x\xi_x,\\
&\,\mathcal{P}_i(r)-A_t(r)\mathcal{J}_i(r)=\langle\delta T^{ti}\rangle-\mu\langle\delta j^i\rangle-sT\partial_t\xi_i,\quad\quad\, \mathcal{J}_i(r)=\langle \delta j^i\rangle-\rho\partial_t\xi_i,\\
&\,\mathcal{T}_-(r)=\langle\delta T^{xx}\rangle-\langle\delta T^{yy}\rangle,\quad\quad\quad\quad\quad\quad\quad\quad\quad\quad\quad\,\, \mathcal{T}_{\times}(r)=\langle\delta T^{xy}\rangle,\\
&\,\mathcal{T}_+(r)=\langle\delta T^{xx}\rangle+\langle\delta T^{yy}\rangle-2sT\partial_t\xi_t-2\rho\left(\delta a_t+\partial_t\lambda\right).
\end{aligned}
\end{equation}
These objects are the bulk variables that conceptually encode the perturbations of entropy density, charge density, heat current, current density, and the components of the stress tensor. Consistency of the solution will later require us to fix the large gauge transformations $\xi_\mu$ and $\lambda$ in terms of sources and expectation values of field theory operators.

{}In addition to the radial conservation equations, the equations of motion imply that these fields also must obey 5 further equations which ensures the field theory obeys the appropriate Ward identities. The first
\begin{equation}
\begin{aligned}
\label{eq:bulktraceidentity}
0=&\,-\frac{C'}{C}\left(-\frac{\mathcal{T}_+(r)}{2}+\frac{1}{2}(sT+\rho A_t)\tilde{h}_{tt}-\rho\tilde{a}_t\right)-\frac{D'}{2D}\left(\mathcal{E}(r)-\frac{1}{2}(sT+\rho A_t)\tilde{h}_+\right)\\
&\,+A_t'\mathcal{Q}(r)-\frac{1}{2}\rho A_t'\tilde{h}_+-2D\sqrt{\frac{C}{B}}\frac{d}{dr}\left(\frac{C'}{\sqrt{BCD}}\right)\frac{d}{dr}\left(\frac{\sqrt{B}}{\Phi'}\tilde{\phi}\right),
\end{aligned}
\end{equation}
implements the Ward identity for the trace of the energy-momentum tensor. To see this, we can use the holographic dictionary to evaluate \eqref{eq:bulktraceidentity} near the boundary and obtain
\begin{equation}
\begin{aligned}
\langle\delta T^{tt}\rangle-\langle\delta T^{xx}\rangle-\langle\delta T^{yy}\rangle+J\langle\delta O_\phi\rangle=0,\\
\end{aligned}
\end{equation}
which is simply the linear perturbation of the trace Ward identity $\langle T^\mu_{\;\;\mu}\rangle=J\langle O_\phi\rangle$ when perturbations of the field theory metric and scalar source vanish. 

{}The remaining 4 equations are
\begin{equation}
\begin{aligned}
\label{eq:bulkwardconservations}
&\,\partial_t\mathcal{E}+\partial_i\mathcal{P}_i=0,\\
&\,\partial_t\left(\mathcal{P}_x+(sT+\rho A_t)\tilde{h}_{xt}+\rho\frac{D}{C}\tilde{a}_x\right)+\frac{D}{C}\partial_x\left(\frac{\mathcal{T}_++\mathcal{T}_-}{2}\right)=0,\\
&\,\partial_t\left(\mathcal{P}_y+(sT+\rho A_t)\tilde{h}_{yt}+\rho\frac{D}{C}\tilde{a}_y\right)+\frac{D}{C}\partial_x\mathcal{T}_{\times}=0,\\
&\,\partial_t\mathcal{Q}+\partial_i\mathcal{J}_i=0.
\end{aligned}
\end{equation}
Ultimately we will see that these implement the Ward identities for the local conservation of energy, momentum and $0$-form charge. For now, we will focus on finding solutions to the 10 equations \eqref{eq:bulkradialconservationeqs} and \eqref{eq:bulktraceidentity} which will yield the hydrodynamic constitutive relations.

\subsubsection{Solutions for $\tilde{h}_-$ and $\tilde{h}_{xy}$}

{}In equations \eqref{eq:bulkradialconservationeqs} and \eqref{eq:bulktraceidentity}, the field $\tilde{h}_-$ decouples and obeys the equation
\begin{equation}
C\sqrt{\frac{D}{B}}\tilde{h}_{-}'=\langle\delta T^{xx}\rangle-\langle\delta T^{yy}\rangle.
\end{equation}
It is straightforward to integrate this to obtain the general solution
\begin{equation}
\begin{aligned}
\label{eq:hminustildesol}
\tilde{h}_-=&\,2\partial_x\xi_x+\left(\langle\delta T^{xx}\rangle-\langle\delta T^{yy}\rangle\right)\int^r_0\sqrt{\frac{B(\bar{r})}{D(\bar{r})}}\frac{1}{C(\bar{r})}d\bar{r}.
\end{aligned}
\end{equation}
This depends on two integration constants which we have labelled using the holographic dictionary. Consistency will later require us to fix $\xi_x$ in terms of the sources and expectation values of field theory operators. 

{}The solution \eqref{eq:hminustildesol} is valid up to corrections of $O(\partial^2)$. In other words, the integration constants $\partial_x\xi_x$ and $\langle\delta T^{xx}\rangle - \langle\delta T^{yy}\rangle$ present in the solution \eqref{eq:hminustildesol} are not necessarily constants in the field theory spacetime coordinates, but are slowly varying functions of $(t,x)$, Specifically, in Fourier space they are linear functions of $(\omega,k)$. 

{}The field $\tilde{h}_{xy}$ also decouples in equations \eqref{eq:bulkradialconservationeqs} and \eqref{eq:bulktraceidentity} and has the solution
\begin{equation}
\tilde{h}_{xy}=\partial_x\xi_y+\langle\delta T^{xy}\rangle\int^r_0\sqrt{\frac{B(\bar{r})}{D(\bar{r})}}\frac{1}{C(\bar{r})}d\bar{r},
\end{equation}
where the integration constants $\partial_x\xi_y$ and $\langle\delta T^{xy}\rangle$ can be taken to be linear functions of $\omega$ and $k$ in Fourier space for the same reason as above.

\subsubsection{Solution for $\tilde{h}_{it}$ and $\tilde{a}_i$}

{}In equations \eqref{eq:bulkradialconservationeqs} and \eqref{eq:bulktraceidentity}, the fields $\tilde{h}_{it}$ and $\tilde{a}_i$ are coupled to one another (but decoupled from all others) and obey the equations
\begin{equation}
\begin{aligned}
\sqrt{\frac{D^3}{B}}\frac{d}{dr}\left(\frac{\tilde{h}_{it}}{D/C}\right)+\sqrt{\frac{D}{B}}ZA_t\left(\tilde{a}_i'+\frac{A_t'}{(D/C)}\tilde{h}_{it}\right)&\,=-\langle\delta T^{ti}\rangle+\mu\langle\delta j^i\rangle+sT\partial_t\xi_i,\\
 \sqrt{\frac{D}{B}}Z\left(\tilde{a}_i'+\frac{A_t'}{(D/C)}\tilde{h}_{it}\right)&\,=\langle \delta j^i\rangle-\rho\partial_t\xi_i.
\end{aligned}
\end{equation}
Solving these and using the holographic dictionary to appropriately name the integration constants gives
\begin{equation}
\begin{aligned}
\label{eq:hxttildesol}
\tilde{h}_{it}=&\,\partial_t\xi_i+\frac{D}{C}\partial_i\xi_t+\frac{(\mu-A_t)\langle \delta j^i\rangle}{sT+\rho A_t}+\frac{\langle\delta T^{ti}\rangle}{\varepsilon+p}\left(\frac{D}{C}-\frac{\varepsilon+p}{sT+\rho A_t}\right)\\
&\,-\left(-\langle\delta T^{ti}\rangle+\frac{\varepsilon+p}{\rho}\langle \delta j^i\rangle\right)\frac{D}{C}\int^r_0\frac{C(\bar{r})}{D(\bar{r})}\frac{d}{d\bar{r}}\left(\frac{1}{sT+\rho A_t(\bar{r})}\right)d\bar{r},\\
\tilde{a}_i=&\,\delta \bar{A}_i+\partial_i(\lambda-A_t\xi_t)+(\mu-A_t)\frac{\langle\delta T^{ti}\rangle}{\varepsilon+p}\\
&\,+\left(-\langle\delta T^{ti}\rangle+\frac{\varepsilon+p}{\rho}\langle \delta j^i\rangle\right)\left(A_t+\frac{sT}{\rho}\right)\int^r_0\frac{C(\bar{r})}{D(\bar{r})}\frac{d}{d\bar{r}}\left(\frac{1}{sT+\rho A_t(\bar{r})}\right)d\bar{r}.
\end{aligned}
\end{equation}
Due to the same argument as around equation \eqref{eq:hminustildesol}, the perturbed sources, expectation values and the derivatives of the large gauge transformations in this solution are not strictly constants but slowly varying functions of the field theory spacetime coordinates (specifically, linear functions of $(\omega,k)$ in Fourier space).

\subsubsection{Solution for $\tilde{h}_{tt}$ and $\tilde{h}_+$, $\tilde{a}_t$ and $\tilde{\phi}$}

{}We now turn to the solutions of \eqref{eq:bulkradialconservationeqs} and \eqref{eq:bulktraceidentity} for the remaining fields $\tilde{h}_{tt}$ and $\tilde{h}_+$, $\tilde{a}_t$ and $\tilde{\phi}$. Although we cannot write down exact closed form solutions for these fields, their existence and asymptotic properties will ultimately be enough to obtain the appropriate theory of relaxed hydrodynamics.

{}As the perturbation equations are linear, it will be helpful to consider the solutions for these fields as the sum of a `thermodynamic' and a `dissipative' solution i.e.~
\begin{equation}
\begin{aligned}
\tilde{h}_{tt}(r)=\tilde{h}_{tt}^{\text{th}}(r)+\tilde{h}_{tt}^{\text{dis}}(r),\quad\quad\quad\quad \mathcal{E}(r)=\mathcal{E}^{\text{th}}(r)+\mathcal{E}^{\text{dis}}(r),
\end{aligned}
\end{equation}
and similarly for the other fields. The thermodynamic solution is analogous to the one described in \cite{Donos:2017ihe} and captures thermodynamic properties of the field theory such as static susceptibilities of heat and charge. The dissipative solution captures the dissipative properties including the bulk viscosity.

{}\textit{Thermodynamic solution}

{}We can infer the existence and properties of the thermodynamic solution from a straightforward argument. For a given field theory metric and scalar source, there will be a two-parameter family of black hole solutions that we can characterise by the entropy density $s$ and charge density $\rho$. So there will exist a family of solutions characterised by $C(r,s,\rho)$ (and similar for $B$, $D$, $A_t$ and $\Phi$) where there is a smooth dependence on $s$ and $\rho$. 

{}This means that, given a black hole solution with a specific $s$ and $\rho$, the linear perturbation of this spacetime
\begin{equation}
\label{eq:perturbedspacetimedefns}
\delta g_{xx}=\delta C\equiv\left(\frac{\partial C}{\partial s}\right)_{\rho,r}\delta s+\left(\frac{\partial C}{\partial \rho}\right)_{s,r}\delta\rho,
\end{equation}
(and analogously for the other perturbations) will be a solution to the linearized gravitational equations at $O(\partial^0)$ for any arbitrary constants $\delta s$ and $\delta\rho$. Using the definitions of the tilded variables \eqref{eq:splitfields}, we can therefore write down the thermodynamic solution
\begin{equation}
\begin{aligned}
\label{eq:thermosol}
\tilde{h}_{tt}^{\text{th}}&\,=\frac{\delta D}{D}-\frac{D'}{2D\sqrt{B}}\int^r_0\frac{\delta B(\bar{r})}{\sqrt{B(\bar{r})}}d\bar{r},\quad\quad\quad \tilde{h}_{+}^{\text{th}}=\frac{2\delta C}{C}-\frac{C'}{C\sqrt{B}}\int^r_0\frac{\delta B(\bar{r})}{\sqrt{B(\bar{r})}}d\bar{r},\\
\tilde{a}_{t}^{\text{th}}&\,=\delta A_t-\frac{A_t'}{2\sqrt{B}}\int^r_0\frac{\delta B(\bar{r})}{\sqrt{B(\bar{r})}}d\bar{r},\quad\quad\quad\,\,\,\, \tilde{\phi}^{\text{th}}=\delta\Phi-\frac{\Phi'}{2\sqrt{B}}\int^r_0\frac{\delta B(\bar{r})}{\sqrt{B(\bar{r})}}d\bar{r}.\\
\end{aligned}
\end{equation}
By construction, the thermodynamic solution satisfies equations \eqref{eq:bulkradialconservationeqs} and \eqref{eq:bulktraceidentity} for any constants $\delta s$ and $\delta\rho$. Since the corrections to these equations appear at $O(\partial^2)$, we can promote $\delta s$ and $\delta\rho$ to slowly varying functions of $t$ and $x$ (linear functions in Fourier space) while still satisfying equations \eqref{eq:bulkradialconservationeqs} and \eqref{eq:bulktraceidentity}.

{}This two-parameter family of solutions is the thermodynamic solution, and the slowly varying parameters $\delta s$ and $\delta\rho$ will eventually become the hydrodynamic variables. We define the slowly varying perturbation of the energy density in terms of these parameters as
\begin{equation}
\begin{aligned}
\label{eq:thermoperturbdefns}
\delta\varepsilon \equiv \left(\frac{\partial\varepsilon}{\partial s}\right)_{\rho}\delta s+\left(\frac{\partial\varepsilon}{\partial\rho}\right)_{s}\delta\rho,
\end{aligned}
\end{equation}
and similarly for $\delta p$, $\delta T$, $\delta\mu$, $\delta\mathcal{O}$ etc. It is straightforward to verify using the holographic dictionary that $\langle\delta j^t\rangle=\delta\rho$ and $\langle\delta T^{tt}\rangle=\delta\varepsilon$ for this solution.

{}Although we do not have closed form expressions for the equilibrium functions appearing in the thermodynamic solution \eqref{eq:thermosol}, our knowledge of their near-boundary and horizon properties will be sufficient for what follows. For now we will use this to derive three useful identities that we will impose from now on. First, by evaluating the equation \eqref{eq:bulktraceidentity} for the thermodynamic solution near the boundary we find
\begin{equation}
-\delta\varepsilon +2\delta p = J \delta\mathcal{O},
\end{equation}
which is the perturbed version of the equilibrium Ward identity \eqref{eq:eqmTraceIdentity}, where the quantities appearing in it are now slowly varying functions of space and time. The other two useful identities are found from the radial conservation laws \eqref{eq:bulkradialconservationeqs} for $\mathcal{E}^{\text{th}}-A_t\mathcal{Q}^{\text{th}}$ and $\mathcal{T}_+^{\text{th}}$. Evaluating these quantities at the horizon and boundary and equating them gives 
\begin{equation}
\label{eq:firstlawholo}
\delta\varepsilon=T\delta s+\mu\delta\rho,\quad\quad\quad\text{and}\quad\quad\quad \delta p=s\delta T+\rho\delta\mu,
\end{equation}
respectively. These are the first law of thermodynamics and the Gibbs-Duhem relation, generalised to the case where the perturbations are slowly varying functions of space and time.

{}\textit{Dissipative solution}

{}The solution above is thermodynamic in nature and so does not capture the dissipative effects of bulk viscosity in the trace of the stress tensor. To construct the dissipative solution that captures this, it is helpful to consider the field
\begin{equation}
\mathcal{Q}_{\text{inc}}(r)\equiv\rho\left(\mathcal{E}(r)-A_t(r)\mathcal{Q}(r)\right)-Ts\mathcal{Q}(r),
\end{equation}
which is radially conserved (up to terms of $O(\partial^2)$) under the equations of motion. Using the expressions \eqref{eq:bulkhydroconstants}, this field captures the perturbation of the field theory's incoherent charge density \cite{Davison:2015taa}. The missing dissipative solution has $\mathcal{Q}_{\text{inc}}=0$: this is because in hydrodynamics perturbations of the incoherent charge density decouple from perturbations of the trace of the stress tensor \cite{Davison:2018nxm}. Under the condition $\mathcal{Q}_{\text{inc}}=0$, equations \eqref{eq:bulkradialconservationeqs} and \eqref{eq:bulktraceidentity} can be combined to derive the equation (up to corrections of $O(\partial^2)$)
\begin{equation}
\label{eq:bulkviscosityeqn}
\frac{d}{dr}\left(C\sqrt{\frac{D}{B}}\frac{C^2{\phi'}^2}{{C'}^2}\psi'(r)\right)-\frac{d}{dr}\left(\frac{1}{\sqrt{BD}}\frac{d}{dr}\left(\frac{C^3(D/C)'}{C'}\right)\right)\psi(r)=0,
\end{equation}
for the field
\begin{equation}
\psi\equiv \frac{1}{2}\tilde{h}_+-\frac{C'}{C\Phi'}\tilde{\phi}-\frac{\mathcal{E}-A_t\mathcal{Q}}{sT}.
\end{equation}
This equation is a gauge-invariant version of the bulk viscosity equation in \cite{Gubser:2008sz}, generalised to charged black holes in AdS$_4$.

{}We already know one solution to the equation \eqref{eq:bulkviscosityeqn}. For the thermodynamic solution described previously, $\mathcal{Q}_{\text{inc}}=T(\rho\delta s-s\delta\rho)$. Thus if we set $\delta\rho=\rho\delta s/s$ in the thermodynamic solution, the fields will satisfy equations \eqref{eq:bulkradialconservationeqs} and \eqref{eq:bulktraceidentity} with $\mathcal{Q}_{\text{inc}}=0$. We label this solution $\psi_{\text{inc}}$ where
\begin{equation}
\psi_{\text{inc}}\equiv \frac{1}{2}\tilde{h}^{\text{inc}}_+-\frac{C'}{C\Phi'}\tilde{\phi}^{\text{inc}}-\frac{\mathcal{E}^{\text{inc}}-A_t\mathcal{Q}^{\text{inc}}}{sT},
\end{equation}
where from equations \eqref{eq:perturbedspacetimedefns} and \eqref{eq:thermosol}
\begin{equation}
 \tilde{h}_+^{\text{inc}}=\frac{2\delta C_{\text{inc}}}{C}-\frac{C'}{C\sqrt{B}}\int^r_0\frac{\delta B_{\text{inc}}(\bar{r})}{\sqrt{B(\bar{r})}}d\bar{r},
\end{equation}
and
\begin{equation}
\delta C_{\text{inc}}=\left(\left(\frac{\partial C}{\partial s}\right)_{\rho}+\frac{\rho}{s}\left(\frac{\partial C}{\partial\rho}\right)_s\right)\delta s\,,\quad \delta B_{\text{inc}}=\left(\left(\frac{\partial B}{\partial s}\right)_{\rho}+\frac{\rho}{s}\left(\frac{\partial B}{\partial\rho}\right)_s\right)\delta s\,.
\end{equation}

{}Given $\psi_{\text{inc}}(r)$, we can use the Wronskian method to obtain an integral expression for the other solution $\psi_{\text{dis}}(r)$ to the equation \eqref{eq:bulkviscosityeqn}
\begin{equation}
\begin{aligned}
\label{eq:psidisdefn}
\psi_{\text{dis}}(r)&\,= \delta\alpha\,\psi_{\text{inc}}(r)\int^r_0\frac{1}{\psi_{\text{inc}}(\bar{r})^2}\sqrt{\frac{B(\bar{r})}{D(\bar{r})}}\frac{1}{C(\bar{r})}\frac{C'(\bar{r})^2}{C(\bar{r})^2\Phi'(\bar{r})^2}d\bar{r},\\
\delta\alpha&\, =-\frac{\delta s}{2s}\left(\langle \delta T^{xx}\rangle+\langle\delta T^{yy}\rangle -2\delta p\right).
\end{aligned}
\end{equation}
Due to linearity of the equation \eqref{eq:bulkviscosityeqn}, $\delta\alpha$ is an arbitrary constant which we have relabelled according to its field theory interpretation using the holographic dictionary in Appendix \ref{app:holodictionaryperturbations}. As before, as the corrections to equations \eqref{eq:bulkradialconservationeqs} and \eqref{eq:bulktraceidentity} are $O(\partial^2)$, we can take $\delta\alpha$ to be a slowly varying function of $t$ and $x$ (a linear function in Fourier space).

{}Finally, given the solution $\psi_{\text{dis}}(r)$ in \eqref{eq:psidisdefn}, we can explicitly invert the equations of motion \eqref{eq:bulkradialconservationeqs} and \eqref{eq:bulktraceidentity} to obtain the following dissipative solution for the tilded variables
\begin{equation}
\begin{aligned}
\label{eq:completedissolution}
\tilde{a}_{t}^{\text{dis}}=&\,\delta \bar{A}_t-\delta\mu+\partial_t\left(\lambda-A_t\xi_t\right)+\frac{A_t'\Psi}{2\sqrt{B}}+\frac{A_t}{2}\int^r_0\psi_{\text{dis}}(\bar{r})\frac{d}{d\bar{r}}\left(\frac{C(\bar{r})^2\phi'(\bar{r})^2}{C'(\bar{r})^2}\right)d\bar{r}\\
&\,-\int^r_0\psi_{\text{dis}}(\bar{r})\frac{d}{d\bar{r}}\left(\frac{1}{\sqrt{B(\bar{r})D(\bar{r})}}\frac{d}{d\bar{r}}\left(\frac{\sqrt{B(\bar{r})D(\bar{r})}C(\bar{r})A_t(\bar{r})}{C'(\bar{r})}\right)\right)d\bar{r},\\
\tilde{h}_{tt}^{\text{dis}}=&\,-2\partial_t\xi_t+\frac{D'\Psi}{2D\sqrt{B}}-\frac{C^2{\phi'}^2}{{C'}^2}\psi_{\text{dis}}+\int^r_0\psi_{\text{dis}}(\bar{r})\frac{d}{d\bar{r}}\left(\frac{C(\bar{r})^2\phi'(\bar{r})^2}{C'(\bar{r})^2}\right)d\bar{r},\\
\tilde{h}_{+}^{\text{dis}}=&\,2\partial_x\xi_x+\frac{C'\Psi}{C\sqrt{B}},\quad\quad\quad\quad \tilde{\phi}^{\text{dis}}=\frac{\phi'\Psi}{2\sqrt{B}}-\frac{C\phi'}{C'}\psi_{\text{dis}},\\
\end{aligned}
\end{equation}
where
\begin{equation}
\Psi(r)=\int^r_0\sqrt{B(\bar{r})}\frac{C(\bar{r})^2\Phi'(\bar{r})^2}{C'(\bar{r})^2}\psi_{\text{dis}}(\bar{r})d\bar{r},
\end{equation}
and we have labelled all integration constants using the holographic dictionary.

\subsubsection{Equations of motion}

{}We have now constructed the required solution to the equations of motion \eqref{eq:bulkradialconservationeqs} and \eqref{eq:bulktraceidentity}. In order to obtain the near-equilibrium properties of the field theory we need to impose two conditions on this solution. First that it satisfies the remaining 4 equations of motion \eqref{eq:bulkwardconservations}: this ensures that the field theory obeys the local conservation equations for energy, momentum and $0$-form charge. Second that it is ingoing at the horizon of the black hole: this will fix the hydrodynamic constitutive relations of the field theory and the relaxation equation for the incoherent current.

{}\textit{Conservation equations}

{}We deal first with the remaining equations of motion \eqref{eq:bulkwardconservations}. The form of these equations is highly suggestive that these correspond to the Ward identities for conservation of energy, momentum and $0$-form charge, and this is indeed the case. Substituting in the general solution derived above, these equations yield four $r$-independent equations that are simply the linear perturbations of the local conservation equations
\begin{equation}
\begin{aligned}
\label{eq:hydroconservationequations}
\partial_\nu \langle T^{\mu\nu}\rangle =\bar{F}^{\mu\nu}\langle j_\nu\rangle,\quad\quad\quad\quad\quad\quad \partial_\mu \langle j^\mu\rangle =0.
\end{aligned}
\end{equation}

{}\textit{Constitutive relations}

{}Finally, we impose ingoing boundary conditions on our solution at the black hole horizon: specifically that the fundamental perturbations are regular at $r=r_0$ in the ingoing coordinate system $(v,x,y,r)$ where $dv=dt-\sqrt{\frac{B}{D}}dr$. To impose this, it is convenient to define the field theory fluid velocity as
\begin{equation}
\label{eq:hydroframedefn2}
\delta u^i\equiv\frac{\langle\delta T^{ti}\rangle}{\varepsilon+p}.
\end{equation}

{}To obtain the near-horizon expansions of the fundamental perturbations from the tilded ones, we must first specify the near-horizon behaviour of the pure gauge solutions. These are 
\begin{equation}
\begin{aligned}
\zeta_r(r\rightarrow r_0)=\frac{1}{4\sqrt{\pi T(r_0-r)}}\Biggl(&\,\Psi(r_0)-\int^{r_0}_0\left(\frac{\delta B}{\sqrt{B}}+\frac{\delta r_0}{\sqrt{4\pi T}(r_0-r)^{3/2}}\right)dr\\
&\,-\frac{2\delta r_0}{\sqrt{4\pi Tr_0}}\Biggr)+\ldots,\\
\end{aligned}
\end{equation}
and $\zeta_\mu,\Lambda\sim(r_0-r)^2$ as $r\rightarrow r_0$. 

Having done this, there are two different types of conditions that arise from imposing regularity in ingoing coordinates. The first come from removing the divergent $\sim (r_0-r)^{-1}$ terms in the near-horizon expansions of $\delta g_{rr}$, $\delta g_{ri}$ and $\delta A_r$ (in ingoing coordinates). These fix the large gauge transformations in terms of the sources and expectation values of field theory operators as follows
\begin{equation}
\begin{aligned}
\label{eq:puregaugesolutions}
\partial_t\lambda&\,=\delta\mu-\delta \bar{A}_t+\ldots,\quad\quad\quad\quad\quad\quad\quad \partial_t\xi_t=\frac{\delta T}{T}+\ldots,\\
sT\partial_t\xi_i&\,=\left(\varepsilon+p\right)\delta u^i -\mu\langle \delta j^i\rangle+\ldots,\\
\end{aligned}
\end{equation}
where the $\ldots$ indicate terms proportional to $\delta\alpha$ which will ultimately be subleading in the derivative expansion and so are omitted here for conciseness.
   
{}The second type of conditions impose relations between field theory sources and expectation values. In order to write these compactly, it is helpful to first define the three quantities
\begin{equation}
\begin{aligned}
\label{eq:holotransportcoeffs}
\eta&\,=\frac{s}{4\pi},\quad\quad \zeta=\frac{s}{4\pi}\left(s\left(\frac{\partial\Phi_0}{\partial s}\right)_{\rho}+\rho\left(\frac{\partial\Phi_0}{\partial\rho}\right)_{s}\right)^2,\quad\quad \sigma=\frac{s^2T^2Z(\Phi_0)}{(\varepsilon+p)^2},
\end{aligned}
\end{equation}
which will ultimately be the three first order transport coefficients in the hydrodynamic description of these states. $\Phi_0$ is defined in equation \eqref{eq:BlackHoleNHExpansion}. $\eta$ is the expression for the shear viscosity \cite{Kovtun:2004de}, $\zeta$ is the expression for the bulk viscosity \cite{Eling:2011ms,Eling:2011ct} and $\sigma$ is the expression for the `incoherent' conductivity \cite{Jain:2010ip,Davison:2015taa}. It is also helpful to define the three timescales
\begin{align}
\tau_{\eta}&\,=\int^{r_0}_0\left(\eta\sqrt{\frac{B}{D}}\frac{1}{C}-\frac{1}{4\pi T(r_0-r)}\right)dr, \label{taueta}\\
\tau_{\zeta}&\,=\int^{r_0}_0\left(\frac{s}{4\pi}\left(\frac{\psi_{\text{inc}}(r_0)}{\psi_{\text{inc}}(r)}\frac{C(r_0)\Phi'(r_0){C'(r)}}{C(r){\Phi'(r)}C'(r_0)}\right)^2\sqrt{\frac{B}{D}}\frac{1}{C}-\frac{1}{4\pi T(r_0-r)}\right)dr, \label{tauzeta}\\
\tau_{\sigma}&\,=\int^{r_0}_0\left(\frac{(\varepsilon+p)^2\sigma}{\rho^2}\frac{C}{D}\frac{d}{dr}\left(\frac{1}{sT+\rho A_t}\right)-\frac{1}{4\pi T(r_0-r)}\right)dr,\label{tausigma}
\end{align}
which will correspond respectively to the relaxation times of the difference and sum of the diagonal components of the stress tensor, and of the incoherent current.

{}With these definitions, we can examine the near-horizon expansions of our solutions $\delta g_{xx}-\delta g_{yy}$, $\delta g_{xy}$, $a_i$ and $\phi$, which in the $(t,x,y,r)$ coordinate system are (up to overall proportionality constants)
\begin{equation}
\begin{aligned}
\delta g_{xx}-\delta g_{yy}\rightarrow&\,\left(\langle\delta T^{xx}\rangle-\langle\delta T^{yy}\rangle\right)\left(-\frac{\log\left(1-\frac{r}{r_0}\right)}{4\pi T}+\tau_\eta\right)+2\eta\partial_x\xi_x+\ldots,
\end{aligned}
\end{equation}
\begin{equation}
\begin{aligned}
\delta g_{xy}\rightarrow \langle\delta T^{xy}\rangle\left(-\frac{\log\left(1-\frac{r}{r_0}\right)}{4\pi T}+\tau_\sigma\right)+\eta\partial_x\xi_y+\ldots,
\end{aligned}
\end{equation}
\begin{equation}
\begin{aligned}
a_i\rightarrow&\,\left(\langle\delta j^i\rangle-\rho\delta u^i\right)\left(-\frac{\log\left(1-\frac{r}{r_0}\right)}{4\pi T}+\tau_\sigma\right)+\frac{(\varepsilon+p)\sigma}{sT}\left(\delta \bar{A}_i+\partial_i\lambda+\mu\delta u^i\right)+\ldots,
\end{aligned}
\end{equation}
and
\begin{equation}
\begin{aligned}
\phi\rightarrow&\,\frac{\delta\alpha}{\delta\Phi_{0,\text{inc}}}\left(-\frac{\log\left(1-\frac{r}{r_0}\right)}{4\pi T}+\tau_\zeta\right)+\eta\delta\Phi_0+\ldots,
\end{aligned}
\end{equation}
where
\begin{equation}
 \delta\Phi_{0,\text{inc}}=\left(\left(\frac{\partial \Phi_0}{\partial s}\right)_{\rho}+\frac{\rho}{s}\left(\frac{\partial \Phi_0}{\partial\rho}\right)_s\right)\delta s.
\end{equation}
These solutions will be regular in ingoing coordinates provided that the constant and logarithmic terms in each expansion are related (up to terms of $O(\partial^2)$) by
\begin{equation}
\begin{aligned}
\label{eq:generalconstequations}
&\,\left(\tau_\eta\partial_t+1\right)\left(\langle\delta T^{xx}\rangle-\langle\delta T^{yy}\rangle\right)=-2\eta\partial_x \delta u^x+\ldots,\\
&\,\left(\tau_\eta\partial_t+1\right)\langle\delta T^{xy}\rangle=-\eta\partial_x\delta u^y+\ldots,\\
&\,\left(\tau_\zeta\partial_t+1\right)\left(\langle\delta T^{xx}\rangle+\langle\delta T^{yy}\rangle-2\delta p\right)=-2\zeta\partial_x\delta u^x+\ldots,\\
&\,\left(\tau_\sigma\partial_t+1\right)\left(\langle\delta j^i\rangle -\rho \delta u^i\right)=-\sigma\left(\partial_i\delta\mu-\frac{\mu}{T}\partial_i\delta T+\partial_t\delta \bar{A}_i-\partial_i\delta \bar{A}_t\right)+\ldots.
\end{aligned}
\end{equation}

{}Equations \eqref{eq:hydroconservationequations} and \eqref{eq:generalconstequations} are a closed set of hydrodynamic equations governing the evolution of the field theory's energy-momentum tensor and $0$-form $U(1)$ current in response to an external electromagnetic field. To compare with the theory of relativistic, non-conformal hydrodynamics we should express the \eqref{eq:generalconstequations} as constitutive relations: derivative expansions for the expectation values. To do this, we must move the $\tau\partial_t$ terms on the left hand side to the right hand side, where they produce $O(\partial^2)$ corrections. Ignoring such higher-derivative corrections, \eqref{eq:generalconstequations} agree exactly with the constitutive relations of first order relativistic, non-conformal hydrodynamics \cite{Kovtun:2012rj} to linear order in perturbations, completing the proof of non-conformal fluid-gravity duality to this order.

\subsection{Theory of relaxed hydrodynamics}

{}The calculations in this Section, until now, apply to general non-conformal equilibrium solutions of the action \eqref{eq:bulkEMDaction}. We are now going to specialise to the states described in Section \ref{sec:NonZeroDensityEMD} that at low temperatures flow to scaling geometries near the horizon. We will focus on the cases for which the $0$-form charge density operator is irrelevant in the infrared. We will show that in these cases first order hydrodynamics can be enhanced to a theory of relaxed hydrodynamics that additionally accounts for the slow relaxation of the incoherent current, and describe the resulting properties of these states.

{}As reviewed in Section \ref{sec:NonZeroDensityEMD}, in these cases the IR spacetimes have dynamical critical exponent $z=1$ and can be characterised by $\theta<0$, which controls the violation of hyperscaling in the state, and the dimension $\Delta_A<0$ of the irrelevant coupling induced by the $0$-form charge density in the IR theory. At low temperatures, $\theta$ controls the low temperature scaling of both thermodynamic observables \cite{Davison:2018ofp,Davison:2018nxm}
\begin{equation}
\begin{aligned}
&\,\rho\sim T^{0},\quad\quad s\sim T^{2-\theta},\quad\quad\chi_{\rho\rho}\sim T^{0},\quad\quad\chi_{ss}\sim T^{1-\theta},\quad\quad\chi_{s\rho}\sim T^{2-\theta},
\end{aligned}
\end{equation}
where $\chi$ denote the static susceptibilities of the charge density $\rho$ and entropy density $s$. Using equation \eqref{eq:holotransportcoeffs}, $\theta$ also governs the low temperature behaviour of the shear and bulk viscosities
\begin{equation}
\eta\sim T^{2-\theta}, \quad\quad\quad\quad\quad\quad \zeta\sim T^{2-\theta}.
\end{equation}
Although the coupling $\Delta_A$ is irrelevant, it is crucial for the low energy physics as it controls the low temperature scaling of the two transport coefficients appearing in the constitutive equation \eqref{eq:generalconstequations} for the $0$-form current \cite{Davison:2018ofp,Davison:2018nxm}
\begin{equation}
\sigma\sim T^{2-\theta+2\Delta_A},\quad\quad\quad\quad\quad\quad \tau_\sigma\sim T^{2\Delta_A-1}.
\end{equation}

{}The irrelevance of the IR operator induced by the charge density ensures that the relaxation timescale $T\tau_\sigma\sim T^{2\Delta_A}$ is parametrically large at small temperatures. As we did for the zero density cases, we can therefore enhance our hydrodynamic theory to incorporate this slow relaxation, by considering a generalised derivative expansion in which we take $\partial\sim T^{1-2\Delta_A}\ll T$. Since generic $O(\partial^2)$ corrections to the constitutive relations \eqref{eq:generalconstequations} (including the $\tau_\eta$ and $\tau_\zeta$ terms) are expected to be suppressed in this limit, the resulting theory of relaxed hydrodynamics is comprised of the local conservation equations \eqref{eq:hydroconservationequations}, the constitutive relations $\langle\delta T^{xx}\rangle=\delta p+\ldots$ and $\langle\delta T^{yy}\rangle=\delta p+\ldots$, and an equation governing the slow relaxation of $\langle\delta j^i\rangle -\rho \delta u^i$ over long timescales $\tau_\sigma\gg T^{-1}$
\begin{equation}
\label{eq:slowrelaxationJinceq}
\left(\tau_\sigma\partial_t+1\right)\left(\langle\delta j^i\rangle -\rho \delta u^i\right)=-\sigma\left(\partial_i\delta\mu-\frac{\mu}{T}\partial_i\delta T+\partial_t\delta \bar{A}_i-\partial_i\delta \bar{A}_t\right)+\ldots,
\end{equation}
where $\ldots$ denote subleading terms in the expansion.

{}The theory of relaxed hydrodynamics supports two different sets of collective modes that transport energy. To describe these, it is convenient to define
\begin{equation}
\begin{aligned}
\label{eq:incvariablesdefinition}
\delta\rho_{\text{inc}}=s^2T\delta\left(\frac{\rho}{s}\right),\quad\quad\quad\quad\quad\quad\delta j_{\text{inc}}^\mu=\left(\varepsilon+p\right)\langle \delta j^\mu\rangle-\rho\langle\delta T^{t\mu}\rangle,
\end{aligned}
\end{equation}
where the incoherent current density $\delta j^\mu_{\text{inc}}$ is the part of the $0$-form current that does not drag momentum, and the incoherent charge density $\delta \rho_{\text{inc}}$ is the corresponding density \cite{Davison:2015taa}. {From \eqref{eq:hydroframedefn2}, $\delta T^{ti}=(\varepsilon+p)\delta u^i$, equation \eqref{eq:slowrelaxationJinceq} shows} that it is the incoherent current that relaxes slowly in these states. We also define
\begin{equation}
\chi_{\rho_\inc \rho_\inc}=T^2\left(s^2\chi_{\rho\rho}-2s\rho\chi_{\rho s}+\rho^2\chi_{ss}\right),\quad\quad\quad\chi_{J_\inc J_\inc}=\frac{(\varepsilon+p)^2\sigma}{\tau_\sigma},
\end{equation}
where $\chi_{\rho_\inc \rho_\inc}$ is the susceptibility of the incoherent charge \cite{Davison:2018ofp,Davison:2018nxm} and $\chi_{J_\inc J_\inc}$ is the susceptibility of the incoherent current after taking the limit $\tau_\sigma\partial_t\gg1$. From the low temperature scalings given above,  $\chi_{\rho_\inc \rho_\inc}\sim\chi_{J_\inc J_\inc}\sim T^{3-\theta}$.

\begin{figure}[h]
\begin{center}
\includegraphics[scale=.32]{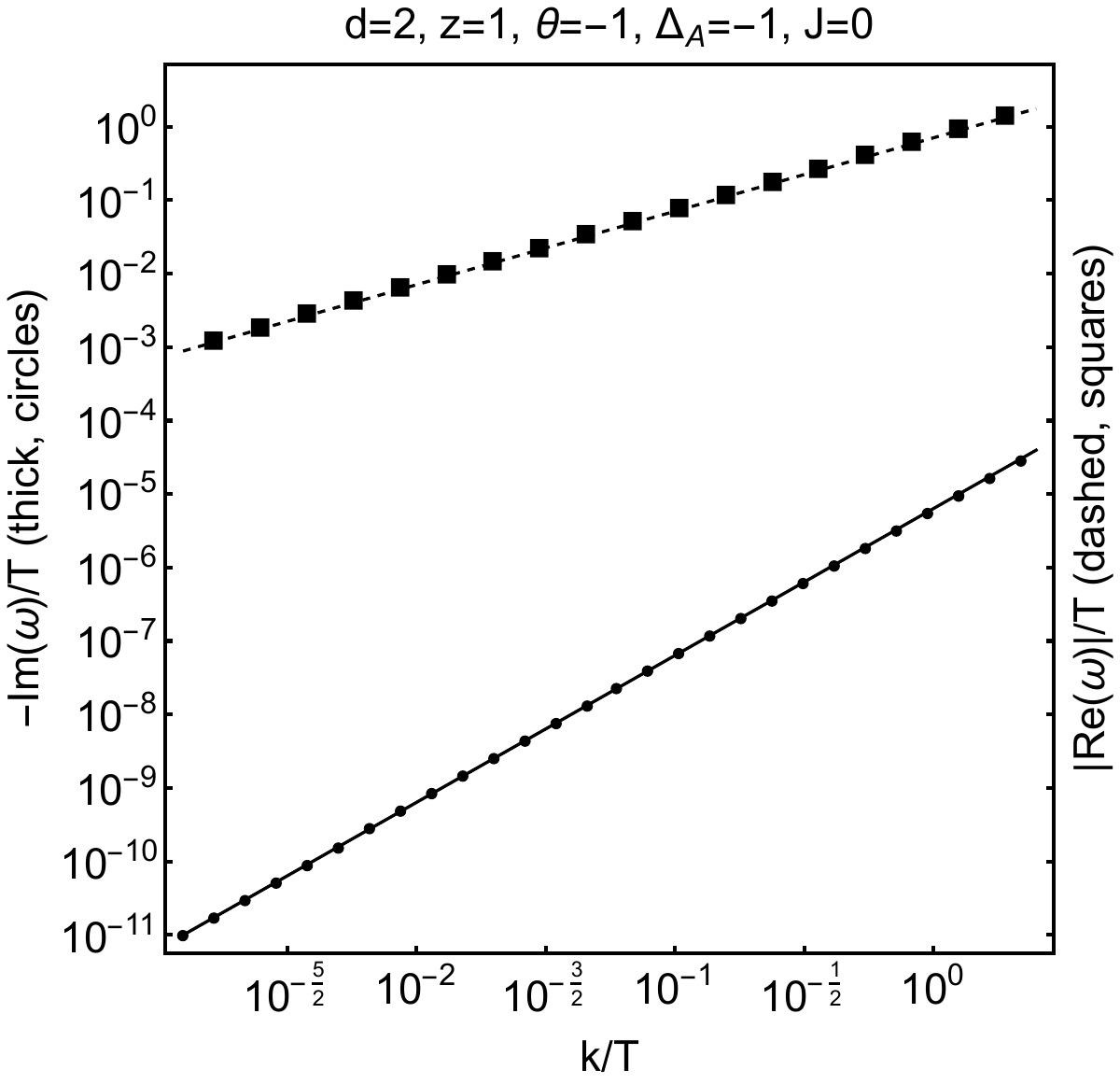}
\includegraphics[scale=.32]{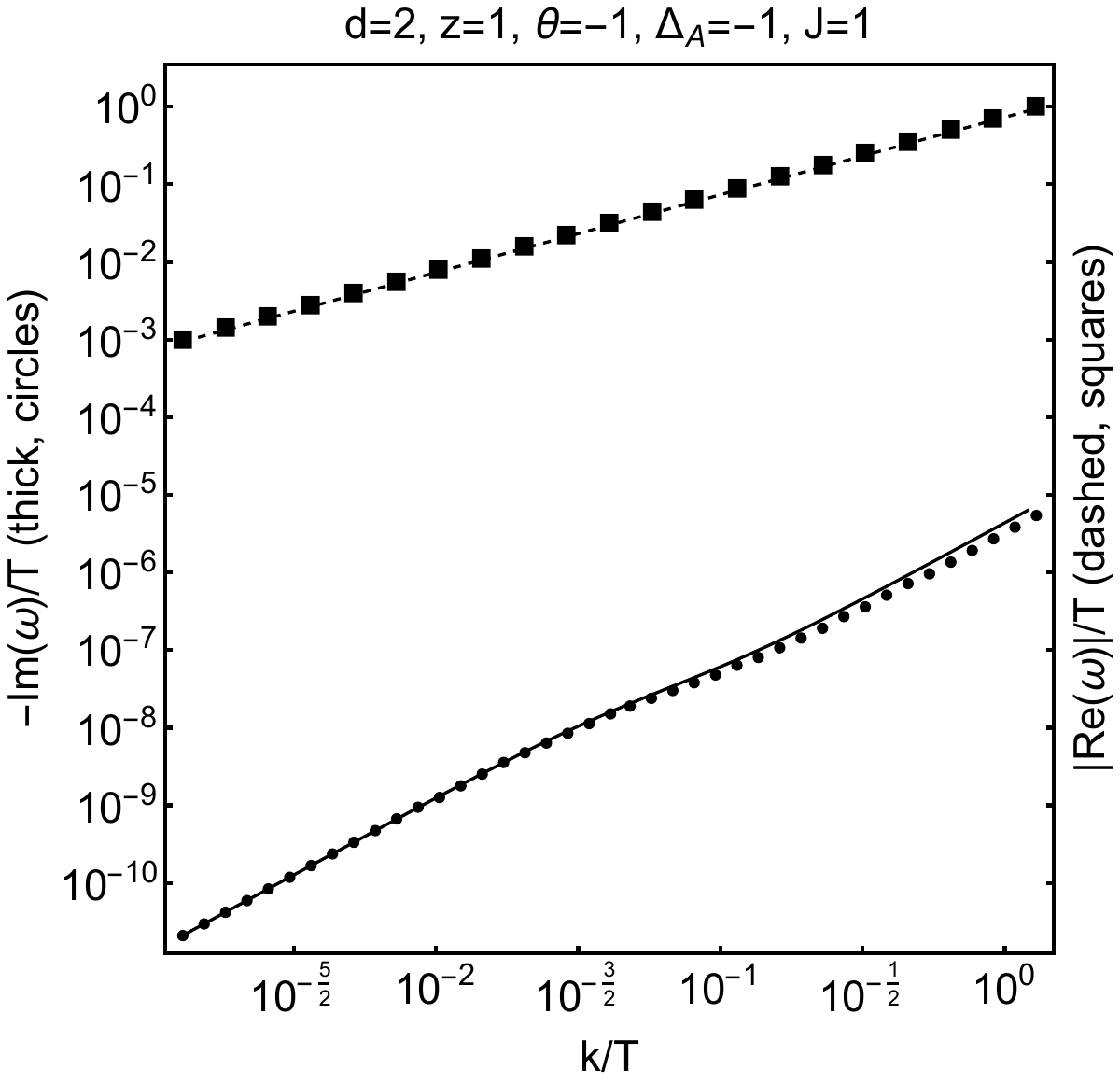}
\caption{\label{fig:SoundModeNonZeroDensity} Comparison between the dispersion relations obtained numerically (points) and the result \eqref{eq:nonzerodensitypropagatingmode} (lines) for states with $d=2$, $z=1$, $\theta=-1$, $\Delta_A=-1$. On the left, we present the results for a state with $J=0$ and $T/\mu = 0.016$ and on the right, the results for a state with $J=1$ and $T/\mu=0.015$. For $k\ll T$, the real part is parametrically larger than the imaginary part and agrees very well with the expression in \eqref{eq:nonzerodensitypropagatingmode} in both cases. Details of the numerical calculations can be found in Appendix \ref{app:numericaldetails_finitedensity}.}
\end{center}
\end{figure}

{}The first set of collective modes are the usual gapless sound waves of charged hydrodynamics, which are insensitive to the slow relaxation of the incoherent current. They have dispersion relations
\begin{equation}
\label{eq:nonzerodensitypropagatingmode}
\omega(k)=\pm v_s k+\ldots,\quad\quad\quad\quad v_s^{-2}=\chi_{\rho_\inc \rho_\inc}^{-1}T^2(\varepsilon+p)\left(\chi_{ss}\chi_{\rho\rho}-\chi_{\rho s}^2\right),
\end{equation}
where $\ldots$ denotes terms that are subleading in the expansion of relaxed hydrodynamics. From the low temperature scalings given above, $v_s\sim T^0$. In Figure \ref{fig:SoundModeNonZeroDensity} we show that for $k\ll T$ the expression \eqref{eq:nonzerodensitypropagatingmode} agrees with the real part of the dispersion relation obtained numerically in two examples. In this Figure we also compare the imaginary part of the dispersion relation with hydrodynamics in the following way. For cases where conformal symmetry is spontaneously broken ($J=0$), the incoherent sector decouples from energy and momentum fluctuations \cite{Davison:2015taa}, and so the leading dissipative corrections to the dispersion relation \eqref{eq:nonzerodensitypropagatingmode} are captured by the usual hydrodynamic expression \cite{Kovtun:2012rj} as shown in the left panel of Figure \ref{fig:SoundModeNonZeroDensity}. On the other hand, when conformal symmetry is explicitly broken ($J\neq0$, $\langle O_\Phi \rangle\ne0$) there is a small deviation from the hydrodynamic prediction (shown on the right panel of Figure \ref{fig:SoundModeNonZeroDensity}) that can be attributed to the long lifetime $\tau_\sigma$. It is straightforward to check that $\tau_\eta$ is not parametrically large in the types of states we are studying, {by checking whether the integral \eqref{taueta} is dominated by the infrared part of spacetime, \cite{Davison:2018nxm}}. We have not explicitly evaluated $\tau_\zeta$ {(the integral in \eqref{tauzeta} is somewhat more involved that those in \eqref{taueta} and \eqref{tausigma})}, but our numerical results in Figure \ref{fig:SoundModeNonZeroDensity} give evidence that $\tau_\zeta$ is not parametrically large either, {as a large $\tau_\zeta$ would imply additional long-lived modes in the spectrum compared to those we observe.}

\begin{figure}[h]
\begin{center}
\includegraphics[scale=.32]{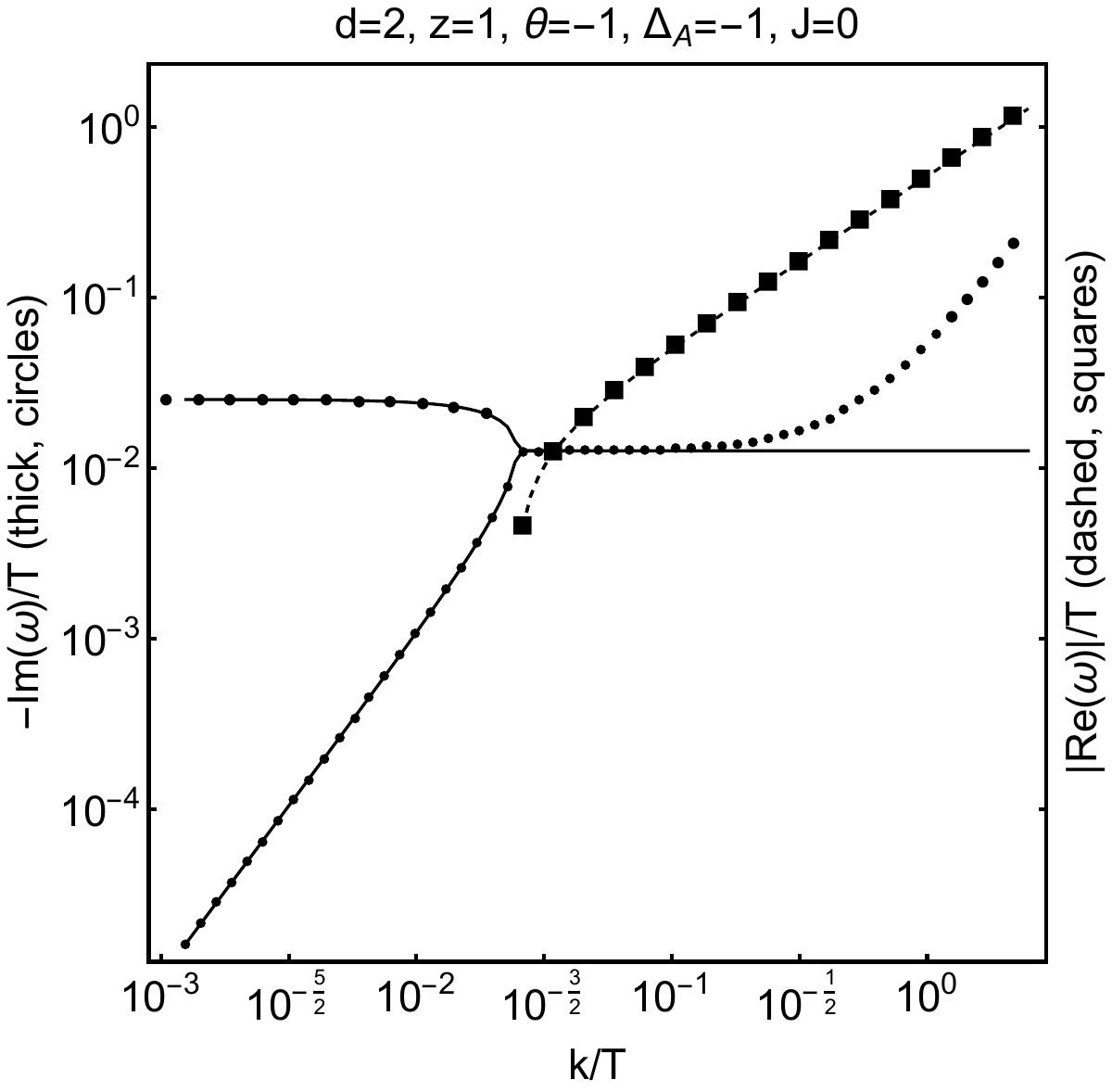}
\includegraphics[scale=.32]{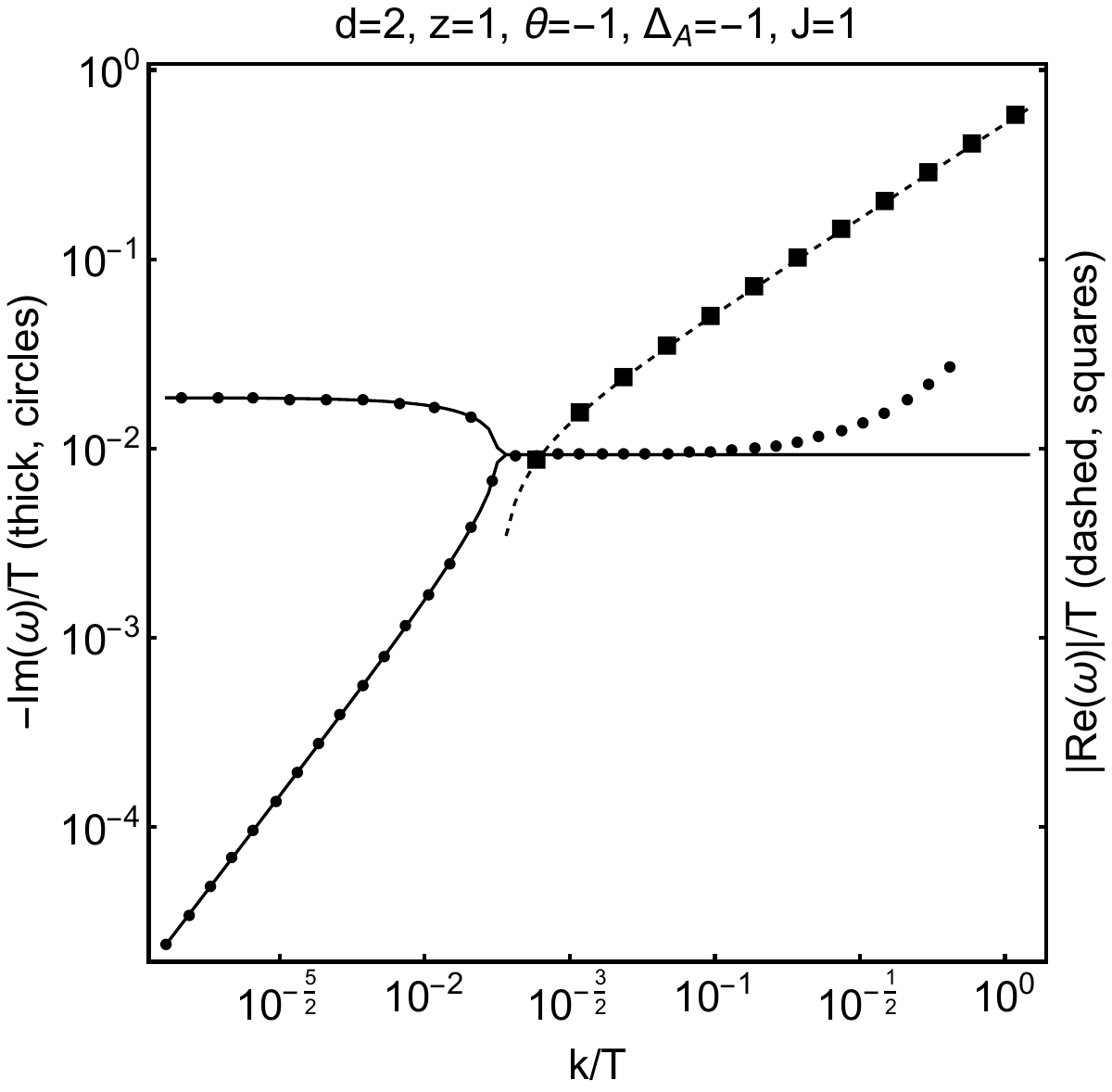}
\caption{\label{fig:IncoherentModesNonZeroDensity} Comparison between the dispersion relations obtained numerically (points) and the expression \eqref{eq:quadraticmodeeqnonzerodensity} (lines) for states with $d=2$, $z=1$, $\theta=-1$, $\Delta_A=-1$. On the left, we present results for a state with $J=0$ and $T/\mu = 0.016$ and on the right, a state with $J=1$ and $T/\mu=0.015$. Details of the numerical calculations can be found in Appendix \ref{app:numericaldetails_finitedensity}.}
\end{center}
\end{figure}

{}In contrast to this, the second set of collective modes are highly sensitive to the slow relaxation of the incoherent current: their dispersion relations are solutions to the quadratic equation
\begin{equation}
\label{eq:quadraticmodeeqnonzerodensity}
\omega^2+i\omega\tau_\sigma^{-1}-v_{\inc}^2k^2=0,\quad\quad\quad\quad\quad v_{\inc}^2=\frac{\chi_{J_\inc J_\inc}}{\chi_{\rho_\inc \rho_\inc}},
\end{equation}
which has the same form as the equation \eqref{eq:zerodensityhydrodispersions} found in the zero density case. At very low frequencies $\omega,k\ll\tau_\sigma^{-1}\ll T$, the two modes are diffusion of the incoherent charge (as in ordinary relativistic hydrodynamics) with dispersion relation 
\begin{equation}
\label{eq:nonzerodensitydiffusionmode}
\omega(k)=-iDk^2+\ldots,\quad\quad\quad\quad\quad\quad D=v_\inc^2 \tau_\sigma,
\end{equation}
as well as the slow relaxation of the incoherent current with dispersion relation $\omega(k)=-i\tau_\sigma^{-1}+\ldots$. From the low temperature scalings given above, the diffusivity $D\sim T^{2\Delta_A-1}$. This is parametrically large compared to expectations from IR dimensional analysis, due to its dependence on the irrelevant coupling \cite{Davison:2018ofp,Davison:2018nxm}. At $\omega,k\sim\tau_\sigma^{-1}$ there is a crossover. Beyond this (for $\tau_\sigma^{-1}\ll\omega,k\ll T$) the incoherent current is approximately conserved and so the modes propagate coherently with dispersion relations
\begin{equation}
\label{eq:incoherentpropdispersion}
\omega(k)=\pm v_{\inc} k-\frac{i}{2}\tau_\sigma^{-1}+\ldots.
\end{equation}
From the low temperature scalings given above, $v_{\inc}\sim T^0$. In fact, in the low temperature limit this speed is a universal quantity given by
\begin{equation}
v_{\inc}^2=\frac{c_{IR}^2}{d-\theta},
\end{equation}
where $c_{IR}$ is the IR speed of light \eqref{eq:cIRdefn}. This verifies the conjecture of \cite{Davison:2018ofp}, where the necessity of this crossover was motivated by causality considerations. In Figure \ref{fig:IncoherentModesNonZeroDensity} we show that the dispersion relations of the collective modes (obtained numerically) agree very well with the solutions to the equation \eqref{eq:quadraticmodeeqnonzerodensity} described above when $k\ll T$.

{}The theory of relaxed hydrodynamics can also be used to determine the thermoelectric conductivities for $\omega\ll T$. Each is composed of two parts \cite{Davison:2015taa}
\begin{equation}
\begin{aligned}
\label{eq:thermoelectricconds}
\sigma_{JJ}(\omega)&\,=\frac{\rho^2}{(\varepsilon+p)}\frac{i}{\omega}+\frac{1}{(\varepsilon+p)^2}\sigma_{\text{inc}}(\omega),\quad\quad \sigma_{J_QJ}(\omega)=\frac{\rho sT}{(\varepsilon+p)}\frac{i}{\omega}-\frac{\mu}{(\varepsilon+p)^2}\sigma_{\text{inc}}(\omega),\\
\sigma_{J_QJ_Q}(\omega)&\,=\frac{s^2T^2}{(\varepsilon+p)}\frac{i}{\omega}+\frac{\mu^2}{(\varepsilon+p)^2}\sigma_{\text{inc}}(\omega),
\end{aligned}
\end{equation}
where $J$ and $J_Q$ here denote the electric and heat currents respectively. The divergent $i/\omega$ contribution is due to the overlap of these currents with the (conserved) momentum, while the finite contribution from $\sigma_{\inc}(\omega)$ is due to processes that do not drag momentum. The latter contribution is sensitive to the slow relaxation of the incoherent current and has the Drude-like form
\begin{equation}
\label{eq:Drudeequationincoherentsec4}
\sigma_{\inc}(\omega)=\frac{\sigma_{\inc}^{dc}}{1-i\omega\tau_{\sigma}},\quad\quad\quad\quad\quad\quad \sigma_{\inc}^{dc}=\chi_{J_\inc J_\inc}\tau_\sigma.
\end{equation}
We emphasize that this Drude-like form has nothing to do with translational invariance and the associated conserved momentum: it arises due to the overlap of the transport currents with the long-lived incoherent current. The incoherent conductivity can be isolated in a single transport measurement by measuring the open circuit thermal conductivity $\kappa=\sigma_{\text{inc}}/(T\rho^2)$.

{}Using the temperature scalings above, the incoherent dc conductivity scales as $\sigma_{\inc}^{dc}\sim T^{2-\theta+2\Delta_A}$ at low temperatures. This scaling is a result of a competition between two effects: the long lifetime $\tau_\sigma$ of the incoherent current tends to produce a divergent conductivity, while the small $\chi_{J_\inc J_\inc}$ tends to produce a vanishing one. Which effect wins depends on the precise values of $d-\theta$ and $\Delta_A$ for the state. The temperature scaling of the dc electrical conductivity is the same as that of $\sigma_{\inc}^{dc}$ and so -- unlike the zero density cases of Section \ref{sec:ZeroDensityHolography} -- it is not possible to determine whether the state supports an anomalously long-lived mode simply from knowledge of its dc conductivity.

\subsection{Comparison with superfluid hydrodynamics}

{}For the zero density states, we described in Section \ref{sec:ZeroDensityEffectiveTheory} how the relaxed hydrodynamics is that of a phase-relaxed superfluid with frozen temperature and velocity fluctuations. For non-zero density states, fluctuations of the $0$-form charge couple to those of temperature and velocity. In this Section we will show that the relaxed hydrodynamics of the non-zero density states is different than that of a phase-relaxed superfluid (including temperature and velocity fluctuations), primarily because of the different ways in which the emergent long-lived mode couples to temperature fluctuations.

{}Consider first the limit $\tau_\sigma^{-1}\ll\omega,k\ll T$, in which the relaxation term on the right hand side of equation \eqref{eq:slowrelaxationJinceq} can be neglected. In this limit, there are two sets of collective modes reminiscent of the sound and second sound modes of a superfluid. However, an analysis of the thermoelectric conductivities in this limit brings to light the key difference from a superfluid. While the supercurrent in a superfluid transports charge but not heat, it is apparent from the definition \eqref{eq:incvariablesdefinition} that the long-lived incoherent current transports both charge and heat. In fact, at low temperatures it is the transport of the incoherent current that dominates the thermal conductivity in our states \eqref{eq:thermoelectricconds}. 

{}This observation is a consequence of the fact that the emergent long-lived mode couples to temperature perturbations differently in our non-zero density states than it does in a superfluid. This discrepancy can be seen clearly by formulating the hydrodynamic theory in this limit in the context of hydrodynamics of a higher-form symmetry. The equation for the incoherent current \eqref{eq:slowrelaxationJinceq} becomes
\begin{equation}
\label{eq:anomalouseqnonzerodensity}
d\star \mathcal{K}=-\bar{F},
\end{equation}
after defining the two-form $\mathcal{K}$ via
\begin{equation}
\label{eq:higherformconstnonzerodensity}
\left(\star\,\mathcal{K}\right)^t=\delta\mu-\frac{\mu}{T}\delta T,\quad\quad\quad\quad\quad \left(\star\,\mathcal{K}\right)^i=(\varepsilon+p)\chi_{J_\inc J_\inc}^{-1}\delta j_\inc ^i.
\end{equation}
Equation \eqref{eq:anomalouseqnonzerodensity} is an anomalous $2$-form conservation law where the anomaly is a mixed anomaly with the $0$-form $U(1)$ symmetry, as $\bar{F}=d\bar{A}$ is the field strength of the external source for the $1$-form current. This equation, in additional to the local conservation laws \eqref{eq:hydroconservationequations} of $0$-form charge, energy and momentum are the equations of motion of the higher-form formulation of relativistic superfluid hydrodynamics, including the temperature and velocity fluctuations \cite{Delacretaz:2019brr}. However, the constitutive relation \eqref{eq:higherformconstnonzerodensity} for $(\star\,\mathcal{K})^t$ has an additional $\delta T$ term not present in the superfluid case. In the higher-form formulation of the hydrodynamic equations, it is this extra term that is responsible for the fact that the emergent long-lived mode transports heat in addition to charge.

{}The result that $\left(\star\,\mathcal{K}\right)^t=\delta\mu$ for small perturbations around a state with no normal or superfluid velocity (i.e.~with zero density of the $2$-form charge) follows as a limit of a more general constraint: that the local version of the second law of thermodynamics is valid for states with an arbitrary density of $2$-form charge \cite{Delacretaz:2019brr}. This suggests that the complete non-zero density holographic theories in fact do not possess an emergent $2$-form symmetry, but that the anomalous conservation equation \eqref{eq:anomalouseqnonzerodensity} is an artifact of the particular states and limit that we are studying. We will return to this in the next subsection.

{}Nevertheless, we can interpret the properties of these particular states in terms of the anomalous conservation equation \eqref{eq:anomalouseqnonzerodensity}. In the hydrodynamic limit under consideration the temperature is always non-zero and thus the relaxation time $\tau_\sigma$ is always finite. As in the zero density cases, this finite relaxation time corresponds to an explicit violation of the conservation law \eqref{eq:anomalouseqnonzerodensity} which has important effects at low energies $\omega\lesssim\tau^{-1}$: the crossover from the propagating modes \eqref{eq:nonzerodensitypropagatingmode} to diffusion \eqref{eq:nonzerodensitydiffusionmode} and the broadening of the incoherent contribution to the thermoelectric conductivities \eqref{eq:thermoelectricconds} into a Drude-like form \eqref{eq:Drudeequationincoherentsec4}. The dc component of the incoherent conductivity in equation \eqref{eq:Drudeequationincoherentsec4} can be expressed as $\sigma_\inc^{dc}=\chi_{\mathcal{K}\mathcal{K}}^{-1}\tau_\sigma$ where $\chi_{\mathcal{K}\mathcal{K}}=(\varepsilon+p)^2\chi_{J_\inc J_\inc}^{-1}$ is the static susceptibility of $(\star\,\mathcal{K})^i$. Phrased in this way, the competition in $\sigma^{dc}_\inc$ between the diverging $\tau_\sigma$ and vanishing $\chi_{\mathcal{K}\mathcal{K}}^{-1}$ at low temperatures is an example of the phenomenon of critical drag described in \cite{Else_2021,Else:2021dhh}.

{}Although the relaxed hydrodynamics of our non-zero density states are in general different from those of a superfluid, there is a closer connection for cases where the equation of state is $\varepsilon=2p$. Due to the trace Ward identity \eqref{eq:eqmTraceIdentity}, states in which the scalar operator $O_\Phi$ breaks the UV conformal symmetry spontaneously ($J=0$, $\langle O_\Phi\rangle\ne0$) will automatically have this equation of state. For these cases, the variables $\left\{\delta\rho_{\text{inc}},\delta j_{\text{inc}}^\mu\right\}$ and $\left\{\delta p,\delta u^\mu\right\}$ decouple \cite{Davison:2015taa} with the former obeying the equations
\begin{equation}
\begin{aligned}
\label{eq:quasihydroeqs-spontaneouscase}
&\,\partial_t\delta\rho_{\text{inc}}+\partial_i\delta j_{\text{inc}}^i=0,\quad\,\,\, \partial_t\delta j_{\text{inc}}^i+\frac{\chi_{J_\inc J_\inc}}{\chi_{\rho_\inc \rho_\inc}}\partial_i\delta\rho_{\text{inc}}=-\frac{\delta j_{\text{inc}}^i}{\tau_\sigma}+\chi_{J_\inc J_\inc}\delta \bar{F}_{it}^{\text{inc}},
\end{aligned}
\end{equation}
where $\delta \bar{F}^{\text{inc}}_{\mu\nu}=(\varepsilon+p)^{-1}\delta \bar{F}_{\mu\nu}$ is the field strength of the external source for the incoherent current density, in the absence of a source for momentum density. These equations have the same structure as the corresponding equations \eqref{eq:zerodensityhydroeq1} in the zero density case and so the dynamics of incoherent charge mirror those of charge in a superfluid with frozen temperature and velocity fluctuations. 

\subsection{Zero temperature dynamics}

{}Thus far we have studied the non-zero density states only in the hydrodynamic limit $\omega,k\ll T$, where they exhibit a novel propagating mode \eqref{eq:incoherentpropdispersion} associated to an approximate higher-form conservation equation. We will now go beyond the hydrodynamic regime and explore the fate of this mode at $T=0$. Unlike in the zero density case, we will show that the propagating mode \eqref{eq:incoherentpropdispersion} exists only within the hydrodynamic regime.

\begin{figure}[t]
\begin{center}
\includegraphics[scale=.35]{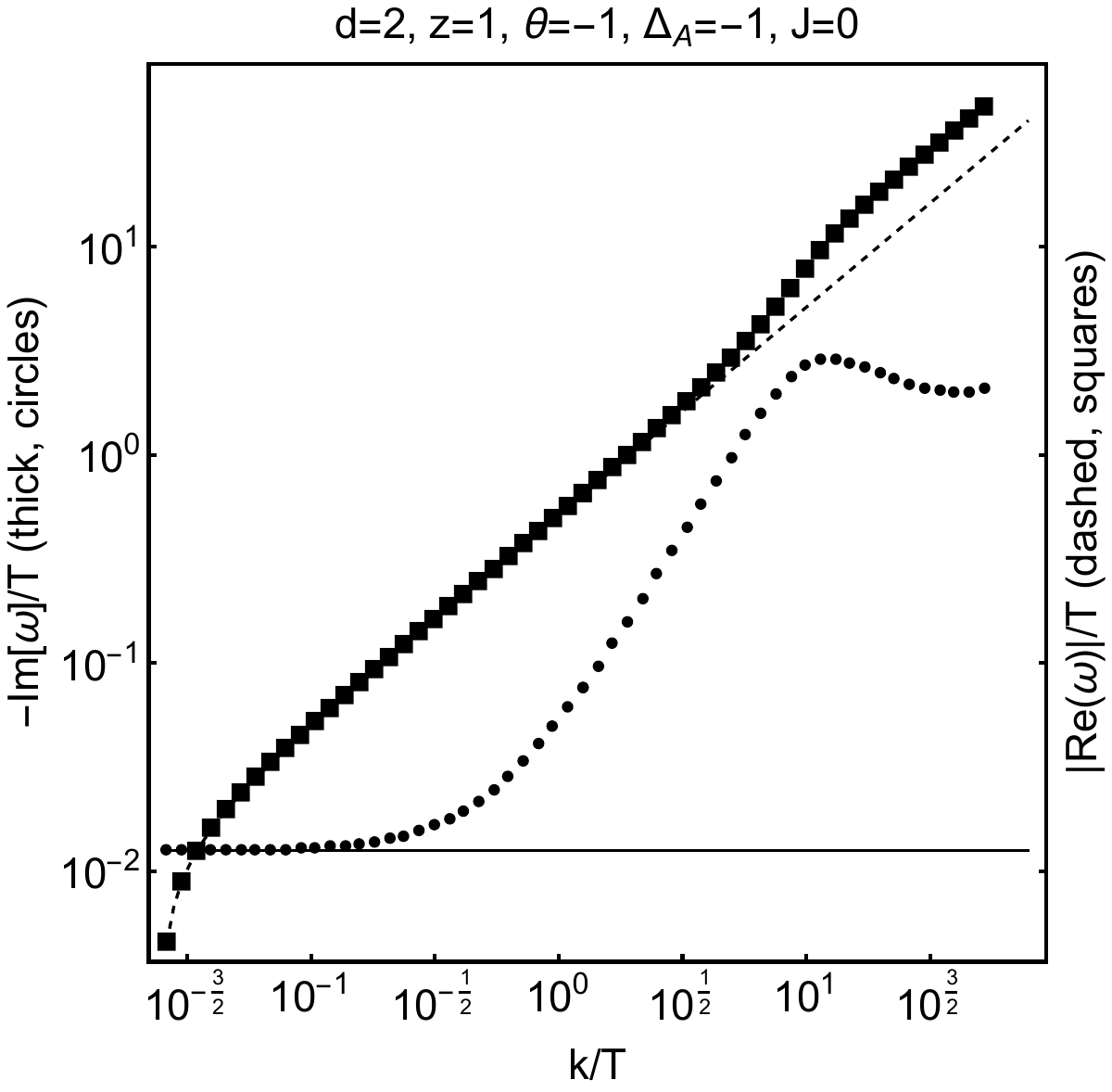}
\includegraphics[scale=.35]{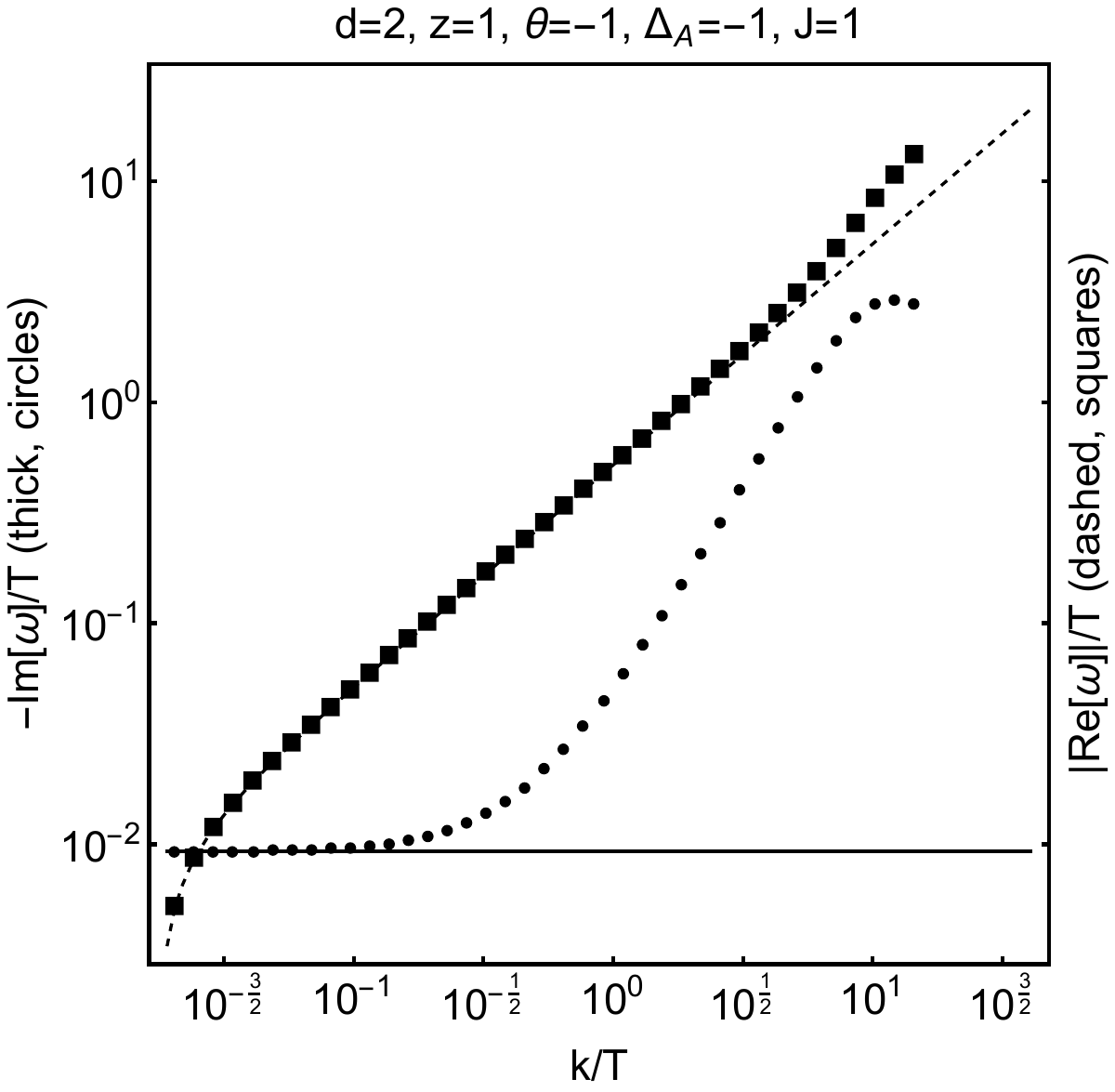}
\caption{\label{fig:IncoherentModesHydroExit} Comparison of the dispersion relations obtained numerically (dots) to the hydrodynamic result \eqref{eq:incoherentpropdispersion} (lines) for two examples of non-zero density states with $d=2$, $z=1$, $\theta = -1$, and $\Delta_{A} = -1$. On the left, we present results for a state with $J=0$ at $T/\mu =0.016$ and on the right, a state with $J=1$ at $T/\mu =0.015$. {The deviation between the numerical results and the hydrodynamic approximation when $k\gtrsim T$ signals that the regime of validity of the hydrodynamic result does not extend to $T\ll k\ll \mu$}. Details of the numerical calculations can be found in Appendix \ref{app:numericaldetails_finitedensity}.}
\end{center}
\end{figure}

{}We access the properties of the $T=0$ collective modes using numerical computations (see Appendix \ref{app:numericaldetails} for more details). More precisely, we determine the collective modes within the regime $\tau_\sigma^{-1}\ll T\ll\omega,k\ll\mu$. In the limit $T/\mu\rightarrow0$, we expect that the modes in this regime connect smoothly onto the low energy modes of the $T=0$ state. Although the $T/\mu\rightarrow0$ limit of the hydrodynamic dispersion relation \eqref{eq:incoherentpropdispersion} is a gapless mode propagating with speed $v_\inc$, this result only strictly applies in a different limit: $\tau_\sigma^{-1}\ll\omega,k\ll T\ll\mu$. To determine whether the propagating mode survives at $T=0$ we therefore need to examine its fate as the hydrodynamic regime is exited at $\omega,k\sim T$.

{}In Figure \ref{fig:IncoherentModesHydroExit} we show the dispersion relations across these two regimes, for two examples of non-zero density states. In both cases there is a qualitative change in the properties of the propagating collective mode at $k\sim T$, signalling that the $T=0$ state does not support a gapless mode with speed $v_\inc$. These plots can be contrasted with the corresponding plots for the zero density case (Figure \ref{intermediatescaling_fig}), where the propagating mode does survive in the $T=0$ state.

\begin{figure}[t]
\begin{center}
\includegraphics[scale=.45]{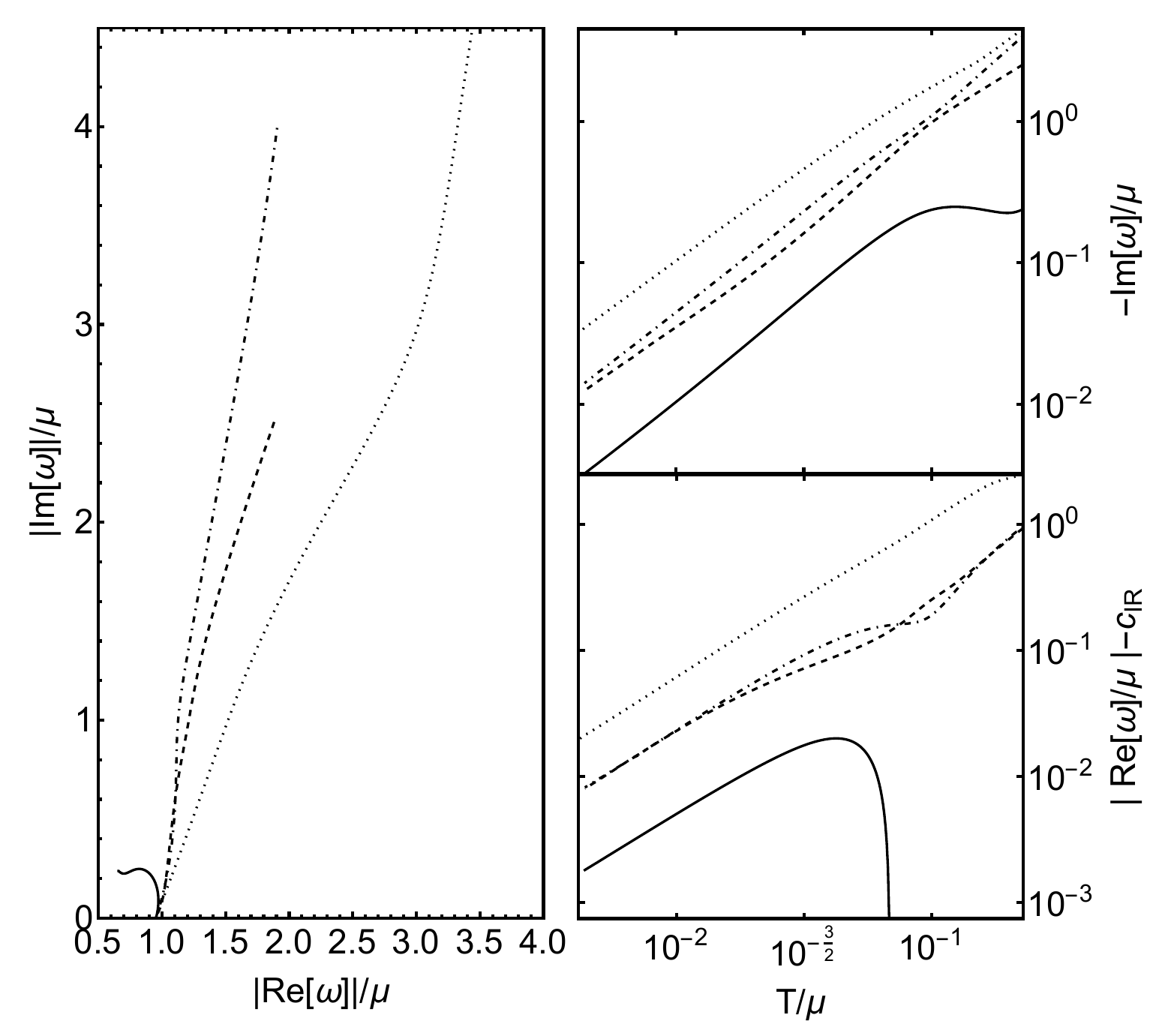}
\caption{\label{branchcutfigure} The formation of a branch cut as $T\to 0$ is demonstrated by the motion of four long-lived collective modes at fixed $k/\mu = 1$ as a function of temperature from $T/\mu = 0.228$ to $T/\mu = 0.004$ in a phase with $z=1$, $\theta=-1$, $\Delta_A = -5/2$, and $J=0$. As shown on the left, the modes coalesce as $T\to 0$ at $\omega/\mu = c_{IR}(k/\mu)$. On the right, we demonstrate that each mode has a form $\omega_i/\mu = \pm c_{IR} (k/\mu) + \pm \gamma_i^R (T/\mu)^{\beta_i^R}-i\gamma_i^I (T/\mu)^{\beta_i^I}$ for some positive powers $\beta_i^R,\beta_i^I$ and positive coefficients $\gamma_i^R, \gamma_i^I$ which are functions of $k/\mu$. For numerical reasons we consider only a finite number of modes, but because the $\gamma_i^R$ and $\gamma_i^I$ are all distinct, our results suggest that were we to include an infinite number of modes they would form branch cuts as $T\rightarrow0$ with branch points at $\omega = \pm c_{IR}k$. Details of the numerical calculations can be found in Appendix \ref{app:numericaldetails_finitedensity}.}
\end{center}
\end{figure}

{}Although we do not yet have an explicit description of the dynamics of the non-zero density states in terms of Goldstone-like fields (analogous to the zero density case described in Sections \ref{sec:GoldstoneActionFiniteT} to \ref{sec:ZeroDensityZeroTemperature}), we can anticipate qualitatively what its properties will be. At non-zero $T$ there will be a Goldstone-like field (a bulk Wilson line-like object) interacting with horizon degrees of freedom, and for energies $\tau_\sigma^{-1}\ll\omega\ll T$ this interaction will be weak leading to the new long-lived propagating mode. However, at $\omega\sim T$ the interaction with the horizon degrees of freedom will become important and change the nature of the mode. At $T=0$, a description in terms of the Goldstone-like field interacting with the infrared quantum critical degrees of freedom represented by the scaling spacetime \eqref{eq:IRmetricFiniteT} near the horizon will be dominated by the latter degrees of freedom, with no trace of the propagating mode from the Goldstone-like field remaining. For motivation for this, notice that the Goldstone-like terms in the action at zero density \eqref{eq:renormalisedgoldstoneaction} are proportional to $\chi_{\rho\rho}\sim T^0$ and $\chi_{JJ}\sim T^0$. At non-zero density we expect the analogous role to be played by $\chi_{\rho_\inc \rho_\inc}$ and $\chi_{J_\inc J_\inc}$. These both vanish as $T\rightarrow0$ and so the zero temperature dynamics should be dominated by the infrared quantum critical degrees of freedom.

{}Based on this reasoning, we expect that the $T=0$ response functions of non-zero density states should exhibit branch points at $\omega=\pm c_{IR} k$ that are characteristic of the infrared quantum critical degrees of freedom \eqref{eq:IRmetricFiniteT} with $z=1$ (for example, consider the result \eqref{eq:zerodensitybranchcut} or the case of perturbations of the energy-momentum tensor in Schwarzschild-AdS \cite{Gubser:1998bc}). In Figure \ref{branchcutfigure} we show how the longest-lived collective modes evolve as the temperature is lowered: they migrate towards the real axis and bunch up towards the points $\omega=\pm c_{IR}k$, as expected for the finite temperature resolution of a zero temperature branch cut \cite{Edalati:2010hk,Edalati:2010pn,Moore:2018mma,Grozdanov:2018gfx}. We leave the detailed analytic construction of the Goldstone-like theory and its coupling to the quantum critical degrees of freedom to future work.

\begin{acknowledgments}

We are grateful to Dominic Else for very helpful discussions and to Yongjun Ahn, Keun-Young Kim and Nick Poovuttikul for earlier collaboration on related topics. The work of R.~D.~is supported by the STFC Ernest Rutherford Grant ST/R004455/1. The work of B.~G.~is supported by the European Research Council (ERC) under the European Union's Horizon 2020 research and innovation programme (grant agreement No758759). The work of E.~M.~was supported in part by NSERC and in part by the European Research Council (ERC) under the European Union's Horizon 2020 research and innovation programme (grant agreement No758759). We all gratefully acknowledge Nordita's hospitality during the Nordita program `Recent developments in strongly-correlated quantum matter' where part of this work was carried out.

 \end{acknowledgments}

\appendix

\section{Details of zero density calculations}

\subsection{Bound on propagating velocity}
\label{app:zerodensityspeedbound}

{}In this Appendix, we show that the speed $v$ of the propagating mode for the states in Section \ref{sec:zerodensityholoquasihydro} never exceeds the speed of light. For cases with $z=1$ we prove a lower bound on the speed that ensures that the propagating mode at zero temperature in these states is stable.

{}In the low $T$ limit where the relaxed hydrodynamic theory of Section \ref{sec:ZeroDensityEffectiveTheory} applies, $v^2=\chi_{JJ}/\chi_{\rho\rho}$ is $T$-independent. Using the expressions for these quantities given in Section \ref{sec:zerodensityholoquasihydro} gives
\begin{equation}
\begin{aligned}
\label{eq:zeroTspeed}
v^2(T\rightarrow0)&\,=\frac{\int^{r_0}_0\frac{D}{C}\sqrt{\frac{B}{D}}\frac{dr}{C^{d/2-1}Z}}{\int^{r_0}_0\sqrt{\frac{B}{D}}\frac{dr}{C^{d/2-1}Z}}+\ldots,
\end{aligned}
\end{equation}
where $\ldots$ denotes terms that vanish as $T\rightarrow0$.

{}To bound the speed, we examine $D/C$ of the black hole solution. Using the Einstein equation
\begin{equation}
\frac{d}{dr}\left(\frac{D}{C}\right)=-\frac{\sqrt{BD}(sT+\rho A_t)}{C^{1+d/2}},
\end{equation}
following from the action \eqref{eq:bulkEMDaction}, this can be expressed as
\begin{equation}
\label{eq:DCintegral}
\frac{D(r)}{C(r)}=\text{constant}-\int^r \frac{(sT+\rho A_t(\bar{r}))\sqrt{B(\bar{r})D(\bar{r})}}{C(\bar{r})^{1+d/2}}d\bar{r}.
\end{equation}
We assume that $B$, $C$, $D$ are positive between the horizon and the boundary, as is $sT+\rho A_t$. This last condition can be understood as follows. At low $T$ the momentum susceptibility is $sT+\rho A_t(r=0)\rightarrow \rho A_t(r=0)$ and so we require $\rho A_t(r=0)>0$ for stability. Since we also must have $A_t(r=r_0)=0$ then provided $A_t$ has no zeroes between the horizon and the boundary, we will also have $sT+\rho A_t(r)>0$. 

{}We can now use positivity of the integral in \eqref{eq:DCintegral}  to bound $D/C$ and therefore $v^2$. The most general bound is to fix the integration constant by using $D/C\rightarrow1$ as $r\rightarrow0$:
\begin{equation}
\frac{D(r)}{C(r)}=1-\int^r_0 \frac{(sT+\rho A_t(\bar{r}))\sqrt{B(\bar{r})D(\bar{r})}}{C(\bar{r})^{1+d/2}}d\bar{r}\leq1,
\end{equation}
where the inequality follows because the integrand is positive. Substituting this into equation \eqref{eq:zeroTspeed} yields the upper bound
\begin{equation}
v^2(T\rightarrow0)\leq 1.
\end{equation}

{}For states that flow to IR fixed points with $z=1$, we can also obtain a non-trivial lower bound. At zero temperature, the metric \eqref{eq:IRmetricFiniteT} has $D/C\rightarrow c_{IR}^2$ as $r\rightarrow\infty$. Therefore, from equation \eqref{eq:DCintegral}
\begin{equation}
\frac{D(r)}{C(r)}=c_{IR}^2+\int^\infty_r \frac{\rho A_t(\bar{r})\sqrt{B(\bar{r})D(\bar{r})}}{C(\bar{r})^{1+d/2}}d\bar{r}\geq c_{IR}^2,\quad\quad\quad z=1,
\end{equation}
where the integrand is positive between the horizon and boundary given the assumptions above. Substituting this into the equation \eqref{eq:zeroTspeed} yields a lower bound on the speed
\begin{equation}
c_{IR}^2\leq v^2(T=0),\quad\quad\quad\quad z=1.
\end{equation}
This inequality ensures that imaginary term in the dispersion relation \eqref{eq:z1dispersionzeroT} is always negative and therefore the corresponding mode is stable.

\subsection{The IR Green's function}

\label{app:zerodensityIRGreens}

{}In this Section we will show how to compute the $T=0$ IR Green's function $G_{\text{IR}}(\omega,k)$ used in Section \ref{sec:ZeroDensityZeroTemperature}, which controls the conductivity and the attenuation of the propagating mode at zero temperature. Using the expression \eqref{eq:IRoperatordefns} for the IR operator $\mathcal{O}^x$, we have that
\begin{equation}
\label{eq:IRGreensdefnappendix}
G_{\text{IR}}(\omega,k)=\frac{Z_0L_x^{d-2}L_t}{\tilde{L}}\left(\Delta_\chi+2(z-1)\right)\left.\frac{a_x^{(1)}(\omega,k)}{a_x^{(0)}(\omega,k)}\right|_{a_t^{(0)}=0},
\end{equation}
where $a_\mu$ are the coefficients in the expansion \eqref{eq:ZeroTIRgaugefieldexps} of the ingoing solutions to the equation of motion in the IR region. 

{}Since the IR spacetime has a scaling symmetry, it is convenient to work with the rescaled radial coordinate $\tilde{R}$, frequency $\tilde{\omega}$ and wavenumber $\tilde{k}$ defined by
\begin{equation}
\tilde{R}\equiv \tilde{\omega}^{1/z}\left(\frac{R}{L}\right),\quad\quad\quad\quad\tilde{\omega}\equiv\frac{\tilde{L}\omega}{L_t},\quad\quad\quad\quad\tilde{k}\equiv\frac{\tilde{L} k}{L_x}.
\end{equation}
As explained in the main text, when $z>1$ it is $G_{\text{IR}}(\omega,0)$ that controls the leading term in the attenuation of the $T=0$ propagating mode. For the case $k=0$, the field $A_x$ has the equation of motion
\begin{equation}
\frac{d}{d\tilde{R}}\left(\tilde{R}^{1-(\Delta_\chi+2(z-1))}\frac{dA_x}{d\tilde{R}}\right)+\tilde{R}^{1-\Delta_\chi}A_x=0,
\end{equation}
whose ingoing solution is
\begin{equation}
A_x\propto \tilde{R}^{\frac{1}{z}(\Delta_\chi+2(z-1))}H_{\frac{\Delta_\chi+2(z-1)}{2z}}\left(\frac{\tilde{R}^z}{z}\right),
\end{equation}
where $H$ is the Hankel function of the first kind. By expanding this function as $\tilde{R}\rightarrow0$, we then use equation \eqref{eq:IRGreensdefnappendix} to obtain the IR Green's function
\begin{equation}
\begin{aligned}
G_{\text{IR}}(\omega,k=0)=&\frac{L_x^{d-2}L_tZ_0}{\tilde{L}}2\pi i z\left(\frac{\tilde{\omega}}{2z}\right)^{\frac{\Delta_\chi+2(z-1)}{z}}\frac{1+i\cot\left(\frac{\pi}{2z}\left(\Delta_\chi+2(z-1)\right)\right)}{\Gamma\left(\frac{\Delta_\chi+2(z-1)}{2z}\right)^{2}}.
\end{aligned}
\end{equation} 
For the purposes of extracting the dispersion relation of the propagating mode from \eqref{eq:zeroTdispersionlocus}, it is more convenient to work with the inverse of the IR Green's function. Using Gamma function identities, this may be written as
\begin{equation}
\begin{aligned}
G_{\text{IR}}^{-1}(\omega,k=0)=-\frac{i\pi}{L_x^{d-2}Z_0}\frac{1}{\omega}\left(\frac{\tilde{\omega}}{2z}\right)^{-\frac{\Delta_\chi+z-2}{z}}\frac{1-i\cot\left(\frac{\pi}{2z}\left(\Delta_\chi+2(z-1)\right)\right)}{\Gamma\left(\frac{2-\Delta_\chi}{2z}\right)^2},
\end{aligned}
\end{equation}
which gives the dispersion relation \eqref{eq:zne1dispersionzeroT} quoted in the main text. Similarly, the coefficient $h(0)^{-1}$ that appears in the correction to the zero temperature conductivity \eqref{eq:zeroTconductivitycorrection} can be written
\begin{equation}
h(0)^{-1}=-\frac{i\pi}{L_x^{d-2}Z_0}\left(\frac{\tilde{L}}{2zL_t}\right)^{-\frac{\Delta_\chi+z-2}{z}}\frac{1-i\cot\left(\frac{\pi}{2z}\left(\Delta_\chi+2(z-1)\right)\right)}{\Gamma\left(\frac{2-\Delta_\chi}{2z}\right)^2}.
\end{equation}

{}To determine the dispersion relation of the mode when $z=1$, we require the full IR Green's function $G_{\text{IR}}(\omega,k)$ rather than just its $k=0$ limit. In order to determine this it is convenient to work in with the gauge-invariant bulk field $\mathcal{E}(R,\omega,k)\equiv\omega a_x+ka_t$. Writing the $R\rightarrow0$ expansion of the ingoing solutions for this field as $\mathcal{E}\rightarrow\mathcal{E}^{(0)}+\mathcal{E}^{(1)}(R/L)^{\Delta_\chi}+\ldots$, the IR Green's function for $z=1$ can then be expressed as
\begin{equation}
\label{eq:z1IRgreensrel}
G_{\text{IR}}(\omega,k)=\frac{L_x^{d-2}L_tZ_0\Delta_\chi}{\tilde{L}}\frac{\omega^2}{\omega^2-c_{IR}^2k^2}\frac{\mathcal{E}^{(1)}}{\mathcal{E}^{(0)}}.
\end{equation}
The field $\mathcal{E}$ has the equation of motion
\begin{equation}
\frac{d}{d\bar{R}}\left(\bar{R}^{1-\Delta_\chi}\frac{d\mathcal{E}}{d\bar{R}}\right)+\bar{R}^{1-\Delta_\chi}\mathcal{E}=0,
\end{equation}
where we have defined a new rescaled radial coordinate by $\bar{R}\equiv \sqrt{\tilde{\omega}^2-\tilde{k}^2} (R/L)$. The ingoing solution to this equation is $\mathcal{E}\propto H_{\Delta_\chi/2}(\bar{R})\bar{R}^{\Delta_\chi/2}$ where $H$ is again the Hankel function of the first kind. Expanding this near the boundary and then substituting into the expression \eqref{eq:z1IRgreensrel} gives the result
\begin{equation}
\label{eq:zerodensitybranchcut}
G_{\text{IR}}(\omega,k)=\frac{L_x^{d-2}Z_0\tilde{L}}{L_t}\left(\tilde{\omega}^2-\tilde{k}^2\right)^{\frac{\Delta_\chi}{2}-1}\omega^2\frac{i\pi\left(1+i\cot\left(\frac{\pi\Delta_\chi}{2}\right)\right)}{2^{\Delta_\chi-1}\Gamma\left(\frac{\Delta_\chi}{2}\right)^2}.
\end{equation}
Again, for the purposes of obtaining the dispersion relation of the propagating mode, it is more convenient to work with the inverse of the IR Green's function. Using Gamma function identities this can be expressed as
\begin{equation}
G_{\text{IR}}^{-1}(\omega,k)=-\frac{i}{\omega^2}\frac{L_t}{\tilde{L}Z_0L_x^{d-2}}\left(\frac{\tilde{\omega}^2-\tilde{k}^2}{4}\right)^{1-\frac{\Delta_\chi}{2}}\frac{2\pi}{\Gamma\left(1-\frac{\Delta_\chi}{2}\right)^2}\left(1-i\cot\left(\frac{\pi\Delta_\chi}{2}\right)\right).
\end{equation}
Substituting this into equation \eqref{eq:zeroTdispersionlocus} and solving for the leading real and imaginary parts of the dispersion relation at small wavenumbers yields the result \eqref{eq:z1dispersionzeroT} quoted in the main text.

\section{Details of non-zero density calculations}
\label{app:nonzerodensityappendix}

\subsection{Linear perturbation equations}
\label{app:perturbationeoms}

{}There are 15 field equations for the tilded variables defined in equation \eqref{eq:splitfields} in the main text. To obtain the theory of relaxed hydrodynamics, we will solve these equations in a derivative expansion. To the order that we work, we can neglect terms of $O(\partial^2)$ in the equations of motion. To this order, it is tedious but straightforward to verify that not all 15 field equations are independent: the equation of motion arising from varying the scalar field is automatically satisfied provided the other 14 equations are. We will therefore not need to explicitly consider it. The remaining 14 equations can be split into two sets. Up to terms of $O(\partial^2)$, the first set are the 10 equations of motion
\begin{equation}
\begin{aligned}
\label{eq:dressedEoMs}
0=&\frac{d}{dr}\left[C\sqrt{\frac{D}{B}}\left(\sqrt{\frac{D}{C}}\frac{d}{dr}\left(\frac{\tilde{h}_+}{\sqrt{D/C}}\right)+\Phi'\tilde{\phi}\right)+A_t\left(\frac{CZ}{\sqrt{BD}}\left(\tilde{a}_t'-\frac{1}{2}A_t'(\tilde{h}_{tt}-\tilde{h}_+)\right)+\frac{A_t'\dot{Z}}{Z}\tilde{\phi}\right)\right],\\
0=&\frac{d}{dr}\left[C\sqrt{\frac{D}{B}}\left(\tilde{h}_{tt}'+\frac{1}{2}\tilde{h}_{+}'+\Phi'\tilde{\phi}\right)-\frac{1}{2}(sT+\rho A_t)\tilde{h}_{tt}+\rho\tilde{a}_t\right],\\
0=&\frac{d}{dr}\left[\sqrt{\frac{D^3}{B}}\frac{d}{dr}\left(\frac{\tilde{h}_{it}}{D/C}\right)+A_t\sqrt{\frac{D}{B}}Z\left(\tilde{a}_i'+\frac{A_t'}{D/C}\tilde{h}_{it}\right)\right],\\
0=&\frac{d}{dr}\left[C\sqrt{\frac{D}{B}}\tilde{h}_-'\right],\\
0=&\frac{d}{dr}\left[C\sqrt{\frac{D}{B}}\tilde{h}_{xy}'\right],\\
0=&\frac{d}{dr}\left[\frac{CZ}{\sqrt{BD}}\left(\tilde{a}_t'-\frac{1}{2}A_t'(\tilde{h}_{tt}-\tilde{h}_+)\right)+\frac{A_t'\dot{Z}}{Z}\tilde{\phi}\right],\\
0=&\frac{d}{dr}\left[\sqrt{\frac{D}{B}}Z\left(\tilde{a}_i'+\frac{A_t'}{(D/C)}\tilde{h}_{it}\right)\right],\\
0=&\frac{C'}{C}\left(\tilde{h}_{tt}'+\frac{1}{2}\tilde{h}_+'+\Phi'\tilde{\phi}\right)+\frac{D'}{2D}\left(\tilde{h}_+'+\Phi'\tilde{\phi}\right)+\frac{ZA_t'}{D}\left(\tilde{a}_t'-\frac{A_t'}{2}\tilde{h}_{tt}+\frac{\dot{Z}A_t'}{Z}\tilde{\phi}\right)\\
&+2\sqrt{\frac{D}{C}}\frac{d}{dr}\left(\frac{C'}{\sqrt{BCD}}\right)\frac{d}{dr}\left(\frac{\sqrt{B}\tilde{\phi}}{\Phi'}\right),
\end{aligned}
\end{equation}
where primes denote derivatives with respect to $r$ and dots denote derivatives with respect to $\Phi$. These are all $O(\partial^0)$ at leading order and in terms of the alternative gauge-invariant perturbation variables \eqref{eq:bulkhydrofieldsdefn} become the radial conservation equations \eqref{eq:bulkradialconservationeqs} and the trace Ward identity \eqref{eq:bulktraceidentity}. Up to terms of $O(\partial^2)$, the second set are the 4 equations of motion
\begin{equation}
\begin{aligned}
\label{eq:bulkconseqns}
0=&\partial_t\left[\sqrt{\frac{D}{C}}\frac{d}{dr}\left(\frac{\tilde{h}_+}{\sqrt{D/C}}\right)+\Phi'\tilde{\phi}\right]-\partial_x\left[\frac{D}{C}\frac{d}{dr}\left(\frac{\tilde{h}_{xt}}{D/C}\right)\right],\\
0=&\partial_t\left[\tilde{h}_{xt}'+\frac{ZA_t'}{C}\tilde{a}_x\right]+\frac{D}{C}\partial_x\left[\frac{1}{\sqrt{D/C}}\frac{d}{dr}\left(\sqrt{\frac{D}{C}}\tilde{h}_{tt}\right)+\frac{1}{2}\tilde{h}_+'+\Phi'\tilde{\phi}-\frac{ZA_t'}{D}\tilde{a}_t-\frac{1}{2}\tilde{h}_-'\right],\\
0=&\partial_t\left[\tilde{h}_{yt}'+\frac{ZA_t'}{C}\tilde{a}_y\right]-\partial_x\left[\frac{D}{C}\tilde{h}_{xy}'\right],\\
0=&\partial_t\left[\tilde{a}_t'+\frac{A_t'}{2}\left(\tilde{h}_+-\tilde{h}_{tt}\right)+\frac{\dot{Z}A_t'}{Z}\tilde{\phi}\right]-\partial_x\left[\frac{D}{C}\tilde{a}_x'+A_t'\tilde{h}_{xt}\right].
\end{aligned}
\end{equation}
These are all $O(\partial)$ at leading order. In terms of the alternative variables \eqref{eq:bulkhydrofieldsdefn} these become the Ward identities \eqref{eq:bulkwardconservations} for the local conservation of energy, momentum and $0$-form $U(1)$ charge.

\subsection{Holographic dictionary for linear perturbations}
\label{app:holodictionaryperturbations}

{}To obtain the near-equilibrium field theory properties captured by the perturbed black hole solution constructed in Section \ref{sec:NonZeroDensityHydroSection}. we use the holographic dictionary described in \cite{Caldarelli:2016nni}. For our solution, perturbations of the field theory metric $\eta_{\mu\nu}$ and of the scalar source $J$ vanish, while there is a non-zero external $U(1)$ potential $\delta\bar{A}_\mu$. The perturbations in the expectation values of the energy momentum tensor $T^{\mu\nu}$ and $U(1)$ current density $j^\mu$ operators around their equilibrium values \eqref{eq:equilibriumexpectationvalues} are related to the $r\rightarrow0$ expansions of the gravitational solutions in Fefferman-Graham coordinates by
\begin{equation}
\begin{aligned}
&\,h_{tt}\rightarrow -\frac{1}{3}\left(\langle\delta T^{tt}\rangle+J\langle\delta O_\phi\rangle\right)r^3+\ldots,\quad h_{it}\rightarrow -\frac{1}{3}\langle\delta T^{ti}\rangle r^3+\ldots,\\
&\,h_{xx}\rightarrow\frac{1}{3}\left(\langle\delta T^{xx}\rangle-J\langle\delta O_\phi\rangle\right)r^3+\ldots,\quad\,\, h_{yy}\rightarrow\frac{1}{3}\left(\langle\delta T^{yy}\rangle-J\langle\delta O_\phi\rangle\right)r^3+\ldots, \\
&\,h_{xy}\rightarrow\frac{1}{3}\langle\delta T^{xy}\rangle r^3+\ldots,\quad\quad\quad\quad\quad\quad\quad\,\,\,\, \phi \rightarrow \langle\delta O_\phi\rangle r^2+\ldots,\\
&\,a_{t}\rightarrow\delta\bar{A}_t-\langle\delta j^t\rangle r+\ldots,\quad\quad\quad\quad\quad\quad\quad a_{i}\rightarrow \delta\bar{A}_i+\langle\delta j^i\rangle r+\ldots,
\end{aligned}
\end{equation}
where we have neglected terms of $O(\partial^2)$. All perturbations are functions of the spacetime coordinates $(t,x)$.

\section{Numerical calculations}
\label{app:numericaldetails}

{}In this Appendix we give further details on how the numerical results in the main text were calculated. The black hole spacetimes were obtained by solving the Einstein equations using pseudospectral methods over a Chebyshev grid \cite{Dias:2015nua}. To obtain sufficiently accurate results for the perturbations, it was often necessary to go beyond machine precision (which is efficiently performed within Mathematica). Grid sizes varied, though a grid size of $N=200$ points was typically sufficient for our desired accuracy (except for the low temperature results in Figure \ref{branchcutfigure}, where a grid size of $N=700$ points was used). We used the potentials
\begin{equation}
V(\Phi) = 6\cosh\left(\frac{\Phi}{\sqrt{3}}\right),\quad\quad\quad\quad\quad Z(\Phi) = \cosh\left(\frac{a_\phi \Phi}{\sqrt{3}}\right),
\end{equation}
where $a_\phi$ is a constant. As $\Phi\to 0$, $V(\Phi\rightarrow0) \rightarrow 6 + \Phi^2 +...$ implying that the scalar field is dual to an operator with UV dimension $\Delta_\Phi = 2$. If $a_\phi \geq 5$, the solution will flow in the IR to a spacetime with $z=1, \theta = -1$, whereas for $a_\phi <5$, the solution flows to a spacetime with $z>1, \theta<-1$.

{}We used a metric ansatz of the form
\begin{equation}
ds^2 = \frac{1}{z^2}\left[-\chi(z)^2f(z)\left(1-\frac{z}{z_0}\right)dt^2 + \frac{dz^2}{f(z)\left(1-\frac{z}{z_0}\right)} + \chi(z)^2(dx^2+dy^2)\right],
\end{equation}
with $z$ the radial coordinate chosen so that the UV boundary is at $z=0$ and the horizon at $z=z_0$. The UV conditions imposed were
\begin{equation}
f(0) = 1, \quad\quad\quad \chi(0) = 1, \quad\quad\quad \Phi'(0) = J,\quad\quad\quad A_t(0) = \mu,
\end{equation}
while at the horizon we imposed regularity of the metric and $\Phi$ as well as $A_t(1) = 0$.

{}We then moved to a dimensionless radial coordinate $z= z_0 y$ so that our grid was defined over the range $y \in [0,1]$. $z_0$ still appears in the equations of motion and in the formulae for the thermodynamic functions, for instance the temperature
\begin{equation}
T = \frac{\chi f}{4\pi z_0}\biggr|_{y=1}.
\end{equation}
This expression was used to move back and forth from the fixed $(\mu, z_0)$ ensemble and the fixed $(\mu, T)$ ensemble. Thermodynamic derivatives, such as $\chi_{\rho\rho}, \chi_{\rho s},$ and $\chi_{ss}$ were calculated by discretizing $\mu$ and $z_0$ over Chebyshev grids and using the appropriate pseudospectral derivative matrices. 

\subsection{Zero density cases}
\label{app:numericaldetails_zerodensity}

{}As explained in the main text, zero density refers to states having zero density of an additional $0$-form $U(1)$ charge, although they have a non-zero density of the original $0$-form $U(1)$ charge which supports the background and allows for an IR scaling region. The black hole spacetimes were obtained as explained just above (see also \cite{Goldstein:2009cv,Charmousis:2010zz,Gouteraux:2011ce,Huijse:2011ef,Gouteraux:2012yr,Gouteraux:2013oca,Davison:2018nxm}), while the dynamics we are interested in is that of an additional probe Maxwell field in these spacetimes. We took the gauge coupling of the additional Maxwell field (labelled $B_\mu$ in this appendix) to be of the form
\begin{equation}
Z_{B}(\Phi) = \cosh\left(\frac{b_\Phi \Phi}{\sqrt{3}}\right),
\end{equation}
so that in the scaling region in the IR
\begin{equation}
Z_{B} \to Z_0\left(\frac{R}{L}\right)^{d-\theta-z+2\frac{\theta}{d}-\Delta_\chi}.
\end{equation}
The value of $\Delta_\chi$ was then tuned by changing the value of $b_\Phi$. 

{}The collective excitations of the $0$-form $U(1)$ density of this additional charge correspond to quasinormal modes of the probe Maxwell field and can be efficiently found by using gauge-invariant variables \cite{Kovtun:2005ev}. After Fourier transforming
\begin{equation}
B_\mu = b_\mu(z) e^{-i\omega t+ ikx},
\end{equation}
we can form the gauge-invariant field $Z_0=kb_t+\omega b_x$. The quasinormal mode boundary conditions appropriate for this field are that it vanishes at the UV boundary and is ingoing at the black hole horizon. Using these boundary conditions, we solved for the frequencies of the quasinormal modes as a generalised eigenvalue problem over the Chebyshev grid \cite{Dias:2015nua}. For Figure \ref{branchcutfigure} this method was too time intensive at the grid size necessary for accurate low temperature results, so we instead promoted $\omega(k)$ to a function of the radial coordinate and used relaxation methods to simultaneously solve for the quasinormal modes and their frequencies.

\subsection{Finite density cases}
\label{app:numericaldetails_finitedensity}

{}The background spacetimes at non-zero density were obtained exactly as for the zero density case explained above. However, in these cases the gauge field fluctuations couple to those of the other fields. Nevertheless, the numerical method for obtaining the quasinormal mode spectrum corresponding to collective excitations of the $0$-form $U(1)$ charge density was identical. After Fourier transforming (with frequency $\omega$, and wavenumber $k$ in the $x$-direction), we define a set of gauge-invariant perturbations
\begin{equation}
\begin{aligned}
Z_1 &= k a_t + \omega a_x + k\rho h_{yy}\frac{\sqrt{DB}}{ZC'},\\
Z_2 &= -k^2\frac{D}{C}h_{tt}+2\omega k h_{tx} + \omega^2 h_{xx} +\left(k^2\frac{D'}{C'}-\omega^2\right)h_{yy},\\
Z_3 & = \phi - \frac{C\Phi'}{C'}h_{yy},
\end{aligned}
\end{equation}
where a prime means a derivative with respect to the radial coordinate of the metric. The equations of motion for these fields are a set of three coupled second order differential equations. These are easily obtained but their explicit form is uninformative and very lengthy and so we do not present them here. The appropriate quasinormal mode boundary conditions are then $Z_i(0) = 0$ and ingoing boundary conditions on $Z_i$ at the horizon. Using these boundary conditions, we numerically solved for the quasinormal modes as a generalised eigenvalue problem for the frequencies $\omega$.

\bibliographystyle{jhep.bst}
\bibliography{biblio}
\end{document}